\documentclass[11pt]{article}
\usepackage[T1]{fontenc}
\usepackage{mathtools}
\usepackage{amssymb}
\usepackage[mathscr]{euscript}
\usepackage{dsfont}
\usepackage{slashed}
\usepackage{fullpage}
\usepackage{cite}
\usepackage{currvita}
\usepackage[colorlinks=true,linkcolor=blue,citecolor=magenta,linktocpage=true]{hyperref}
\usepackage{braket}
\usepackage{titlesec}
\titleformat*{\section}{\normalsize\bfseries}
\titleformat*{\subsection}{\normalsize\bfseries}
\titleformat*{\subsubsection}{\normalsize\bfseries}
\makeatletter
\renewcommand{\@dotsep}{1000}

\def\be#1\ee{\begin{align}#1\end{align}}
\def\nn{\nonumber}
\def\q{\qquad}
\def\f{\frac}
\def\eps{\varepsilon}

\def\lb{\big\lbrace}
\def\rb{\big\rbrace}

\def\ip{\lrcorner\,}
\def\ipp{\ip\!\!\!\ip}

\def\st{{\star}}
\def\as{{*}}

\def\Tr{\mathrm{Tr}}

\def\De{\mathrm{D}}
\def\de{\mathrm{d}}
\def\i{\mathrm{i}}
\def\k{\mathrm{k}}

\def\G{\mathcal{G}}
\def\H{\mathcal{H}}

\def\J{\mathcal{J}}

\def\O{\mathcal{O}}
\def\Q{\mathcal{Q}}

\def\S{\mathcal{S}}

\def\Z{\mathcal{Z}}
\numberwithin{equation}{section}

\def\sr{\text{\tiny{R}}}
\def\sl{\text{\tiny{L}}}

\begin{document}

\title{\Large{\textbf{\sffamily Extended actions, dynamics of edge modes,\\ and entanglement entropy}}}
\author{\sffamily Marc Geiller$^1$ \& Puttarak Jai-akson$^{2,3}$}
\date{\small{\textit{
$^1$Univ Lyon, ENS de Lyon, Univ Claude Bernard Lyon 1,\\ CNRS, Laboratoire de Physique, UMR 5672, F-69342 Lyon, France\\
$^2$Perimeter Institute for Theoretical Physics,\\ 31 Caroline Street North, Waterloo, Ontario, Canada N2L 2Y5\\
$^3$Department of Physics and Astronomy, University of Waterloo,\\ 200 University Avenue West, Waterloo, ON, N2L 3G1, Canada\\}}}

\maketitle

\begin{abstract}
In this work we propose a simple and systematic framework for including edge modes in gauge theories on manifolds with boundaries. We argue that this is necessary in order to achieve the factorizability of the path integral, the Hilbert space and the phase space, and that it explains how edge modes acquire a boundary dynamics and can contribute to observables such as the entanglement entropy. Our construction starts with a boundary action containing edge modes. In the case of Maxwell theory for example this is equivalent to coupling the gauge field to boundary sources in order to be able to factorize the theory between subregions. We then introduce a new variational principle which produces a systematic boundary contribution to the symplectic structure, and thereby provides a covariant realization of the extended phase space constructions which have appeared previously in the literature. When considering the path integral for the extended bulk + boundary action, integrating out the bulk degrees of freedom with chosen boundary conditions produces a residual boundary dynamics for the edge modes, in agreement with recent observations concerning the contribution of edge modes to the entanglement entropy. We put our proposal to the test with the familiar examples of Chern--Simons and BF theory, and show that it leads to consistent results. This therefore leads us to conjecture that this mechanism is generically true for any gauge theory, which can therefore all be expected to posses a boundary dynamics. We expect to be able to eventually apply this formalism to gravitational theories.
\end{abstract}

\thispagestyle{empty}
\newpage
\setcounter{page}{1}

\hrule
\tableofcontents
\vspace{0.7cm}
\hrule


\section{Motivations}

Gauge theories defined on manifolds with boundaries, be they asymptotic or at finite distance, exhibit emergent boundary degrees of freedom, sometimes referred to as edge modes. This well-established fact has been investigated in depth mostly for theories with no propagating bulk degrees of freedom, such as 3-dimensional gravity and Chern--Simons theory, where the edge modes posses an explicit boundary dynamics and encode physical properties of e.g. black holes \cite{Banados:1998ta,Carlip:2005zn,Afshar:2017okz} or condensed matter systems \cite{Wen:2004ym,Wen:1992vi,Tong:2016kpv}. There is however no doubt that edge modes also encode non-trivial physics in theories with local degrees of freedom, such as e.g. 4-dimensional gravity and electromagnetism, although in this context no systematic investigation of the nature of the edge modes and of their boundary dynamics has been carried out, and the literature remains scarce (relevant references will be cited below). This stems from the obvious difficulties and subtleties in identifying the edge modes, and in disentangling their dynamics from that of the bulk degrees of freedom.

Recently, many new results revealing important insights into the role of edge modes have been obtained, both at finite distance and at infinity. At finite distance, there have been successful definitions of quasi-local holography through the path integral for quantum gravity \cite{Asante:2018kfo,Asante:2019ndj,Dittrich:2018xuk,Dittrich:2017hnl,Dittrich:2017rvb,Riello:2018anu,Goeller:2019zpz}, leading to boundary models which can be thought of as capturing the dynamics of the edge modes. There have also been efforts to characterize, for local subsystems, the most general boundary symmetry algebras spanned by the edge modes \cite{Grumiller:2016pqb,Donnelly:2016auv,Grumiller:2017sjh,Geiller:2017xad,Speranza:2017gxd,Geiller:2017whh,Freidel:2019ofr,Freidel:2018fsk}, with potentially important consequences for quantum gravity \cite{Freidel:2018pvm,Freidel:2019ees}. Another important development at finite distance has been the realization that a proper treatment of the edge modes is crucial even when dealing with fictitious entangling interfaces, which has consequences in the computations of entanglement entropy \cite{Buividovich:2008gq,Donnelly:2011hn,Casini:2013rba,Casini:2014aia,Donnelly:2014fua,Donnelly:2015hxa,Agarwal:2016cir,Pretko:2018nsz,Blommaert:2018rsf,Lin:2018bud,Belin:2019mlt}. At infinity on the other hand, a lot of work has been dedicated towards understanding the intricate infrared properties of theories with massless excitations, and there a central role is played by large gauge transformations and soft modes \cite{Strominger:2017zoo}. While the connection between all these aspects is far from being understood\footnote{For example, there is no clear understanding of the relationship between the edge modes at finite distance and the soft modes at infinity, appart possibly from \cite{Blommaert:2018rsf,Freidel:2019ohg}.}, a unifying thread is that of having degrees of freedom supported on the boundary. Therefore, a natural question to ask is whether these edge modes can be unequivocally identified, and whether there exists a framework for studying their dynamics. In this note we would like to take a step in this direction, and show that there is a unified treatment of the boundary dynamics of edge modes which furthermore sheds light on their contribution to the entanglement entropy. This agrees in particular with the proposal recently put forward in \cite{Blommaert:2018oue,Blommaert:2018iqz} (see also \cite{Mathieu:2019lgi} for a homological viewpoint), and frames it in a slightly more general setup.

To understand how this construction comes about, let us first explain how the edge modes can be understood as arising from an extension of a given gauge theory. On a spatial hypersurface, the physical phase space and Hilbert space of a gauge theory both fail to be factorizable due to the presence of the gauge constraints\footnote{In the case of a continuum scalar field the factorization does actually already fail due to the requirement of continuity of the field across the entangling cut \cite{Witten:2018zxz}, but this is usually bypassed by resorting to a cutoff (or a lattice regularization). In the case of gauge theories, even the lattice construction fails because of the gauge constraints.} and the resulting inherent non-locality of gauge-invariant observables. Aside from being a conceptual issue for the definition of local subsystems \cite{Donnelly:2016auv}, this also represents an a priori technical obstruction to computing quantities such as the entanglement entropy of gauge fields across a fictitious interface between two regions \cite{Lin:2018bud}. This difficulty can however be bypassed by resorting to a so-called extended Hilbert space. The idea of this construction is that $\H_{\Sigma\cup\bar{\Sigma}}$ can be factorized into factors attached to $\Sigma$ and $\bar{\Sigma}$ provided that we extend these by attaching edge modes living on $S=\partial\Sigma=\partial\bar{\Sigma}$ and transforming under the action of a boundary symmetry group $\G_S$. Denoting the resulting extended Hilbert space by $\H_{\Sigma,S}$, one can then realize the total Hilbert space as a subspace $\H_{\Sigma\cup\bar{\Sigma}}\subset\H_{\Sigma,S}\otimes\H_{\bar{\Sigma},S}$, where the factorized right-hand side allows for the definition of a reduced density matrix. The total physical Hilbert space of gauge-invariant states is then recovered as $\H_{\Sigma\cup\bar{\Sigma}}=\H_{\Sigma,S}\otimes_{\G_S}\H_{\bar{\Sigma},S}$, where $\otimes_{\G_S}$ denotes an entangling product which identifies and gets ``rid of'' the extra boundary degrees of freedom. This construction has proven very useful in computations of entanglement entropy in lattice gauge theory \cite{Buividovich:2008gq,Buividovich_2009,Donnelly:2011hn,Casini:2013rba,Casini:2014aia,Delcamp:2016eya,Lin:2018bud}, and in the case of Chern--Simons theory can also be made precise in the continuum \cite{Fliss:2017wop,Wong:2017pdm}. In \cite{Donnelly:2018ppr} it has also been shown that the extended Hilbert space of 2-dimensional Yang--Mills theory naturally relates to the structure of an extended TQFT. Our description of the boundary dynamics of edge modes will take as its starting point the extended phase space, which is the classical counterpart of the extended Hilbert space, and which we will introduce in section \ref{sec:2}.

Although the edge modes enabling for the definition of an extended Hilbert space may seem to be purely auxiliary and non-physical, they do actually contribute to the entanglement entropy. The computation of this latter being a notoriously subtle issue, a more precise statement is that there exist many ways of computing entanglement entropy (giving the same result but using different technical routes and tools, see e.g. \cite{Lin:2018bud}), with only some of these making manifest the role of the edge modes. In lattice gauge theory, the contribution of the edge modes of the extended Hilbert space to the entanglement entropy was noted in \cite{Buividovich_2009,Donnelly:2011hn}. In Abelian Chern--Simons theory, the entanglement entropy (say for two spatial disc-like regions glued along a circle) has a universal and cut-off independent contribution which can be traced back to (the zero modes of) the edge modes \cite{Fendley:2006gr,Dong:2008ft,Das:2015oha,Wen:2016snr,Wong:2017pdm,Fliss:2017wop}. This is the so-called topological entanglement entropy \cite{Kitaev:2005dm,Levin:2006zz}, which is an important physical quantity since it can serve as an order parameter for topological phases. Similarly, the entanglement entropy of Maxwell fields has a contribution sometimes referred to as the Kabat contact term \cite{Kabat:1995eq,Kabat:1995jq,Kabat:2012ns}, which can be traced back to the contribution of edge modes in the form of normal electric field configurations \cite{Casini:2014aia,Donnelly:2014fua,Donnelly:2015hxa,Agarwal:2016cir,Pretko:2018nsz}.

The following question naturally arises: Given a gauge theory, what are the Lagrangian and Hamiltonian descriptions of the edge modes which contribute to the entanglement entropy and constitute the boundary dynamics? In other words, what is the action and the symplectic structure for the edge modes, and which freedom is there in their construction? Our proposal for answering this will be guided by the example of Chern--Simons theory, for which the dynamics of the edge modes and their contribution to the entanglement entropy is known. This will show that, as far as edge modes are concerned, generic gauge theories are not different from Chern--Simons theory in the sense that they do all posses a non-trivial boundary dynamics\footnote{This is to be contrasted with some statements in the literature, which sometimes place Chern--Simons theory on a different footing, and trace back the origin of its boundary dynamics to the gauge non-invariance of its action.}. We will present here a framework for studying this dynamics, and give some preliminary examples of the subtleties and differences which arise for different gauge theories (e.g. depending on boundary conditions and Hamiltonians, and on whether the theory is topological or not). Briefly speaking, the symplectic structure of the edge modes will come from the above-mentioned extended phase space, which can be constructed in a systematic manner as a gauge-invariant boundary extension of the bulk phase space of a theory. To study the dynamics of the edge modes, we will propose a new action principle which includes the edge modes in a boundary Lagrangian and then naturally reproduces the extended phase space and its symplectic structure. Then, we will explain how integrating out the bulk degrees of freedom in a subregion produces an effective boundary action which will contribute to the entanglement entropy.

Our construction is summarized schematically on figure \ref{figure}. Consider two spacetime manifolds $M$ and $\bar{M}$ with respective time-like boundaries $\partial M$ and $\partial\bar{M}$. A gauge theory on each manifold is defined by bulk fields, but also by boundary degrees of freedom. These edge modes are introduced via a boundary Lagrangian, which couples in a gauge-invariant manner the bulk gauge fields and the edge modes to a boundary current (which can be thought of as the edge mode's conjugate momentum). The presence of these edge modes is precisely what allows for the splitting of the path integral over $M\cup\bar{M}$ into two factors. This is the covariant analogue of the factorization in terms of extended Hilbert spaces, and it requires to relax the boundary conditions and to allow for open Wilson lines to end on the boundary. In a path integral context, one can then manipulate the factorized path integrals over $M$ and $\bar{M}$ in two ways: $i)$ integrating out the edge modes living on $\partial M$ and $\partial\bar{M}$ (with suitable matching constraints) will glue the theories defined on the subregions and lead to the path integral over $M\cup\bar{M}$, while $ii)$ integrating out the bulk fields of region $\bar{M}$ will produce an effective boundary theory on $\partial\bar{M}$. This second point is very important. It means that integrating out the bulk degrees of freedom in $\bar{M}$, when taking properly into account the presence of the edge modes on the boundary $\partial M=\partial\bar{M}$, does \textit{not} reproduce the path integral defined on $M$ only: there is a residual contribution on the boundary due to the dynamics of the edge modes, and this will contribute in particular to the entanglement entropy.

\begin{figure}[h]\label{figure}
$$\vcenter{\hbox{\includegraphics[scale=1.1]{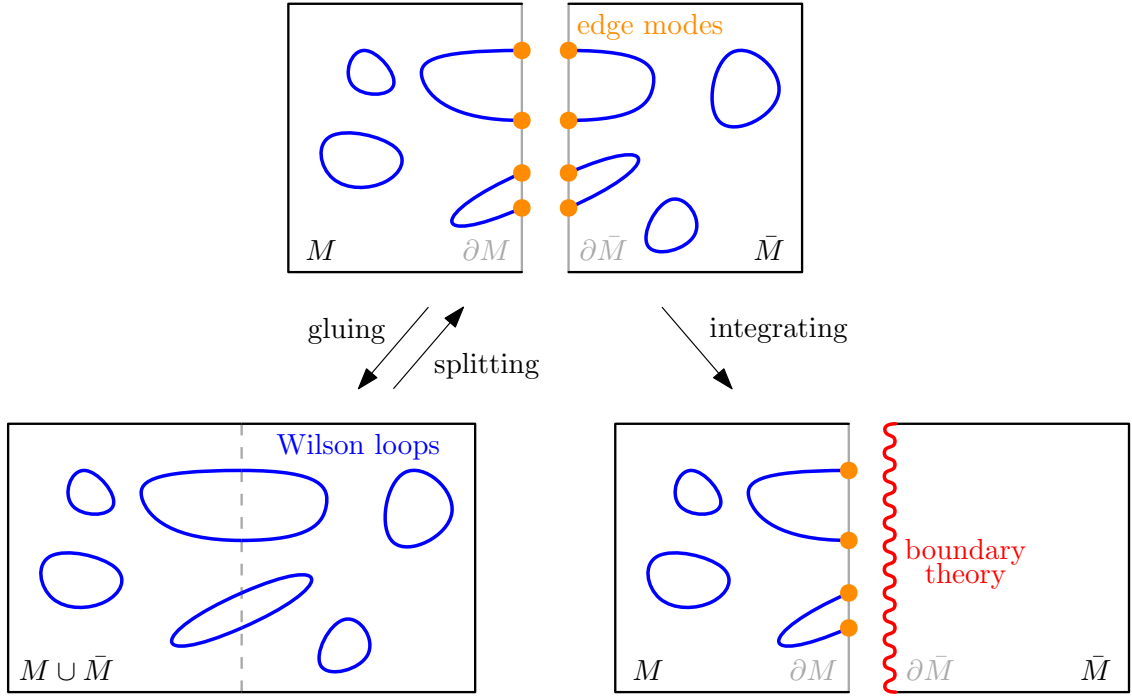}}}$$
\caption{On the top, we have a same gauge theory defined on two neighboring manifolds with boundaries. Each factor contains bulk fields and edge modes. Integrating out the edge modes leads to the theory defined over $M\cup\bar{M}$ and, conversely, the splitting of $M\cup\bar{M}$ into subregions requires to introduce edge modes on which Wilson lines can end. Once the theory is split into factors associated with the regions $M$ and $\bar{M}$, one bulk region can be integrated out, thereby leading to an effective boundary dynamics for the edge modes. This latter will in turn be seen by and contribute to the entanglement entropy.}\label{figure}
\end{figure}

One can clearly see the fundamental role played by the edge modes in this construction: they appear once we split a theory (i.e. its Hilbert space, phase space, or path integral), dictate how regions should be glued along an interface, and encode a leftover boundary dynamics once one of the bulk regions is integrated out. There is yet another elegant argument, due to Carlip \cite{Carlip:1995cd}, which justifies and highlights the role of these edge modes. Consider a theory defined on the total manifold $M\cup\bar{M}$, and whose path integral $\Z$ can be computed to find a functional determinant. By defining the theory separately on the subregions, where one can compute $\Z_M$ and $\Z_{\bar{M}}$, it is however not possible to reconstruct the total $\Z$ by simply multiplying the results for the subregions. This is because the functional determinants do not factorize, but instead satisfy a relation of the form $\Z=\mathscr{K}\,\Z_M\Z_{\bar{M}}$, sometimes referred to as the Forman--Burghelea--Friedlander--Kappeler (FBFK) gluing formula \cite{Forman1987,BURGHELEA199234,doi:10.1063/1.4936074,PARK_2006,Kirsten2018ThePA,Kirsten:2019xel}. Although this argument was initially formulated for a scalar field theory\footnote{Our interpretation therefore suggests that even the scalar field has ``scalar edge modes''. Interestingly, such scalar edge modes have been identified at null infinity through the soft theorem / asymptotic symmetry correspondence of Strominger's infrared triangle \cite{Campiglia:2017dpg}, and can be given a gauge theory interpretation in terms of the dual 2-form theory \cite{Campiglia_2019,Henneaux_2019}.}, where the factor $\mathscr{K}$ is related to the so-called Poisson kernel of the Laplacian, it is clear that a similar relation should hold for gauge theories as well (although the determinants might be complicated to evaluate). This was made explicit in \cite{Agarwal:2016cir} in the case of 3-dimensional Maxwell theory (which is actually dual to a scalar field), where it was indeed shown that the boundary contributions obtained after integrating over the bulk of the subregions are necessary in order to reproduce the FBFK gluing formula once the two regions are glued back together. This has led the authors to identify the gluing factor $\mathscr{K}$ as the Kabat contact contribution to the entanglement entropy. However this was done without explicitly introducing edge modes, but instead via a Green decomposition of the gauge field and its momentum between their bulk and boundary parts. The authors of \cite{Blommaert:2018oue} have presented a more systematic and dimension-independent study of the boundary dynamics of Maxwell--Yang--Mills theory, and also argued that the boundary contribution obtained after going on-shell in the bulk is the Kabat contact term seen by the entanglement entropy. Here we will present a framework confirming the generality of these results and applicable to any gauge theory, and show explicitly how it provides a covariant realization of the extended Hilbert space (or phase space) construction.

For this, we will first recall in section \ref{sec:2} how the edge modes can conveniently be parametrized in a Hamiltonian setting by using an extended phase space containing a boundary symplectic structure, then construct in section \ref{sec:3} compatible boundary actions, and finally present in section \ref{sec:4} examples and applications. Section \ref{sec:5} will describe prospects for future work. Some of our notations concerning the covariant phase space are gathered in appendix \ref{appendix:1}, and subsequent appendices contain various details of calculations. We will always refer to Chern--Simons and Maxwell--Chern--Simons theory respectively as CS and MCS theory.

\section{Extended phase spaces}
\label{sec:2}

For a given gauge theory, the extended phase space is the classical analogue of the above-mentioned extended Hilbert space. The extension consists in adding to the bulk phase space, for each type of gauge transformations in the theory, a corresponding edge mode field (which is nothing but a Stueckelberg field) living on the boundary. This idea was introduced and implemented in \cite{Donnelly:2016auv} for Yang--Mills theory and metric gravity, in \cite{Geiller:2017xad} for non-Abelian BF and CS theories, in \cite{Speranza:2017gxd} for higher curvature gravity, in \cite{Geiller:2017whh} for 3-dimensional gravity in first order connection-triad variables, in \cite{Balasubramanian:2018axm} for open string field theory, and in \cite{Setare:2018mii} for Einstein--Maxwell theory. It has also been used in \cite{Freidel:2018fsk} to describe magnetic charges and electromagnetic duality.

The construction of the extended phase space takes place in the covariant Hamiltonian formalism, and exploits a well-known corner (i.e. co-dimension 2) ambiguity \cite{Iyer:1994ys,Jacobson:1993vj}, which is that of supplementing the (pre-symplectic) potential $\theta$ by a total exterior derivative $\de\vartheta$. By adding edge mode fields living on the boundary $S=\partial\Sigma$ of spatial hypersurfaces and transforming in a particular way under gauge transformations\footnote{We focus here on internal gauge transformations, the treatment of diffeomorphisms being more subtle.} (we will provide examples below), one can construct in a minimal way an extended potential $\theta_\text{e}=\theta+\de\vartheta$ such that the associated symplectic structure
\be\label{symplectic structure}
\Omega=\int_\Sigma\delta\theta_\text{e}=\int_\Sigma\delta\theta+\int_S\delta\vartheta
\ee
disentangles in a natural manner the role of gauge transformations from that of boundary symmetries. This extended symplectic structure is indeed such that gauge transformations are generated by constraints which vanish on-shell and have no Hamiltonian charge, while boundary symmetries are generated by surface observables which satisfy a boundary symmetry algebra, and this without the need to impose boundary conditions on the dynamical fields or on the parameters of gauge or symmetry transformations\footnote{One can of course construct directly the extended symplectic structure which satisfies these desired properties. However, following \cite{Donnelly:2016auv,Geiller:2017xad}, it actually turns out to be easier and more systematic to derive this by considering finite and field-dependent gauge transformations of the potential.}. The role of the edge mode fields appearing with their canonical momenta in the boundary symplectic structure is two-fold: $i)$ to restore the seemingly broken gauge-invariance due to the presence of the boundary, $ii)$ to parametrize the boundary symmetries and observables.

Representing a gauge transformation by a tangent vector $\delta_\alpha$ in field space, one has in other words that the field space contraction\footnote{See appendix \ref{appendix:1} for our notations and conventions concerning the covariant phase space.} $\delta_\alpha\ipp\Omega$ is integrable and vanishing on-shell. This is nothing but the familiar Hamiltonian generator of the transformation $\delta_\alpha$, which is however stripped from its usual boundary charge because this latter has been cancelled by the contribution of the boundary symplectic structure containing the edge modes. This is a first advantage of the extended phase space: gauge transformations are null directions of the extended symplectic structure even when they have support on the boundary. Similarly, a boundary symmetry can be represented by a tangent vector $\Delta_\alpha$, and is characterized by a generator $\Delta_\alpha\ipp\Omega$ which is integrable, gauge-invariant in the sense that $\delta_\alpha\ipp\Delta_\beta\ipp\Omega=0$, equal to a boundary integral, and satisfies a boundary symmetry algebra $\Delta_\alpha\ipp\Delta_\beta\ipp\Omega$. 

It has been shown in \cite{Geiller:2017xad,Geiller:2017whh} that the generators of such boundary symmetries $\Delta_\alpha$ are the usual Hamiltonian boundary observables introduced in \cite{Regge:1974zd,Balachandran:1991dw,Balachandran:1992qg,Balachandran:1994up,Balachandran:1995qa,Husain:1997fm}, in which however the bulk fields are ``dressed'' in a gauge-invariant manner by the new edge mode fields which have been introduced on the boundary. This is a second advantage of the extended phase space: the edge modes which have been added through the boundary symplectic structure are now part of the phase space and parametrize the boundary observables and their symmetry algebra. While this description may seem formal at this point, we will provide explicit examples in section \ref{sec:4}.

The natural next step is to search for a dynamical description of these edge modes, and to conceive them not as living only on the boundary $S$ of a spatial slice, but on the whole time-like boundary $S\times\mathbb{R}$. This is a familiar situation in CS theory, where the time-like boundary is known to carry a gapless chiral theory \cite{Witten:1988hf,MR1025431,Carlip:1991zm}. However, the construction of the boundary dynamics in CS theory typically relies on studying the behavior under gauge transformations of the action itself. This explains the difference of treatment which has subsisted so far between e.g. Maxwell--Yang--Mills and CS theory: the former has a gauge-invariant action while the latter does not. From this, one would (wrongly) conclude that Maxwell--Yang--Mills theory does not posses a boundary dynamics. However, as we have argued above, the study of gauge (non)-invariance should instead be carried out at the level of the symplectic structure. There, one can easily motivate the need to work with an extended phase space containing edge mode fields. Let us now describe how their boundary symplectic structure can be obtained from a boundary action.

\section{Extended actions}
\label{sec:3}

Let us consider for simplicity that the $d$-dimensional spacetime manifold is of the form $M=\Sigma\times\mathbb{R}$, where the time-like boundary is $\partial M=S\times\mathbb{R}$. The extended symplectic structure described above can be thought of as arising from a 0-1-extended field theory, where the co-dimension 0 (i.e. bulk) and co-dimensional 1 (i.e. boundary) submanifolds each posses a Lagrangian, equations of motion, and a (pre-symplectic) potential. In order to see this, let us write the extended action and its variation in the form
\be\label{variation of action}
S=\int_ML_M+\int_{\partial M}L_{\partial M},\q\q\delta S=\int_M\text{EOM}_M+\int_{\partial M}\theta+\delta L_{\partial M}.
\ee
This is of course a familiar step in field theory and in the covariant phase space formalism, where it identifies the potential $\theta$ as the total exterior derivative term arising from the integrations by parts isolating the bulk equations of motion. In usual constructions of the covariant phase space \cite{Iyer:1994ys,Barnich:2001jy,Compere:2018aar}, the introduction of a boundary Lagrangian $L_{\partial M}$ is simply understood as resulting in a shift $\theta\mapsto\theta+\delta L_{\partial M}$ of the potential. The boundary conditions defining the variational principle are then taken to be $(\theta+\delta L_{\partial M})\big|_{\partial M}\stackrel{!}{=}0$, and one concludes that the boundary Lagrangian cannot affect the symplectic structure since upon taking a second variation to obtain the symplectic current one has $\delta\theta\mapsto\delta\theta+\delta^2 L_{\partial M}=\delta\theta$ by virtue of the property $\delta^2=0$.

However, this viewpoint turns out to be unnecessarily restrictive, and one can be more general by realizing that this ambiguity in the boundary term fits perfectly well with the above-mentioned corner ambiguity. In other words, there is a natural way in which the boundary Lagrangian may provide a corner term. This is what was done already in \cite{Geiller:2017whh} with the Gibbons--Hawking--York boundary term\footnote{Although the result is unfortunately hidden in appendix B.} (see also \cite{Wieland:2017zkf}), and what we will now explain in full generality. Coincidentally, while the present article was being written, along with an application of the formalism to 4-dimensional first order gravity \cite{Freidel:2020xyx,Freidel:2020svx}, Harlow and Wu have constructed in \cite{Harlow:2019yfa} a version of the covariant phase space formalism which precisely describes and formalises the step we are going to take\footnote{An important difference is that Harlow and Wu are not concerned with edge modes and extended phase spaces. They describe how a boundary Lagrangian can provide a corner term and discuss at length the example (among others) of Einstein--Hilbert gravity with the Gibbons--Hawking--York term, but do not consider Lagrangians which include edge mode fields. Appart from this conceptual difference, our constructions are the same.}. The idea is simply to realize that acceptable boundary conditions can be more generally taken to be\footnote{We allow ourselves to change the sign convention for $c$ with respect to \cite{Harlow:2019yfa} in order to match the definition of the extended symplectic structure given above.}
\be\label{general boundary conditions}
(\theta+\delta L_{\partial M})\big|_{\partial M}\stackrel{!}{=}-\de c.
\ee
Interestingly, this fits nicely with our desire to encode the dynamics of the edge mode fields in the boundary Lagrangian. Indeed, if this latter contains derivatives, upon taking a field space variation one can then integrate by parts to isolate boundary equations of motion and a boundary pre-symplectic potential. We can then suggestively rewrite the variation of the action in \eqref{variation of action} as
\be\label{general variation of action}
\delta S=\int_M\text{EOM}_M+\int_{\partial M}\text{EOM}_{\partial M}-\de c,
\ee
where on the boundary we have now explicitly combined the potential $\theta$ of $L_M$ with part of the variations of $L_{\partial M}$ to get the boundary equations of motion, and also kept the total exterior derivative containing the potential $c$ of $L_{\partial M}$. In this picture, the boundary conditions \eqref{general boundary conditions} are just rewritten as the requirement that $\text{EOM}_{\partial M}\stackrel{!}{=}0$. As we will see in explicit examples below, this requirement will generally translate into several conditions, which can be fulfilled either by fixing some variables on $\partial M$ (e.g. Dirichlet boundary conditions in gravity), or by imposing boundary equations of motion.

As explained in \cite{Harlow:2019yfa}, the correct potential to consider for the construction of a conserved symplectic structure is then $\theta+\de c$, and therefore naturally includes a corner contribution. In our more general setup, where the boundary Lagrangian may contain edge mode fields, we will see that the correct extended symplectic potential $\theta_\text{e}=\theta+\de\vartheta$ will be obtained once we explicitly rewrite $\theta+\de c$ on-shell of (some of) the boundary equations of motion which identify $c\approx\vartheta$. This is of course fine since in any case the covariant phase space formalism is on-shell, and since going on-shell of the boundary equations of motion is simply enforcing part of the boundary conditions \eqref{general boundary conditions} defining the variational principle. More precisely, we will see that in the whole set $\text{EOM}_{\partial M}$ we will have to explicitly use the boundary equations of motion involving the initial potential $\theta$. This is a desired feature, since it means that instead of holding fixed a field configuration on the boundary (e.g. the gauge potential of Maxwell theory), we are relaxing this condition by imposing the conjugated boundary equations of motion instead. Once again, this should become crystal clear in the following section where we present concrete examples.

In summary, in order to achieve our construction relating the boundary Lagrangian $L_{\partial M}$ (which is the object we are trying to identify) to the extended symplectic structure \eqref{symplectic structure} defining the extended phase space (which is the object we already know from the various constructions \cite{Donnelly:2016auv,Geiller:2017xad,Speranza:2017gxd,Geiller:2017whh,Balasubramanian:2018axm,Setare:2018mii}), we simply have to look for a boundary Lagrangian whose potential $c$ is such that the extended potential is obtained as
\be\label{extended potential}
\theta+\de c\approx\theta+\de\vartheta=\theta_\text{e}.
\ee
Our formalism and that of \cite{Harlow:2019yfa} guarantee that this is possible, and we will give illustrative examples in the next section. A few comments are now in order before going on.

$i)$ In this construction the boundary Lagrangian is more than a mere boundary term: it contains derivatives, and therefore a potential, which is then combined with the bulk potential in order to get the extended symplectic structure. As we have argued, this falls outside of the usual covariant Hamiltonian formalism of e.g. \cite{Iyer:1994ys,Barnich:2001jy,Compere:2018aar}, and fixes unambiguously the corner contribution $c$. Furthermore, adding edge modes into the boundary Lagrangian achieves more than a simple change of polarization: it allows to completely relax the boundary conditions by replacing them with boundary equations of motion.

$ii)$ One can be puzzled by the apparent sign mismatch between the boundary potential in \eqref{general variation of action} and its contribution to the extended potential in \eqref{extended potential}. This follows of course from the compatibility of the symplectic current (more precisely the conservation of the symplectic structure) with the boundary conditions \eqref{general boundary conditions}. A more heuristic way to understand this is to remember that we are trying to match the corner terms constructed in \cite{Donnelly:2016auv,Geiller:2017xad,Speranza:2017gxd,Geiller:2017whh,Balasubramanian:2018axm,Setare:2018mii} by reaching the corner from the space-like hypersurface $\Sigma$, to the corner terms inherited from the boundary Lagrangian, and which therefore reach the corner from the time-like boundary $\partial M$. One can therefore understand the sign difference as coming from the sign of the bi-normal to the co-dimension 2 corner $S$, which depends on whether the corner is reached from a space-like slice or from a time-like boundary.

$iii)$ We will see on the examples below that the boundary equations of motion which are used to write $c\approx\vartheta$ are, in the language of \cite{Donnelly:2016auv}, gluing constraints which determine the extended phase space by soldering together, via a classical fusion product, the boundary symplectic structure to the bulk one. The boundary Lagrangian $L_{\partial M}$ contains initially the edge mode fields and their unspecified conjugate momenta, and the boundary condition obtained through the boundary equation of motion involving $\theta$ identifies these momenta with part of the initial bulk fields.

$iv)$ The minimal requirement which we have imposed so far on the boundary Lagrangian does only specify the symplectic structure for the edge mode fields, and not their dynamics. In order to access this later, we will have to resort to an on-shell evaluation of the bulk action, thereby leading to an effective boundary action. We are also free to add to the boundary Lagrangian terms which do not change the symplectic structure and which are compatible with gauge-invariance. Such terms are in fact boundary Hamiltonians, i.e. they affect the boundary conditions (or equations of motion), but not the symplectic structure. The details of this procedure will depend on the theory under consideration, so let us now finally discuss some examples.

$v)$ It is important to appreciate that there are two notions of ``boundary dynamics'' in the framework which we are proposing and which we have outlined above. First, there are boundary equations of motion which appear in \eqref{general variation of action} when varying the extended bulk + boundary action. These can be seen as continuity conditions relating the bulk and boundary fields. However, these equations alone do not determine the boundary dynamics of the edge mode fields. As we have mentioned above, this latter is obtained when evaluating the bulk fields on-shell. It will become clear in the examples discussed below that these two levels of equations of motion are different\footnote{One can think of this in analogy with first order theories, where one replaces a second order equation of motion by two first order equations. One can focus on one single first order equation, but this may not determine completely the dynamics of a dynamical variable, which is only obtained when going on-shell of the other first order equation.}.

Note that a general framework for analyzing the boundary dynamics based on a proper identification of the boundary Lagrangian and action principle was recently proposed also in \cite{Rubalcava-Garcia:2020bcp}, although without explicitly introducing edge modes. The example of Abelian Chern--Simons theory was also extensively discussed in this reference, reaching the same conclusions as our construction.

\section{Examples}
\label{sec:4}

We now present some relevant examples of extended bulk + boundary actions. This will illustrate in particular formulas \eqref{general variation of action} and \eqref{extended potential}, and reproduce the extended symplectic structures which have been studied before in the literature. It will also enable us to identify and discuss some remaining ambiguities in the characterization of the boundary dynamics, and to give more details on the factorization and the gluing of path integrals. We will also be able to establish a connection with previous results on the edge mode contributions to the entanglement entropy in various theories. We focus here on Abelian theories, and the discussion of the extended actions and phase spaces for non-Abelian theories is deferred to appendix \ref{appendix: non-Abelian}.

\subsection{Chern--Simons theory}

Let us focus on the Abelian case for simplicity, and describe in details all the steps of the calculations. As usual, the theory is defined in the bulk by a connection 1-form $A$, transforming under gauge transformations as $\delta_\alpha A=\de\alpha$, and with curvature $F=\de A$. On the boundary, we now add a 0-form $a$ transforming as $\delta_\alpha a=-\alpha$, and a gauge-invariant 1-form $j$. With this field content, we can then form the action
\be\label{CS extended action}
S=S_M+S_{\partial M}=\int_M A\wedge F+\int_{\partial M} aF+j\wedge\De a\pm\f{1}{2}\as j\wedge j,
\ee
where the Abelian covariant derivative is $\De a\coloneqq\de a+A$, where $\as$ is the Hodge dual on the boundary, and where we have dropped for clarity the usual coupling constant $\k/(4\pi)$. The first term on the boundary, which is not gauge-invariant by itself, compensates for the gauge non-invariance of the bulk term, and the full action is therefore gauge-invariant. The presence of the last term, which requires to use the metric and therefore breaks the topological invariance of the theory, will be explained momentarily. This term is a boundary Hamiltonian $h[j]$, whose choice does not affect the boundary symplectic structure, but does change the boundary dynamics.

\subsubsection{Extended phase space}

Following the discussion of the previous section, let us now see what the introduction of the two fields $a$ and $j$ via the boundary Lagrangian implies. The variation of the action can be written in the form \eqref{variation of action} as
\be
\delta S=\left(2\int_M\delta A\wedge F+\int_{\partial M}\delta A\wedge A\right)+\delta S_{\partial M},
\ee
where one can see that the potential coming from the bulk is $\theta=\delta A\wedge A$. Writing explicitly the variation of the boundary action now leads to the form of expression \eqref{general variation of action}, which is
\be\label{CS variation extended action}
\delta S=2\int_M\delta A\wedge F+\int_{\partial M}\delta A\wedge(\De a-j)+\delta j\wedge(\De a\mp\as j)+\delta a(\de j-F)-\de(j\delta a-a\delta A),
\ee
where on the boundary the first three terms identify the boundary equations of motion, and the last term identifies the boundary potential $c$. To access the bulk equations of motion, we need to impose the vanishing of the first term on the boundary. Conveniently, this can be done by imposing the boundary equation of motion $j=\De a$ instead of fixing the variation $\delta A$ of the gauge potential to be vanishing. This boundary equation of motion is precisely the one involving the potential $\theta$ coming from the bulk Lagrangian. With this, the extended potential \eqref{extended potential} becomes
\be\label{CS extended potential}
\theta_\text{e}=\theta+\de(j\delta a-a\delta A)=\delta A\wedge A+\de(j\delta a-a\delta A)\approx\delta A\wedge A+\de(\De a\delta a-a\delta A),
\ee
where we have been careful about the sign when including the boundary potential as our corner term, and then in the last equality used the boundary equation of motion involving $\theta$. This result is interesting, as it reproduces precisely the extended potential which was derived in \cite{Geiller:2017xad} for Abelian CS theory, thereby proving that the extended phase space structure can be recovered from the boundary Lagrangian introduced in \eqref{CS extended action} and the construction outlined in the previous section.

With this extended potential we have all the desirable properties mentioned in section \ref{sec:2}. In particular, the extended symplectic structure \eqref{symplectic structure} is given by
\be
\Omega=\int_\Sigma\delta\theta_\text{e}=-\int_\Sigma\delta A\wedge\delta A+\int_S\delta(\De a)\delta a-\delta a\delta A,
\ee
and is such that for gauge transformations the generator defined by $\delta_\alpha\ipp\Omega$ is integrable and vanishing on-shell. Indeed, this is
\be
\delta_\alpha\ipp\Omega=-2\int_\Sigma\de\alpha\wedge\delta A+2\int_S\alpha\delta A=2\int_\Sigma\alpha\delta F.
\ee
The transformation $\delta_\alpha$ is therefore a true gauge transformation, even when it has support on the boundary, and as such it has no Hamiltonian charge. In addition, the transformation acting as $\Delta_\alpha A=0$ and $\Delta_\alpha a=\alpha$, which we will now call boundary symmetry as opposed to gauge transformation, has an integrable generator given by the manifestly gauge-invariant boundary integral
\be\label{CS boundary observables}
\Q[\alpha]=2\int_S\alpha\De a,
\ee
and these generators satisfy the Abelian Ka\v c--Moody commutation relation
\be
\lb\Q[\alpha],\Q[\beta]\rb=\Delta_\alpha\ipp\Delta_\beta\ipp\Omega=2\int_S\alpha\de\beta.
\ee
As is well-known, these are the boundary symmetries of CS theory on a spatial disc. One can see, as explained above, that their generator is a gauge-invariant ``dressed'' version of the usual Hamiltonian charge of $\delta_\alpha$, where the dressing corresponds to the finite gauge transformation of $A$ by the edge mode field $a$.

\subsubsection{Boundary dynamics}

The Ka\v c--Moody commutation relations which we have derived on the extended phase space result from the presence of a chiral scalar field, which is evidently the edge mode field $a$. To access the dynamics of this scalar field, we will write and manipulate the path integral for the extended action \eqref{CS extended action}, following \cite{MR1025431}. The key point of this derivation is to expand the components of the gauge field in the action and to carefully perform the path integral. For this, we assume that the spacetime has the topology $M=\mathbb{R}\times D$ of a cylinder, with coordinates $x^\mu=(t,r,\phi)$ such that $\eps^{tr\phi}=1$ and $\phi$ is compact, and that the space-like normal to the boundary cylinder at finite radius $r$ is $s=(0,1,0)$. The Hodge dual is then such that $\as j\wedge j=(\as j)_\phi j_t-(\as j)_tj_\phi=j_t^2-j_\phi^2$. After integrations by parts in the bulk, the total action \eqref{CS extended action} can be written explicitly as
\be\label{CS in components}
S
&=\int_M2A_t(\partial_rA_\phi-\partial_\phi A_r)+A_\phi\partial_tA_r-A_r\partial_tA_\phi\nn\\
&\phantom{=\ }+\int_{\partial M}a(\partial_\phi A_t-\partial_tA_\phi)+j_\phi(A_t+\partial_ta)-j_t(A_\phi+\partial_\phi a)-A_tA_\phi\pm\f{1}{2}j_t^2\mp\f{1}{2}j_\phi^2.
\ee
It is then clear that $A_t$ plays the role of a Lagrange multiplier. Path integrating\footnote{As the details do not matter for our purposes so far, we will not explicitly write the path integrals and the various pre-factors coming from the integrations, but simply the resulting effective actions.} over $A_t$ imposes the bulk and boundary relations
\be
\partial_rA_\phi-\partial_\phi A_r=0,\q\q j_\phi=A_\phi+\partial_\phi a,
\ee
which are part of the equations of motion imposed by $\delta A$ (i.e. the bulk equation of motion and the corresponding boundary condition). The first constraint can be solved by writing
\be
A_r=\partial_r\alpha,\q\q A_\phi=\partial_\phi\alpha.
\ee
With this, after integrations by parts the bulk piece of the action becomes a boundary term, as
\be
A_\phi\partial_tA_r-A_r\partial_tA_\phi=\partial_\phi\alpha\partial_t\partial_r\alpha-\partial_r\alpha\partial_t\partial_\phi\alpha=\partial_r(\partial_\phi\alpha\partial_t\alpha)-\partial_\phi(\partial_r\alpha\partial_t\alpha),
\ee
and \eqref{CS in components} reduces to the boundary action
\be\label{CS intermediate effective action}
S_\text{edge}=\int_{\partial M}\partial_t\varphi\partial_\phi\varphi-j_t\partial_\phi\varphi\pm\f{1}{2}j_t^2\mp\f{1}{2}(\partial_\phi\varphi)^2,
\ee
where we have introduced the gauge-invariant scalar combination $\varphi\coloneqq a+\alpha$. We recognize as the first term the canonical term of a chiral field. This is to be expected since so far the current $j$ has in a sense played no role, and we have reproduced the classic calculation of \cite{MR1025431} (where the authors obtain instead the boundary theory $\partial_t\alpha\partial_\phi\alpha-A_t\partial_\phi\alpha$, since neither $a$ not $j$ have been introduced there). In \cite{MR1025431}, one then uses the boundary conditions $A_t\propto A_\phi=\partial_\phi\alpha$ in order to get a Hamiltonian for the chiral field. In our setup however, the last step is to perform the Gaussian integral over the current $j_t$ to finally obtain the effective action
\be\label{chiral boundary action}
S_\text{edge}=\int_{\partial M}\partial_t\varphi\partial_\phi\varphi\mp(\partial_\phi\varphi)^2,
\ee
which is known as the Floreanini--Jackiw action. Its equations of motion are that of a chiral field, i.e. $\partial_t\varphi=\pm\partial_\phi\varphi$. This is the boundary dynamics of Abelian CS theory, and we have recovered it from the on-shell evaluation of the path integral for the extended bulk + boundary action \eqref{CS extended action}. The authors of \cite{Blommaert:2018oue,Blommaert:2018iqz} have also presented a derivation of the edge mode dynamics of CS theory, but we believe that our presentation follows more closely the original construction presented in \cite{Blommaert:2018oue} for Maxwell theory. Moreover, we have shown explicitly the link between the extended action and the extended phase space.

The last step of the above calculation makes clear the role of the quadratic $j$ term introduced in \eqref{CS extended action}. Without this term, the construction of the extended potential \eqref{CS extended potential} would have of course gone through, but the derivation of the boundary dynamics would not have provided a desirable Hamiltonian for the chiral field after \eqref{CS intermediate effective action}. This shows, as announced above, that the last term in \eqref{CS extended action} plays the role of a Hamiltonian: it does not affect the extended symplectic structure, but it changes the boundary dynamics. In this simple example of Abelian $\text{U}(1)$ CS theory, this change of dynamics is equivalent to changing the velocity of the chiral bosons. It would be interesting to study richer Abelian theories, such as $\text{U}(1)^N$, which admit topological boundary conditions \cite{Kapustin:2010hk} and gluing along heterointerfaces \cite{Fliss:2017wop}, and eventually the non-Abelian case and the classification of all possible boundary theories\footnote{The extended phase space of non-Abelian CS theory is derived form the extended action in appendix \ref{appendix: non-Abelian}.}.

As a subtlety, one can observe in \eqref{CS variation extended action} that the two boundary equations of motion obtained by varying $A$, $j$, when combined, simply imply that $\De a=\pm\as\De a$, meaning that $a$ is a gauged chiral field. This is essentially the same equation of motion as that derived from the effective boundary action \eqref{chiral boundary action}. From this, one could conclude that the boundary dynamics is in a sense already encoded in the initial bulk + boundary action \eqref{CS extended action}. This is however just a coincidence due to our choice of boundary Hamiltonian. Indeed, if we choose instead $h=(j_t\mp j_\phi)j_\phi$, it is easy to see that replacing the last two terms in \eqref{CS intermediate effective action} by $j_t\partial_\phi\varphi\mp(\partial_\phi\varphi)^2$ and then path integrating over $j_t$ leads once again to \eqref{chiral boundary action}, while, however, the boundary equations of motion give $\De_ta=(\pm2-1)\De_\phi a$. This last equation is once again that of a chiral field, but now the two chiralities have a different velocity. This is a known fact in CS theory and condensed matter, namely that the velocity is an external input which can be tuned by changing the Hamiltonian \cite{Tong:2016kpv}. However, this example illustrates clearly the fact that there is a slight quantitative difference between the boundary equations of motion derived from the bulk + boundary action \eqref{CS extended action} and that derived from the on-shell evaluation of the action. For topological theories, these two views on the boundary dynamics are in a sense equivalent (at least qualitatively, as we have just seen), since on-shell bulk configurations are simply gauge transformations. For non-topological theories however, the on-shell evaluation of the action is crucial since it imprints on the boundary a left-over dynamics from the bulk (which is not just a gauge transformation). We will see with the example of Maxwell theory that the derivation of the boundary dynamics requires an on-shell evaluation of the action, and cannot be read off the initial extended action alone.

Finally, as a curiosity, and in order to make contact with previous literature on the subject, one can insert the boundary equation of motion $j=\De a$ back into the action to obtain
\be\label{gauged chiral action}
S=\int_M A\wedge F+\int_{\partial M} aF\pm\f{1}{2}\as\De a\wedge\De a,
\ee
which we recognize as the action for CS theory coupled to a gauged chiral field on the boundary \cite{Balachandran:1994ik}. As in \cite{Arcioni:2002vv}, this constitutes the off-shell and gauge-invariant description of the boundary dynamics of CS theory, in the sense that it leads to the equations of motion of a chiral field without having to evaluate the bulk theory on-shell. However, a subtle yet important point is that variation with respect to $a$ on the boundary of \eqref{gauged chiral action} leads to the equation of motion $\De a=\mp\as\De a$, which is the opposite chirality to what we have derived from \eqref{CS extended action}. This is to be expected since the manipulations leading to \eqref{gauged chiral action} are different from that leading to the effective action \eqref{chiral boundary action}. In particular, obtaining \eqref{gauged chiral action} does not require the on-shell evaluation of the bulk fields. As it will become clear below, it is indeed this on-shell evaluation which one should carry out in order to access the effective boundary dynamics, and this latter cannot simply be read off from the boundary equations of motion using \eqref{CS variation extended action} and \eqref{gauged chiral action}.


\subsubsection{Gluing of subregions}
\label{sec: CS gluing}

Referring to figure \ref{figure}, we have so far described the operations of splitting and of integrating. Splitting CS theory on $M\cup\bar{M}$ requires to consider for each subregion with boundary the extended actions \eqref{CS extended action}. Integrating over the bulk gauge field of a subregion leads to a path integral over boundary fields only, and the dynamics of the boundary edge mode field $a$ is that of a chiral theory.

We can now describe the operation of gluing of two subregions, which will involve getting rid of the edge mode field contributions from the two boundaries. For two boundary theories on $\partial M$ and $\partial\bar{M}$ with opposite chirality, the gluing of $S[A,a,j]$ and $S[\bar{A},\bar{a},\bar{j}]$ is then given by
\be\label{CS gluing}
\Z
&=\int\mathscr{D}[A,\bar{A},a,\bar{a},j,\bar{j}]\,\delta(a+\bar{a})\,\delta(j+\bar{j})\cr
&\phantom{=\int}\exp\left(\i\int_M A\wedge F+\i\int_{\bar{M}}\bar{A}\wedge\bar{F}+\i\int_{\partial M} aF+\bar{a}\bar{F}+j\wedge\De a+\bar{j}\wedge\bar{\De}\bar{a}\pm\f{1}{2}\as j\wedge j\mp\f{1}{2}\as\bar{j}\wedge\bar{j}\right)\cr
&=\int\mathscr{D}[A,\bar{A},a,j]\exp\left(\i\int_M A\wedge F+\i\int_{\bar{M}}\bar{A}\wedge\bar{F}+\i\int_{\partial M} a(F-\bar{F})+j\wedge(A-\bar{A}+2\de a)\right)\cr
&=\int\mathscr{D}[A,\bar{A},a]\,\delta(A-\bar{A}+2\de a)\big|_{\partial M}\exp\left(\i\int_M A\wedge F+\i\int_{\bar{M}}\bar{A}\wedge\bar{F}+\i\int_{\partial M} a(F-\bar{F})\right)\cr
&=\int\mathscr{D}[A,a]\exp\left(\i\int_{M\cup\bar{M}}A\wedge F\right)\cr
&\propto\int\mathscr{D}[A]\exp\left(\i\int_{M\cup\bar{M}}A\wedge F\right).
\ee
Here we have written the path integral over all the bulk and boundary fields coming from the two subregions and their boundaries, with two delta functions enforcing the identification of the edge modes coming from the two boundaries. After integrating over $\bar{a}$ and the two currents $j$ and $\bar{j}$, in the third equality we are left with a delta function on the boundary, imposing that the gauge fields incoming from the two subregions are equal up to a gauge transformation. Integrating over $\bar{A}\big|_{\partial M}$ then eliminates the last boundary integral, and we are left with the path integral for CS theory over $M\cup\bar{M}$. In the last step we have simply eliminated a redundant integration over $a$ by dropping a gauge volume factor. This gluing operation is the application to CS theory of the gluing described in appendix A of \cite{Blommaert:2018oue}.

\subsubsection{Entanglement entropy}
\label{sec: CS EE}

Finally, let us conclude this section by discussing the role of the edge modes and the extended phase space in the computations of entanglement entropy. In general, the entanglement entropy $\S$ of a spatially bipartite system $\Sigma\cup\bar{\Sigma}$ receives contributions from two sources,
\be
\S=\S_\text{bulk}+\S_\text{edge}.
\ee
The first piece, $\S_\text{bulk}$, comes from physical degrees of freedom in the bulk, while $\S_\text{edge}$ originates from degrees of freedom localized at the boundary, which for the bipartite system is the entangling surface $S=\partial\Sigma=\partial\bar{\Sigma}$ between the two subregions. CS theory being topological, it does not have physical bulk degrees of freedom, and therefore the sole contribution to its entanglement entropy comes from the boundary degrees of freedom, i.e. the edge modes. Although the computation of entanglement entropy in CS theory is already well understood and has been studied by many authors, it is still worth briefly reviewing the different computational techniques in order to emphasize the role of the edge modes. After all, this is the narrative which we are trying to build in the present paper: there is a unified treatment of the extended phase space for all gauge theories, and a Lagrangian description of the corresponding edge modes. In CS theory, it is well accepted (and even tested!) that these edge modes have a dynamics and a contribution to entanglement entropy. This therefore strongly suggests that what is known about edge modes in CS theory is actually a generic feature of \textit{any} gauge theory.

There are essentially three approaches for computing entanglement entropy in CS theory. The first one exploits the knowledge of the surface symmetry algebra, the second one uses a Hamiltonian quantization of the effective boundary action \cite{Cano:2014pya}, and the third one the replica trick calculation \cite{Dong_2008}. We briefly mention the first approach below, and summarize the second one in appendix \ref{appendix:CS-EE}.

The computation of entanglement entropy from the surface symmetry follows from \cite{Wong:2017pdm,Fliss:2017wop,Das:2015oha,Wen:2016snr,Das:2015oha,Belin:2019mlt}. It relies on the extended Hilbert space construction, and on the factorization
\be
\H_\text{ext}=\H_{\Sigma,S}\otimes\H_{\bar{\Sigma},S},
\ee
where $\H_{\Sigma,S}$ denotes the extended Hilbert space on each subregion, containing edge states living on the entangling surface $S$. This extended Hilbert space, though it has the advantage of being factorized, contains in a sense two copies of the edge modes (one coming from each subregion) and therefore many non-physical states. The total Hilbert space of physical gauge-invariant states, which is a subspace of the factorized extended Hilbert space, is obtained as an entangling product
\be
\H_{\Sigma\cup\bar{\Sigma}}=\H_{\Sigma,S}\otimes_{\G_S}\H_{\bar{\Sigma},S}\subset \H_{\text{ext}},
\ee
and is spanned by gauge-invariant states $\ket{\psi}_\text{phys}$ satisfying the quantum gluing condition
\be
(\Q[\alpha]\otimes\bar{\mathbb{I}}+\mathbb{I}\otimes\bar{\Q}[\alpha])\ket{\psi}_\text{phys}=0.
\ee
Here, the boundary symmetry generators \eqref{CS boundary observables} derived from the classical theory are promoted to quantum operators, and correspondingly the Poisson brackets are turned into operator commutators. For Abelian CS theory, the algebra is the $\text{U}(1)$ Ka\v c--Moody algebra (with the factor $\k/4\pi$ restored),
\be
\big[\Q[\alpha],\Q[\beta]\big]=\f{\i\k}{2\pi}\int_S\de\phi\,(\alpha\partial_\phi\beta).
\ee
The fact that the boundary of CS theory carries a 2-dimensional chiral boson with corresponding Ka\v c--Moody algebra allows us to use techniques in boundary conformal field theory. In terms of the mode expansions
\be
\Q[\alpha]=\sum_{n\in\mathbb{Z}}\alpha_n\J_n,\q\q\alpha(\phi)=\sum_{n\in\mathbb{Z}}\alpha_ne^{\i n\phi},
\ee
the algebra becomes
\be
[\J_m,\J_n]=\k n\delta_{m+n,0}.
\ee
Identifying $\alpha_n=\overline{\alpha}_{-n}$, the gluing condition, which can now be rewritten as
\be
(\J_n\otimes\bar{\mathbb{I}}+\mathbb{I}\otimes\bar{\J}_{-n})|\psi\rangle_\text{phys}=0,
\ee
tell us that physical states are singlets under the action of left-moving and right-moving current operators on each side of the entangling surface (therefore the entanglement entropy in this case is known as left-right entanglement entropy). The gluing condition is solved by the conformally-invariant Ishibashi states \cite{Ishibashi:1988kg}
\be
|q\rangle\!\rangle=\sum_{N=0}^\infty\sum_{j=1}^{\text{dim}(N)}\ket{q,N,j}\otimes\overline{\ket{q,N,j}},
\ee
where the orthogonal states are labelled by a quasiparticle charge $q$, such that the choice $q=0$ corresponds to the vacuum state, and $q\neq0$ to states with Wilson lines with charge $q$ and $-q$ piercing through $\Sigma$ and $\bar{\Sigma}$. The quantum numbers $N,j$ label the descendants. The Ishibashi states are in general non-normalizable, and therefore need to be appropriately regularized. The regularized Ishibashi states are defined via the CFT modular Hamiltonian as
\be
|q\rangle\!\rangle_\text{reg}=\f{e^{-\epsilon H}}{\sqrt{n_q}}|q\rangle\!\rangle,\q\q H=\f{2\pi}{\ell}\left(\J_0+\bar{\J}_0-\f{c}{12}\right),
\ee
where $\epsilon$ is a cut-off parameter, $\ell$ is the length of the entangling surface $S$, and $c$ is the central charge of the corresponding CFT. With this, the generic edge states are linear combinations of the regularized Ishibashi states, and the entanglement entropy can be computed as the standard von-Neumann entropy. The result for the simplest case of a spherical hypersurface divided into two disks, $S^2=D\cup D$, and without quasiparticles, is given by \cite{Levin:2006zz,Kitaev:2005dm,Dong:2008ft}
\be\label{CS EE}
\S_\text{CS}=\f{\mathbf{A}}{2\pi}\f{\pi}{24\epsilon}-\f{1}{2}\log\k+\O(\ell^{-1}).
\ee
The first term is the non-universal area law, with $\mathbf{A}=2\pi\ell$, and the second term, which is area- and cut-off-independent, is the famous topological entanglement entropy of CS theory.

Another way of computing this entanglement entropy is to apply the standard method of Hamiltonian quantization to the boundary effective action \cite{Cano:2014pya}. This leads to the same topological contribution from the edge modes. In appendix \ref{appendix:CS-EE} we reproduce the details of the calculation for the interested reader.

\subsection{Maxwell theory}
\label{sec: Maxwell}

We now turn to the case of Maxwell theory. To construct the bulk + boundary action, in addition to the bulk gauge field $A$, let us consider on the boundary a 0-form $a$ transforming as $\delta_\alpha a=-\alpha$, and a gauge-invariant 2-form $j$. With this we can form the gauge-invariant action
\be\label{M extended action}
S=-\f{1}{2}\int_M\st F\wedge F+\int_{\partial M}j\wedge\De a+h,
\ee
where once again $\De a\coloneqq\de a+A$ and $h[j]$ is a boundary Hamiltonian depending on the current $j$ only. This simple action for Maxwell theory coupled to boundary currents is also the starting point of \cite{Blommaert:2018oue}, where it is however introduced from the point of view of the gluing of Maxwell theory for two neighboring regions $M$ and $\bar{M}$ (this gluing is strictly analogous to what we have described in section \ref{sec: CS gluing} for CS theory). This action is also motivated in \cite{Pretko:2018nsz} by the need to couple Maxwell theory to currents (or matter fields) in order to achieve its factorizability. The introduction of the boundary edge mode fields allows to factorize the theory between two neighboring subregions, and here we will furthermore show that these edge mode fields reproduce the extended phase space structure of \cite{Donnelly:2016auv}. Following \cite{Blommaert:2018oue}, we will set $h=0$ and show that even in this case there is a non-trivial boundary dynamics. We leave the study of various other possibilities for $h$ and their physical interpretation for future work.

Following \eqref{general variation of action}, the variation of this action reveals the bulk and boundary equations of motion, as well as the boundary potential. This variation is
\be
\delta S=-\int_M\delta A\wedge\de\st F+\int_{\partial M}\delta A\wedge(j-\st F)+\delta j\wedge\De a-\delta a\de j+\de(j\delta a).
\ee
We can observe that the boundary equations of motion imposed by the variation of $A$ and $a$ on the boundary are together consistent with the bulk equations of motion. The boundary condition imposed by $\delta A$ identifies the edge mode momentum $j$ with the normal electric field $\st F$, i.e. states that\footnote{We should always keep in mind that equalities involving $j$ are pulled back to the boundary $\partial M$.} $j=\st F$. With this, the extended potential \eqref{extended potential} becomes
\be\label{Maxwell extended potential}
\theta_\text{e}=\theta-\de(j\delta a)=-\delta A\wedge\st F-\de(j\delta a)\approx-\delta A\wedge\st F-\de(\st F\delta a),
\ee
which is the extended potential derived\footnote{One can also find a similar derivation in \cite{Mathieu:2019lgi}, where the authors treat however the corner terms differently and start from a different boundary Lagrangian.} in \cite{Donnelly:2016auv}. The extended symplectic structure derived from this potential is such that for gauge transformations the generator defined by $\delta_\alpha\ipp\Omega$ is vanishing on-shell. Indeed, this is
\be
\delta_\alpha\ipp\Omega=\int_\Sigma\de\alpha\wedge\delta(\st F)-\int_S\alpha\delta(\st F)=-\int_\Sigma\alpha\delta(\de\st F).
\ee
In addition, the transformation acting as $\Delta_\alpha A=0$ and $\Delta_\alpha a=\alpha$ has an integrable generator given by the ``electric charge''
\be
\Q[\alpha]=\int_S\alpha\st F
\ee
smeared with an arbitrary function $\alpha$. We therefore see how the extended bulk + boundary action allows us to recover the extended phase space structure of Maxwell theory. We can now turn to the boundary dynamics.

The boundary dynamics for the edge mode field $a$ is obtained by integrating out the bulk degrees of freedom in the path integral, as done in \cite{Blommaert:2018oue} and above for CS theory. Since Maxwell theory is quadratic, the integration over the bulk gauge field $A$ produces a determinant of the bulk operator multiplying the path integral for the bulk + boundary action evaluated on-shell. Using the fact that the on-shell bulk Maxwell action is itself a boundary term, and that on the boundary the normal electric field gets identified to the boundary current $j$ according to the boundary equation of motion $j=\st F$, we have that the path integral for \eqref{M extended action} is
\be\label{M path integral}
\int\mathscr{D}[A,a,j]\exp(\i S)
&=(\det\O)^{-1/2}\int\mathscr{D}[a,j]\exp\left(\i\int_{\partial M}j\wedge\De a-\f{1}{2}\st F\wedge A[j]\right)\nn\\
&=(\det\O)^{-1/2}\int\mathscr{D}[a,j]\exp\left(\i\int_{\partial M}j\wedge\left(\f{1}{2}A[j]+\de a\right)\right).
\ee
In this expression, the quantity $A[j]$ refers to the boundary value of the gauge field obtained by solving the bulk Maxwell equations and the boundary conditions, i.e. the solution to
\be\label{M A and j relation}
\de\st F=0,\q\q\st F\big|_{\partial M}=j.
\ee
These equations can either be interpreted in the form given here, i.e. as the free bulk equations of motion with specific boundary conditions, or alternatively as a bulk equations of motion which are not free but sources by boundary currents. The equivalence between these viewpoints is explained in appendix \ref{appendix: BCs}. The evaluation of $A[j]$ depends on the background spacetime geometry under consideration, but will always lead to a linear expression in $j$. The effective action on the right-hand side of \eqref{M path integral} is therefore quadratic in $j$, and integrating this auxiliary current out will therefore produce a boundary action quadratic in the edge mode field $a$. This is the same construction as in \cite{Blommaert:2018oue}, and we have now shown its generality by comparison with the CS construction of the previous section.

To be more concrete, we can solve \eqref{M A and j relation} as an example in the case of 3-dimensional Minkowski spacetime\footnote{The generalization to arbitrary dimension is of course straightforward, provided we keep track of more spacetime indices.}. In the radial gauge $A_r=0$, the boundary condition $j=\st F$ translates into the two conditions
\be
j_t=\sum_k\tilde{j}_t(k)e^{\i k\cdot x}=(\st F)_t=F_{r\phi}=\partial_rA_\phi,\q\q j_\phi=\sum_k\tilde{j}_\phi(k)e^{\i k\cdot x}=(\st F)_\phi=F_{tr}=-\partial_rA_t,
\ee
which can be solved by writing
\be
A_\phi=\sum_k\f{\tilde{j}_t(k)}{\i k_r}e^{\i k\cdot x},\q\q A_t=-\sum_k\f{\tilde{j}_\phi(k)}{\i k_r}e^{\i k\cdot x}.
\ee
Noticing the switch of components between $A$ and $j$, one can see that the term $j\wedge A[j]$ in \eqref{M path integral} is indeed quadratic in $j_t$ and $j_\phi$. In order to satisfy the bulk equations of motion, which in the Lorentz gauge are $\Box A_\mu=0$, we simply need to restrict the sum over momenta $k$ to $k^2=-k_t^2+k_r^2+k_\phi^2=0$. It is then clear that integrating \eqref{M path integral} over $j$ produces a quadratic effective action for the edge mode $a$.

In appendix \ref{appendix: M radial gauge} we give a more generic formula for this, and explain in details how $A[j]$ and the effective boundary action can be obtained in the case of 3-dimensional Maxwell theory in the radial gauge. The result of this calculation is that the effective path integral for the edge modes is
\be\label{M effective edge action 1}
\Z_\text{edge}=\int\mathscr{D}[\varphi]\exp\left(\f{\i}{2}\int\de^2k\,k^2\tilde{\varphi}(k)\tilde{G}(k)^{-1}\tilde{\varphi}(-k)\right),
\ee
where $\varphi$ is simply a field-redefinition\footnote{More precisely, we have $\varphi\coloneqq\alpha+a$, where $\alpha$ comes from the Hodge decomposition \eqref{Hodge main} of the 3-dimensional gauge field $A$. This variable $\varphi$ is therefore gauge-invariant. Alternatively, if one does not use the Hodge decomposition, the on-shell evaluation of the action in \eqref{M path integral} requires as usual a gauge-fixing, and the resulting effective action depends on $a$ instead of $\varphi$. These two viewpoints are of course equivalent.} of the initial edge modes $a$, and where $\tilde{G}(k)$ is the solution to \eqref{M A and j relation}. In appendix \ref{appendix: BCs}, we explain how one can alternatively see the boundary conditions in \eqref{M A and j relation} as boundary sources for the bulk equations of motion. By doing so we obtain an equivalent expression for the effective path integral for the edge modes, which is
\be\label{M effective edge action 2}
\Z_\text{edge}=\int\mathscr{D}[\varphi]\exp\left(-\f{\i}{2}\int_{\partial M}\de^2y\,\sqrt{|q|}\int_{\partial M}\de^2y'\,\sqrt{|q|}\,\partial^i\varphi(y)G(0,y-y')^{-1}\partial_i\varphi(y')\right).
\ee
This is the Maxwell analogue of the Poisson kernel integral obtained in \cite{Carlip:1995cd} in the case of a scalar field. There, it was argued that properly splitting and sewing scalar field theory path integrals on manifolds with boundaries requires ``scalar edge modes'' in order to reproduce the FBFK gluing formula for functional determinants\footnote{We come back to this point in section \ref{sec: MCS EE}.} \cite{Forman1987,BURGHELEA199234,doi:10.1063/1.4936074,PARK_2006,Carlip_1990}, and that the corresponding edge scalar partition function on each side of the boundary comes from the boundary term needed in order to have a well-defined variational principle for the bulk scalar field action. As such, this argument would be puzzling when transposed to Maxwell theory, since in Maxwell the bulk action already has a well-defined variational principle without the need to add a boundary term, and one does not see where the Poisson kernel contributions of \cite{Carlip:1995cd} could come from. We have shown that these contributions come from the path integral of the edge modes, whose introduction is natural since in a gauge theory they are needed in order to even have a notion of splitting of the path integral in the first place.


Finally, it is interesting to point out that in \cite{Blommaert:2018oue} it was shown in great details how the effective action \eqref{M path integral}, when evaluated in Rindler space, produces the path integral introduced by Donnelly and Wall \cite{Donnelly:2014fua,Donnelly:2015hxa} in order to explain the origin of the Kabat contact contribution to the entanglement entropy.

\subsection{Maxwell--Chern--Simons theory}
\label{sec:MCS}

In this section we study 3-dimensional MCS theory. This is equivalent to a theory of massive photons, where the so-called topological mass is provided by the CS term (and at the difference with higher dimensional Stuckelberg-like theories does not require to introduce new fields). This theory has a wide range of applications in condensed matter physics. Its boundary dynamics has been analyzed previously in \cite{balach1993MCS,Blasi:2010gw,Agarwal:2016cir,Maggiore:2018bxr} in flat space, and in \cite{Andrade:2011sx} in the context of AdS holography, and reveals the presence of a chiral edge mode, just like in pure CS theory. However, these references have conceptually different ways of introducing the edge degrees of freedom, so we believe it is useful to revisit MCS theory in the light of the general framework which we are presenting in this work. In particular, this will confirm the result of \cite{Agarwal:2016cir} concerning the contributions to the entanglement entropy, which will feature a contact term coming from the Maxwell part of the theory, but also a topological term coming from the CS part.

Introducing for convenience the topological mass $m=\k/(4\pi)$, where $\k$ is the coupling of CS theory, the bulk + boundary action is simply a combination of the extended actions \eqref{CS extended action} and \eqref{M extended action}, i.e.
\be\label{MCS extended action}
S=S_M+S_{\partial M}=\int_M-\f{1}{2}\st F\wedge F+mA\wedge F+\int_{\partial M}maF+j\wedge \De a+h,
\ee
where $h$ is a boundary Hamiltonian depending only on $j$, and which we leave unspecified for now. Following the same logic as in the previous sections, we are going to study the extended phase space of this theory, the boundary symmetries, the effective boundary dynamics, and the edge mode contribution to the entanglement entropy.

\subsubsection{Extended phase space}

The variation of the action is
\be
\delta S
&=\int_M\delta A\wedge(2m F-\de\st F)\nn\\
&\phantom{=\ }+\int_{\partial M}\delta A\wedge(m\De a-\st F-j)+\delta j\wedge(\De a+\delta_jh)+\delta a(\de j+mF)-\de(j\delta a-ma\delta A),
\ee 
from which we can read the bulk equations of motion
\be
\de\st F-2mF=0,
\ee
together with the boundary equations of motion
\be\label{MCS boundary EOMs}
j=m\De a-\st F,\q\q\De a=-\delta_jh,\q\q\de j=-mF.
\ee
The first thing one can notice is that, similarly to the case of pure Maxwell theory, the first and last set of boundary equations of motion (i.e. the ones obtained by varying $A$ and $a$) are together consistent with the bulk equations of motion. Furthermore, acting on these bulk equations of motion with the operator $\st\de\st$ leads to
\be
(\st\de\st\de-4m^2)(\st F)=(\Box-4m^2)(\st F)=0,
\ee
which shows the equivalence of MCS theory with a massive scalar field. We will come back to the precise statement of this relationship below when deriving the effective dynamics.

We can now construct the extended potential following the prescription \eqref{extended potential} and imposing the first boundary equation of motion in \eqref{MCS boundary EOMs}, which gives
\be
\!\!\theta_\text{e}=\delta A\wedge(mA-\star F)+\de\big(j\delta a-ma\delta A\big)\approx\delta A\wedge(mA-\star F)+\de\big((m\De a-\star F)\delta a-ma\delta A\big).
\ee
This in turn leads to the extended symplectic structure
\be
\Omega=\int_\Sigma\delta A\wedge\big(\delta(\star F)-m\delta A\big)+\int_S\big(m\delta(\De a)-\delta(\star F)\big)\delta a-m\delta a\delta A,
\ee
which as expected is that of Maxwell plus ($m$ times) that of CS theory. From this one can now easily check that the generators of gauge transformations obtained as $\delta_\alpha\ipp\Omega$ are indeed vanishing on-shell. For the boundary symmetries $\Delta_\alpha(A,a)=(0,-\alpha)$, one can compute $\Delta_\alpha\ipp\Omega$ to find that this quantity is integrable and has a manifestly gauge-invariant generator given by
\be\label{MCS charges}
\Q[\alpha]=\int_S\alpha(2m\De a-\star F).
\ee
Gauge-invariance of this generator is the statement that $\delta_\alpha\Q[\beta]=\delta_\alpha\ipp\Delta_\beta\ipp\Omega=0$. Finally, the algebra of these boundary charges is again given by the Ka\v c--Moody commutation relations
\be\label{MCS symmetry}
\lb\Q[\alpha],\Q[\beta]\rb=\Delta_\alpha\ipp\Delta_\beta\ipp\Omega=m\int_S\alpha\de\beta.
\ee
This shows that the surface symmetry algebra of MCS theory is identical to that of pure CS theory, even though in both cases the generators are different. This suggests the presence of a chiral boundary field, which we will now identify by evaluating the path integral.

\subsubsection{Boundary dynamics}

We now focus on the effective boundary dynamics of the theory, which as in the previous sections will be obtained by integrating out the bulk degrees of freedom. For clarity we will proceed in three steps of increasing complexity, depending on the type of boundary. First, if the spacetime has no boundary, integrating out the bulk degrees of freedom will lead as expected to the path integral of a massive scalar field. Then, if the spacetime has an outer boundary, i.e. a boundary separating the bulk from an empty manifold, the bulk will give rise to the massive scalar field, and the boundary will carry a chiral field (for the specific Hamiltonian which we choose). Finally, for an entangling boundary within the spacetime, separating the bulk between two regions, the boundary will carry a chiral field but also an additional contact contribution due to the splitting of the path integral measure between the two subregions.

A convenient way to carry out these calculations is to use the temporal gauge $A_t=0$ as well has a Hodge decomposition of the phase space variables. All the details are given in appendix \ref{appendix: MCS} and here we will only summarize the results. Forgetting for the moment about the boundary, we aim at computing the path integral for the bulk part $S_M$ of \eqref{MCS extended action}. Using the $2+1$ decomposition
\be
S_M=\int_M\Pi_a\dot{A}_a-\f{1}{4}(F_{ab})^2-\f{1}{2}(\Pi_a-m\eps_{ab}A_b)^2
\ee
together with the decomposition
\be\label{Hodge main}
A_a=\partial_a\alpha+\eps_{ab}\partial_b\beta,\q\q\Pi_a=\partial_a\xi+\eps_{ab}\partial_b\lambda,
\ee
it is explained in appendix \ref{appendix: MCS} that the path integral reduces to
\be
\Z_M
&=\int\mathscr{D}[A,\Pi]\delta(G)\exp(\i S_M)\nn\\
&=(\det\Delta)^{1/2}\int\mathscr{D}[\beta]\exp\left(\f{\i}{2}\int_M\beta(-\Delta)(\Box-4m^2)\beta\right)\nn\\
&=\big(\det(\Box-4m^2)\big)^{-1/2}.
\ee
Here $\delta(G)$ is imposing the Gauss constraint coming from the use of the temporal gauge, and the factor of $(\det\Delta)^{1/2}$ comes from three contributions: inserting the Hodge decomposition \eqref{Hodge main} in the phase space measure, in the Gauss constraint, and performing a Gaussian integral over $\lambda$. This is, as expected, the path integral for a massive scalar field, and it represents the contribution of the bulk degrees of freedom of MCS theory.

Now we have to discuss how this result will be affected by the presence of a boundary. The effective boundary dynamics will receive contributions from two sources: the boundary action $S_{\partial M}$ in the extended action \eqref{MCS extended action}, but also boundary terms coming from the the Hodge decomposition of the bulk action $S_M$. As shown in appendix \ref{appendix: MCS}, carefully collecting all these terms and imposing the bulk and boundary Gauss constraints due to the temporal gauge, we find that the boundary contributions are given by
\be
S_\text{edge}=\int_{\partial M}B[\beta]+m\partial_t\chi\partial_\phi\chi-j_t\left(\partial_\phi\chi-\f{1}{2m}\partial_t\partial_\phi\beta-\partial_r\beta\right)+h,
\ee
where $B[\beta]$ is given in \eqref{B of beta}, and where we have defined the new field $\chi\coloneqq a+\alpha+\partial_t\beta/(2m)$. Guided by the fact that the boundary symmetries \eqref{MCS symmetry} are that of a chiral field, we can now choose the boundary Hamiltonian to be
\be\label{MCS boundary Hamiltonian}
h=\f{1}{m}(j_t\mp j_\phi)j_\phi,
\ee
where $j_\phi$ is fixed by the temporal gauge to be \eqref{MCS j phi chi}. With this the boundary contributions become
\be\label{MCS effective action with h}
S_\text{edge}=\int_{\partial M}B[\beta]+m\partial_t\chi\partial_\phi\chi+j_t\left(\f{1}{m}\partial_t\partial_\phi\beta+2\partial_r\beta\right)\mp\f{1}{m}\left(m\partial_r\beta+\f{1}{2}\partial_t\partial_\phi\beta+m\partial_\phi\chi\right)^2,
\ee
and path integrating over $j_t$ finally yields the chiral action
\be\label{MCS edge action}
S_\text{edge}=m\int_{\partial M}\partial_t\chi\partial_\phi\chi\mp(\partial_\phi\chi)^2.
\ee
In the limit $m\rightarrow\infty$, which corresponds to isolating the CS piece of the action, we recover consistently \eqref{chiral boundary action}. The result can be summarized by writing the total path integral for \eqref{MCS extended action} as
\be
\Z=\big(\det(\Box-4m^2)\big)^{-1/2}\int\mathscr{D}[\chi]\exp(\i S_\text{edge})=\Z_\text{3d massive scalar}\:\Z_\text{2d chiral scalar}.
\ee
This is consistent with what we have observed from the classical theory, namely that the bulk equations of motion describe a massive scalar field, and that the boundary symmetries are that of a chiral field.

It is interesting to notice that there is a global factor of $m$ in front of the effective boundary action \eqref{MCS edge action}. Naively, this suggests that taking the $m=0$ limit of MCS theory, i.e. going back to pure Maxwell theory, leads to a vanishing effective boundary action. There are however several subtleties with this reasoning. First, one should remember that different boundary Hamiltonians were considered in the previous section for pure Maxwell theory (where we have chosen $h=0$), and in this section for MCS theory (where we have chosen a chiral Hamiltonian). Second, the analysis of pure Maxwell theory in the previous section was done in the radial gauge, while here we have studied MCS theory in the temporal gauge. This means that the effective boundary dynamics \eqref{MCS edge action} of MCS theory cannot be straightforwardly compared in the $m=0$ limit with the effective boundary dynamics \eqref{M effective edge action 2} of Maxwell theory. However, one can still go through the calculations of appendix \ref{appendix: MCS} with $h=0$ and $m=0$, which can then be compared to the results of appendix \ref{appendix: M radial gauge}. This provides the comparison between Maxwell theory in the radial and temporal gauges. As we will see below, it reveals that a crucial difference between the radial and temporal gauges is that in this latter there is a leftover determinant factor coming from the rewriting of the path integral measure, which is precisely the FBFK gluing factor identified in \cite{Agarwal:2016cir}.

Let us now make a few important observations. The first one is that, due to our choice of boundary Hamiltonian, the constraint imposed by $j_t$ in \eqref{MCS effective action with h}, namely $2m\partial_r\beta+\partial_t\partial_\phi\beta=0$, corresponds actually to the vanishing of the normal electric field to the boundary. Indeed, this latter is given in the temporal gauge by $(\star F)_\phi=F_{tr}=\partial_tA_r=\partial_t\partial_r\alpha+\partial_t\partial_\phi\beta=2m\partial_r\beta+\partial_t\partial_\phi\beta$, where for the last step we have used \eqref{MCS boundary Gauss}. In light of this, we can investigate further the boundary equations of motion given the choice of boundary Hamiltonian \eqref{MCS boundary Hamiltonian} we made. Explicitly, the first two sets of boundary equations of motion in \eqref{MCS boundary EOMs} are
\be
\begin{cases}
j_t=m\De_ta-(\star F)_t=m\De_ta-F_{r\phi},\\
j_\phi=m\De_\phi a-(\star F)_\phi=m\De_\phi a-F_{tr},
\end{cases}
\q\q
\begin{cases}
m\De_ta=\pm2j_\phi-j_t,\\
m\De_\phi a=j_\phi.
\end{cases}
\ee
Combining the two equations on the last line leads to $F_{tr}=0$, while combining the ones on the first line and using the boundary Gauss constraint \eqref{MCS boundary Gauss} leads to
\be
2m(\partial_t\chi\mp\partial_\phi\chi)+(\Box-4m^2)\beta=0,
\ee
which features the chiral and massive scalars. Second, let us point out that we can also use the decomposition \eqref{Hodge main} and the boundary Gauss constraint \eqref{MCS boundary Gauss} into the boundary observable \eqref{MCS charges} (which we smear with a function $\epsilon$ since $\alpha$ is the notation used for the Hodge decomposition) to get
\be
\Q[\epsilon]=\int_S\de\phi\,\epsilon\big(2m\De_\phi a-(\star F)_\phi\big)=\int_S\de\phi\,\epsilon(2m\De_\phi a-F_{tr})=2m\int_S\de\phi\,\epsilon\partial_\phi\chi.
\ee
This consistency check show that the boundary chiral field is indeed the variable $\chi$ which we have identified in the computation of the effective boundary action.

The results of this section are in agreement with previous observations in the literature about the fact that adding a Maxwell term to CS theory does not change the boundary symmetry algebra nor affect the presence of a boundary chiral field \cite{Wen:1992vi,Andrade:2005ur,Blasi:2010gw}. However, this does not mean that the entanglement entropy of MCS will only receive a topological contribution from the CS term. As we are going to show, the Maxwell contact term does also appear in MCS theory, although in the form of the FBFK gluing factor identified in \cite{Agarwal:2016cir}.

\subsubsection{Entanglement entropy}
\label{sec: MCS EE}

In order to discuss contributions of the edge modes to entanglement entropy, we need to consider an inner boundary which is separating the spacetime between two subregions. In this case we are on top of figure \ref{figure}, and we want to integrate the bulk degrees of freedom of one subregion.

At first sight, one would think that the result of the previous subsection is enough, and that the entanglement entropy receives contributions from two sources: the massive scalar field in the bulk (i.e. the usual distillable part with its non-universal area law), and the chiral bosons representing the effective boundary theory and providing the same topological contribution as in the pure CS case. However, as pointed out in \cite{Carlip_1990,Carlip:1995cd,Agarwal:2016cir}, one should acknowledge that there is a third contribution coming from the splitting of the path integral measure and the constraint between the two subregions. Indeed, as can be seen in \eqref{MCS path integral steps}, before integrating over the bulk fields the path integral written in terms of the Hodge decomposition and the temporal gauge is
\be
\Z_M=\det\Delta\int\mathscr{D}[\alpha,\beta,\xi,\lambda]\delta(\tilde{G})\exp(\i S_M).
\ee
The determinant factor can be traced back to the change of integration measure and the rewriting of the Gauss law in terms of the Hodge variables following \eqref{MCS Hodge symplectic} and \eqref{MCS Hodge Gauss}. Importantly, one should recognize that this factor is not here in the radial gauge path integral computed in appendix \ref{appendix: M radial gauge}. Crucially, this determinant does not simply split between the two subregions. Instead, according to the FBFK gluing formula \cite{Forman1987,BURGHELEA199234,doi:10.1063/1.4936074,PARK_2006}, we have that
\be
\det\Delta_{M\cup\bar{M}}=\mathscr{K}\det\Delta_M\det\Delta_{\bar{M}},
\ee
where the extra factor $\mathscr{K}\coloneqq\det(K_M+K_{\bar{M}})$ features the so-called Poisson kernels $K_{M,\bar{M}}$, which can be expressed in terms of the normal derivatives of the Green functions for $\Delta$ restricted to $M$ and $\bar{M}$.

In fact, we could have expected the appearance of such a factor $\mathscr{K}$ on physical grounds. Indeed, in the previous subsection we have shown that, when using the Hodge decomposition, the Maxwell fields contribute in the form of (massive) scalars, and the CS term gives rise to a chiral boundary theory. Since the massive scalar is not a gauge theory and does not bring edge modes, it would naively seem that when using the Hodge decomposition we have lost track of some of the edge modes. This is not the case, and the factor of $\mathscr{K}$ precisely keeps track of the pure Maxwell edge modes. This is the contact term identified in \cite{Agarwal:2016cir}. We have already encountered it in the previous section when deriving the boundary dynamics of pure Maxwell theory (both with and without the Hodge decomposition in radial gauge), and here it resurfaces through our change of variables and the corresponding splitting of the Gauss constraint and path integral measure.

Putting all the ingredients together, we get that the pure bulk (i.e. glued) path integral $\Z_{M\cup\bar{M}}$ over $M\cup\bar{M}$ factorizes in terms of extended bulk + boundary actions \eqref{MCS extended action} as\footnote{It should be noted that of course all the fields are integrated over in the path integrals. Here we have simply written the arguments of all the path integrals $\Z$ in \eqref{path integral splitting} in order to keep track of which variables (i.e. the initial gauge fields, the fields of the Hodge decomposition, or the edge modes) are integrated over.}
\be \label{path integral splitting}
\Z_{M\cup\bar{M}}[A,\Pi]
&=\Z[A,\Pi,a,j]\times_\text{glue}\Z[\bar{A},\bar{\Pi},\bar{a},\bar{j}]\nn\\
&=\mathscr{K}\,\Z[\alpha,\beta,\xi,\lambda,a,j]\times_\text{glue}\Z[\bar{\alpha},\bar{\beta},\bar{\xi},\bar{\lambda},\bar{a},\bar{j}]\nn\\
&=\mathscr{K}\,\big(\Z_M[\alpha,\beta,\xi,\lambda]\,\Z_\text{edge}[\alpha,\beta,a,j]\big)\times_\text{glue}\big(\Z_M[\bar{\alpha},\bar{\beta},\bar{\xi},\bar{\lambda}]\,\Z_\text{edge}[\bar{\alpha},\bar{\beta},\bar{a},\bar{j}]\big).
\ee
For the first equality, we have introduced the edge modes $(a,j)$ and $(\bar{a},\bar{j})$ on $\partial{M}=\partial{\bar{M}}$, together with the constraints enforcing that the two left and right path integrals glue together when integrating over the edge modes. This is the step which was described in \eqref{CS gluing} for CS theory. For the second equality, we have simply used the Hodge decomposition, which has produced the factor of $\mathscr{K}$, and for the third equality we have further split the Hodge decomposition into bulk and boundary actions. In the previous subsection we have seen that integrating out the bulk degrees of freedom in a subregion produces a chiral theory on its boundary. The contribution of this chiral theory has been computed in section \ref{sec: CS EE}. We can therefore conclude that the entanglement entropy in MCS theory receives contributions from three sources, i.e.
\be
\S_\text{MCS}=\S_\text{3d massive scalar}+\S_\text{2d left-right bosons}+\log\mathscr{K},
\ee
in agreement with \cite{Agarwal:2016cir}.

\subsection{BF theory}

We now discuss 3-dimensional Abelian BF theory. The generalization to arbitrary dimensions is straightforward, while the non-Abelian case is briefly discussed in appendix \ref{appendix: non-Abelian}. BF theory is an interesting case study because of its relevance for the description of topological phases of matter \cite{Cho:2010rk,Chen_2016,Chen_2017}, which also involves the study of its entanglement entropy, and because in the non-Abelian case it describes 3-dimensional gravity in the first order formulation. Recently there has also been a lot of interest in a 2-dimensional BF theory model known as Jackiw--Teitelboim gravity (although there it appears with the non-Abelian gauge group $\text{SL}(2,\mathbb{R})$) \cite{Lin:2018xkj,Blommaert:2018iqz,Mertens:2018fds,Gonzalez:2018enk,Harlow:2018tqv}. We hope to apply our construction of the boundary dynamics to these more complicated cases in the future. General ideas on the boundary dynamics of 3-dimensional BF theory have already been formulated in \cite{Momen:1996dg,Blasi_2012}, where the authors have identified chiral boundary currents. Here we show that depending on the choice of boundary Hamiltonian it is possible to obtain a chiral or non-chiral boundary scalar field theory.

To construct the bulk + boundary action, in addition to the bulk 1-forms $A$ and $B$ we add on the boundary the 0-forms $a$ and $b$ and a current 1-form $j$, and consider
\be\label{BF extended action 1}
S=\int_MB\wedge F+\int_{\partial M}bF+j\wedge\De a+h.
\ee
The role of the new boundary field $b$ is to make the total action invariant under the so-called shift transformations
\be
\delta_\phi B=\de\phi,\q\q\delta_\phi b=-\phi.
\ee
This is the edge mode field for the shift symmetry. Similarly to the cases studied above, this action produces the corner term which is needed for the extended phase space. To see this, consider the variation
\be\label{BF variation extended action 1}
\delta S
&=\int_M\delta B\wedge F+\delta A\wedge\de B\nn\\
&\phantom{=\ }+\int_{\partial M}\delta A\wedge(B+\de b-j)+\delta j\wedge(\De a+\delta_jh)+\delta a\de j+\delta bF-\de(j\delta a-b\delta A).
\ee
Using the boundary equation of motion $j=B+\de b$, the extended potential becomes
\be
\theta_\text{e}=\delta A\wedge B+\de(j\delta a-b\delta A)\approx\delta A\wedge B+\de\big((B+\de b)\delta a-b\delta A\big).
\ee
Now, notice that there exists an alternative boundary action, related to this one by a change of polarization, and which reads\footnote{In higher-dimensional BF theory writing this action would require to change the form degree of $j$.}
\be\label{BF extended action 2}
S'=\int_MB\wedge F+\int_{\partial M}B\wedge\De a+j\wedge(B+\de b)+h.
\ee
The bulk equations of motion are of course unchanged, and the on-shell variation is
\be\label{BF variation extended action 2}
\delta S'\approx\int_{\partial M}\delta B\wedge(\De a-j)+\delta j\wedge(B+\de b+\delta_jh)+\delta a\de B+\delta b\de j-\de(j\delta b+B\delta a),
\ee
from which one can clearly see the symmetry with \eqref{BF variation extended action 1}. Using the boundary equation of motion $j=\De a$, the extended potential becomes
\be
\theta'_\text{e}=\delta A\wedge B+\de(j\delta b+B\delta a)\approx\delta A\wedge B+\de(\De a\delta b+B\delta a)=\theta_\text{e}+\delta\de(b\De a).
\ee
Notice that with the introduction of the edge modes the boundary equations of motion in the extended action ``reverse'' the polarization. Indeed, in \eqref{BF variation extended action 1} instead of fixing $A$ on the boundary we use the boundary equation of motion to fix $(B,b)$ in terms of $j$. Conversely, in \eqref{BF variation extended action 2} instead of fixing $B$ on the boundary we impose a condition on $(A,a)$.

Since the potentials derived from the two extended actions differ by a total field variation, they lead to the same symplectic structure (although in a discretized setting the change of polarization can lead to inequivalent symplectic structures \cite{Dupuis:2017otn,Delcamp:2018sef,Freidel:2018pbr}), which is
\be
\Omega=-\int_\Sigma\delta A\wedge\delta B-\int_S\delta(\De a)\delta b+\delta B\delta a,
\ee
in agreement with \cite{Geiller:2017xad}. With this extended symplectic structure, we can then show as expected that the ``Lorentz'' and shift gauge generators $\delta_\alpha\ipp\Omega$ and $\delta_\phi\ipp\Omega$ vanish on-shell. In addition, we now also have boundary symmetries acting on the edge modes as
\be
\Delta^\text{g}_\alpha(a,b)=(\alpha,0),\q\q\Delta^\text{t}_\phi(a,b)=(0,\phi),
\ee
and generated by the boundary observables
\be
\Q^\text{g}[\alpha]=\int_S\alpha(B+\de b),\q\q\Q^\text{t}[\phi]=\int_S\phi(A+\de a).
\ee
As expected from CS theory, these generators satisfy a $\text{U}(1)\times\text{U}(1)$ Ka\v c--Moody algebra
\be
\lb\Q^\text{g}[\alpha],\Q^\text{g}[\beta]\rb=0,\q\q\lb\Q^\text{t}[\phi],\Q^\text{t}[\chi]\rb=0,\q\q\lb\Q^\text{g}[\alpha],\Q^\text{t}[\phi]\rb=\int_S\phi\de\alpha.
\ee
Based on this algebra of boundary symmetries, we can expect to find two chiral fields on the boundary.

Now, let us look at the effective boundary dynamics obtained by integrating out the bulk degrees of freedom. This calculation follows closely that in CS theory, since the actions are similar. In components, \eqref{BF extended action 1} becomes
\be
S
&=\int_MB_t(\partial_rA_\phi-\partial_\phi A_r)+A_t(\partial_rB_\phi-\partial_\phi B_r)+B_\phi\partial_tA_r-B_r\partial_tA_\phi\nn\\
&\phantom{=\ }+\int_{\partial M}A_t(j_\phi-B_\phi-\partial_\phi b)+A_\phi\partial_tb+j_\phi\partial_ta-j_t(A_\phi+\partial_\phi a)+h.
\ee 
As usual, the time components $A_t$ and $B_t$ are Lagrange multipliers enforcing the bulk Gauss and flatness constraints
\be
\eps^{ab}\partial_aB_b=0,\q\q\eps^{ab}\partial_aA_b=0,
\ee
and on the boundary the relation
\be
j_\phi=B_\phi+\partial_\phi b.
\ee
Path integrating over $A_t$ and $B_t$ allows us to go on-shell and to write $A_a=\partial_a\alpha$ and $B_a=\partial_a\beta$, and with this the extended action reduces to
\be
S_\text{edge}=\int_{\partial M}\partial_t\varphi\partial_\phi\psi-j_t\partial_\phi\varphi+h,
\ee
where we have introduced the gauge-invariant scalars $\varphi\coloneqq a+\alpha$ and $\psi\coloneqq b+\beta$. Starting instead from the alternative action \eqref{BF extended action 2} we have
\be
S'
&=\int_MB_t(\partial_rA_\phi-\partial_\phi A_r)+A_t(\partial_rB_\phi-\partial_\phi B_r)+B_\phi\partial_tA_r-B_r\partial_tA_\phi\nn\\
&\phantom{=\ }+\int_{\partial M}B_t(j_\phi-A_\phi-\partial_\phi a)+B_\phi\partial_ta+j_\phi\partial_tb-j_t(B_\phi+\partial_\phi b)+h,
\ee 
and path integrating over $A_t$ and $B_t$ gives
\be
S'_\text{edge}=\int_{\partial M}\partial_t\psi\partial_\phi\varphi-j_t\partial_\phi\psi+h.
\ee
The kinetic terms of the two effective boundary actions differ only by an integration by parts, and show that $\varphi$ and $\psi$ are canonically conjugated (with a derivative $\partial_\phi)$. This means that we have the freedom of integrating out one of the fields (together with the current $j$) in order to obtain an effective boundary dynamics for the remaining one. This dynamics will depend on the choice of boundary Hamiltonian.

With the chiral Hamiltonian $h=(j_t\mp j_\phi)j_\phi$ which we have used previously in CS and MCS theory we obtain
\be
S_\text{edge}=\int_{\partial M}\partial_t\varphi\partial_\phi\psi-j_t(\partial_\phi\varphi-\partial_\phi\psi)\mp(\partial_\phi\psi)^2.
\ee
Integrating over $j_t$ then yields the chiral action
\be
S_\text{edge}=\int_{\partial M}\partial_t\varphi\partial_\phi\varphi\mp(\partial_\phi\varphi)^2.
\ee
One can verify that using the alternative action \eqref{BF extended action 2} also leads to this chiral action. Alternatively, we can also use the boundary Hamiltonian $2h=\pm\as j\wedge j=\pm(j_t^2-j_\phi^2)$. In CS theory, this has produced the chiral action \eqref{chiral boundary action}. Here, integrating over $j_t$ leads to
\be
S_\text{edge}=\int_{\partial M}\partial_t\varphi\partial_\phi\psi\mp\f{1}{2}(\partial_\phi\varphi)^2\mp\f{1}{2}(\partial_\phi\psi)^2,
\ee
from which the equations of motion obtained by varying $\varphi$ or $\psi$ are chiral for $\psi$ and $\varphi$ respectively. This is the form of the edge theory which was studied in a condensed matter context in \cite{Chen_2016} (although in four dimensions), where it has been shown that it can also be quantized using the Hamiltonian methods of appendix \ref{appendix:CS-EE}, and leads to a topological contribution to the entanglement entropy of $-\log\k$.

Going one step further, one may also integrate out one of the two chiral fields. For example, integrating out $\psi$ leads to
\be
S_\text{edge}=\pm\f{1}{2}\int_{\partial M}(\partial_t\varphi)^2-(\partial_\phi\varphi)^2,
\ee
which is now a single non-chiral scalar field.

\section{Perspectives}
\label{sec:5}

Many recent papers are revisiting and unraveling the role of edge states in gauge theories and gravity, both quasi-locally and asymptotically. This effort is pushing forward and shedding new light onto ideas which have been around for a long time in condensed matter, pure gauge theory, and the study of black hole entropy \cite{MR1025431,Wen:1992vi,Balachandran:1995iq,Balachandran:1995qa,Carlip:1995cd}, but whose generality has perhaps not yet been fully appreciated. In particular, this has highlighted the important role of edge states in the definition of entanglement entropy \cite{Buividovich:2008gq,Donnelly:2011hn,Lin:2018bud,Blommaert:2018oue}, their relationship with the infrared properties of massless theories \cite{Strominger:2017zoo,Kapec:2015ena,Blommaert:2018rsf}, and has unraveled rich examples of boundary symmetries and dynamics in the gravitational context \cite{Afshar:2019axx,Afshar:2017okz,Grumiller:2016pqb,Grumiller:2017sjh,Donnelly:2016auv,Freidel:2019ofr}. In spite of these remarquable developments, we believe that there is still no unified treatment for describing how exactly the edge modes arise (in relation to e.g. specific choices of boundaries or boundary conditions) and for studying their dynamics. This is of course a notoriously complicated open issue, and for example already in the seemingly simple case of 2-dimensional gravity the boundary dynamics is not fully understood \cite{Blommaert:2018iqz,Harlow:2018tqv,Afshar:2019axx}.

While we do not claim to have solved this issue, in the present work we have taken a small step towards proposing a conceptual general framework for describing the dynamics of edge modes. For this, we have first argued and recalled in the introduction and in section \ref{sec:2} why it is necessary to construct quasi-local extensions of the phase space of a gauge theory to include extra edge mode fields. The edge modes restore the seemingly broken gauge-invariance due to the presence of the boundary and parametrize the boundary symmetries of the theory. This has proven to lead to known and consistent results in the familiar cases of e.g. Chern--Simons and BF theory \cite{Geiller:2017xad,Geiller:2017whh}, but also enabled to unravel some new large sets of boundary symmetries in gravitational theories \cite{Donnelly:2016auv,Freidel:2018fsk,Freidel:2019ees,Freidel:2019ofr,Freidel:2020xyx,Freidel:2020svx}. In section \ref{sec:3}, we have then proposed an extended variational principle (similar to the one proposed in \cite{Harlow:2019yfa}, although in a different context) which supplements the bulk symplectic structure with a boundary symplectic structure including the edge mode fields and descending from a boundary action. It is indeed natural to view the complete definition of a field theory as the specification of a Lagrangian for each submanifold, and fusion products gluing the associated symplectic structures together. The addition of a boundary action with edge modes provides a covariant realization of the extended Hilbert space construction and gives rise to the setup represented schematically on figure \ref{figure}. The edge modes are necessary in order to factorize the Hilbert space, phase space, or path integral of a theory. Most importantly, once a theory has been split between two subregions by introducing edge modes on the boundary together with a boundary current $j$ and a Hamiltonian enforcing boundary conditions, the bulk of a subregion can be evaluated on-shell and a residual dynamics gets imprinted on the boundary. This agrees with the proposal made in \cite{Blommaert:2018oue}, which we have now therefore connected with the extended phase space constructions of \cite{Donnelly:2016auv,Geiller:2017xad,Geiller:2017whh}.

We have put this proposal for deriving the boundary dynamics of edge modes to the test in various examples of theories with Abelian internal gauge symmetries (we plan to come back to the study of the non-Abelian case and diffeomorphisms in future work). In the case of Chern--Simons theory, we have shown that we recover consistently the known chiral boundary theory, and rederived the associated topological contribution of the edge modes to the entanglement entropy. More importantly, we have then applied this construction to Maxwell theory. In this case, the need to include edge modes in order to achieve a factorization of the path integral and the Hilbert space had already been acknowledged in \cite{Casini:2014aia,Blommaert:2018rsf,Blommaert:2018oue,Pretko:2018nsz}. We have now however explicitly shown how the inclusion of edge modes leads to a boundary action describing the extended phase space structure introduced in \cite{Donnelly:2016auv}, and furthermore recovered the boundary dynamics of \cite{Blommaert:2018oue} once the bulk gauge field is integrated out. This confirms the contribution of the edge modes to the entanglement entropy of Maxwell fields. More precisely, in \eqref{M path integral} we have obtained the generic form of the effective path integral for the extended Maxwell theory \eqref{M extended action}, with photons in the bulk and dynamical edge modes on the boundary. This expression is valid in any dimension. Specializing to the 3-dimensional case, we have used the radial gauge to rewrite the effective boundary path integral for the edge modes in the form \eqref{M effective edge action 1} and \eqref{M effective edge action 2}. Then, we have reproduced this calculation in the temporal gauge with the inclusion of a Chern--Simons mass term. In this case, we have shown for a specific choice of boundary Hamiltonian that the boundary dynamics is that of a chiral field, but also that the path integral measure features a determinant factor which needs to be properly split between the two subregions if one wants to factorize the path integral. In agreement with \cite{Agarwal:2016cir}, this is the origin of the contact term contribution to the entanglement entropy when using a Hodge decomposition and the temporal gauge.

Following this introductory work, whose goal was to present the framework and illustrate the connection between the boundary action, the extended phase space, and the on-shell boundary dynamics, many directions remain to be explored. We conclude by listing a few of these directions.
\begin{itemize}
\item \textit{Relationship between boundary conditions, Hamiltonian, and dynamics.} It is now important to study more precisely the nature of the boundary theory obtained for Maxwell theory (first steps in this direction have been taken in \cite{Blommaert:2018oue}), and in particular to investigate its dependency on the choice of boundary Hamiltonian. We also plan on understanding the relationship with other constructions involving the quasi-local edge modes of Maxwell theory \cite{Barnich:2019xhd}. In the case of 2- and 3-dimensional non-Abelian BF theory (which describe first order gravity) on the other hand, studying the boundary dynamics for various choices of boundary Hamiltonians (i.e. boundary conditions) is an important task in order to connect with proposals for the boundary dynamics and the boundary symmetries which have been presented already in the literature \cite{Banados:1998ta,Coussaert:1995zp,Carlip:2016lnw,Grumiller:2016pqb,Grumiller:2017sjh,Afshar:2019axx,Gonzalez:2018enk,Blommaert:2018iqz,Mertens:2018fds}.
\item \textit{Contributions to entanglement entropy.} In this work we have tried to argue from the point of view of the continuum theory that edge modes contribute to the computation of entanglement entropy. The reason for this can be traced back to the need to introduce edge modes in the first place in order to have a notion of decomposition of a region into subregions, and to the fact that a boundary dynamics gets induced once the bulk degrees of freedom of a subregion are integrated out. In the case of topological theories such as Chern--Simons and Abelian BF theory this fact is well established, as we have seen in section \ref{sec: CS EE}. In the case of Maxwell theory, we have argued along similar lines as \cite{Agarwal:2016cir,Blommaert:2018oue,Blommaert:2018rsf} that the edge modes contribute through their effective boundary path integral \eqref{M effective edge action 2}. On the one hand, it was shown in \cite{Blommaert:2018rsf} that this path integral, when evaluated in Rindler space, reproduces the edge mode contribution introduced in \cite{Donnelly:2014fua,Donnelly:2015hxa} in order to explain the origin of the Kabat contact term. It would be interesting to investigate further this relationship already in the case of Minkowski space. On the other hand, the authors of \cite{Agarwal:2016cir} have argued that the Kabat contact term arises from the FBFK gluing factor for functional determinants. This argument fits also nicely with that of \cite{Carlip:1995cd} concerning the contribution of edge modes when factorizing a path integral between two subregions. However, in section \ref{sec:MCS} we were only able to argue that the FBFK gluing factor appears in Maxwell--Chern--Simons theory once we write the theory over $M\cup\bar{M}$ in the temporal gauge with the Hodge decomposition, and then decompose it between two subregions (which then requires to split the path integral measure and to use the FBFK formula). We were not able to compare the results of this calculation with that of pure Maxwell theory because the massless limit of Maxwell--Chern--Simons theory is anomalous. In future work, it would therefore be interesting to look more closely at the 3-dimensional Maxwell theory in the temporal gauge with the Hodge decomposition, and to attempt at obtaining explicitly the FBFK gluing factor from the effective boundary path integrals coming from the two separate subregions. In other words, while the present work has focused on the general framework and provided preliminary examples of the dynamics and the contribution of edge modes, in future work we will focus more precisely on 3-dimensional Maxwell theory in order to compare the various computations of entanglement entropy and contact terms which have appeared in the literature, the various gauge choices (temporal or radial), and the use of the Hodge decomposition to rewrite Maxwell theory as a scalar field. This should be able to provide a comprehensive picture of the relationship between the edge modes and the entanglement entropy for Maxwell fields.
\item \textit{Extension to other theories.} There are essentially two types of theories to which the present work should be extended: non-Abelian gauge theories, and theories with diffeomorphism symmetry. Here we have presented the construction of the extended action and symplectic structure for non-Abelian gauge theories in appendix \ref{appendix: non-Abelian}. This does not present any conceptual difference with the Abelian case. However, the study of the boundary dynamics of non-Abelian theories is for the most part unknown (preliminary steps have been taken in \cite{Blommaert:2018oue}). In the case of diffeomorphism symmetry however, already the introduction of the edge modes in the extended phase space or the boundary action is conceptually different from what we have presented in this work, since it requires embedding variables \cite{Donnelly:2016auv,Speranza:2019hkr}. It would be interesting to construct and study a boundary action which introduces these embedding variables and leads to the extended symplectic structure for metric gravity derived in \cite{Donnelly:2016auv}. Finally, in the case of topological theories it would be interesting to investigate whether the boundary actions used to introduce edge modes on lower-dimensional submanifolds can be understood from the point of view of extended TQFTs \cite{Gelca_1997,tsumura20132categorical,Donnelly:2018ppr,Carqueville:2017fmn}, and (in the case of e.g. 3-dimensional BF theory) if the edge modes introduced in \cite{Geiller:2017xad} have an interpretation in terms of the corresponding extended TQFT data \cite{2010arXiv1004.1533K,Dittrich:2016typ}.
\item \textit{Link with the dynamics of soft modes at asymptotic infinity.} A major open question is to relate the edge modes presented in this work (and in all the references cited in the introduction) to the so-called soft modes which appear in the infrared regime of massless theories \cite{Strominger:2017zoo}. Some steps in this direction have already been taken in \cite{Blommaert:2018rsf,Freidel:2019ohg}, and there have also been proposals for the description of the infrared dynamics itself \cite{Gervais:1980bz,Balachandran:2018jwf}. It is therefore natural to ask whether such proposals can be recovered (or corrected) from the boundary dynamics of Maxwell theory presented in this work, provided we can choose appropriate boundary conditions and push the boundary to infinity.
\item \textit{Link with other treatments of edge modes, gauge-invariance, and gluing.} Finally, we would like to point out that an extensive framework for dealing with gauge symmetries and gauge-invariance in gauge theories has been developed in a series of recent articles \cite{Gomes:2016mwl,Gomes:2018dxs,Gomes:2018shn,Gomes:2019xto,Gomes:2019xhu,Gomes:2019otw,Gomes:2019rgg}. Instead of explicitly introducing edge mode fields in order to restore the seemingly-broken gauge-invariance (as we do here), this framework adopts a more geometrical viewpoint and introduces instead a connection in field space. While there are some sharp conceptual differences between this approach and the one presented here, we would like in future work to investigate further the possible points of convergence, and whether the procedure of gluing/splitting presented here can be made more geometrical and recast in the above-mentioned formulation.
\end{itemize}

We hope that this work will trigger some of these developments, and eventually lead to a coherent picture for including edge modes in gauge theories and deriving their holographic dynamics.

\section*{Acknowledgements}

We would like to thank Glenn Barnich, Andreas Blommaert, William Donnelly, Laurent Freidel, Christophe Goeller, Robert Leigh, Seyed Faroogh Moosavian, and \DJ or\dj e Radi\v cevi\' c for discussions and/or comments. PJ's research is supported by the DPST Grant from the government of Thailand, and Perimeter Institute for Theoretical Physics. Research at Perimeter Institute is supported by the Government of Canada through the Department of Innovation, Science and Economic Development, and by the Province of Ontario through the Ministry of Research \& Innovation.

\appendix

\section{Covariant phase space}
\label{appendix:1}

In the covariant phase space formalism, we make extensive use of the field variations $\delta$ and the notation $\delta_\alpha$ for gauge transformations with field-independent parameter $\alpha$. Following a rather standard nomenclature, we view $\delta$ as a differential in the space of fields. For a Lagrangian defined on a $d$-dimensional spacetime manifold, the potential $\theta$ is therefore a $(d-1,1)$-form, meaning a $(d-1)$-form in spacetime and a 1-form in the space of fields. Repeated action of $\delta$ is understood with anti-symmetrization, and the symplectic structure $\Omega$ is therefore a field space 2-form
\be
\Omega=\int_\Sigma\delta\theta[\delta]=\int_\Sigma\delta_1\theta[\delta_2]-\delta_2\theta[\delta_1].
\ee
Gauge transformations $\delta_\alpha$, if understood as tangent vectors, can then be contracted with field space differential forms. This is the meaning of the notation $\delta_\alpha\ipp\Omega$, which translates explicitly into
\be
\delta_\alpha\ipp\Omega=\int_\Sigma\delta_\alpha\theta[\delta]-\delta\theta[\delta_\alpha].
\ee
By analogy with differential forms on spacetime, the object $\delta_\alpha\ipp\Omega$, which is the contraction of a field space vector with a 2-form, is therefore a field space 1-form.

\section{Maxwell theory in radial gauge}
\label{appendix: M radial gauge}

Inspired by holography, let us work in the radial Hamiltonian formulation, where the radial coordinate $r$ is treated as the Hamiltonian time. The natural gauge fixing corresponding to this choice is then the radial gauge where we set $A_r =0$. This condition, together with the Lorenz gauge $\partial^\mu A_\mu=0$, are the gauge choices used in \cite{Blommaert:2018oue} and in section \ref{sec: Maxwell}.

Let us focus on the 3-dimensional case, and decompose the spacetime coordinates $x^\mu=(t,r,\phi)$ as $x^\mu=(r,y^i)$, with $y^i=(t,\phi)$ the coordinates on the $r=\text{constant}$ hypersurfaces. With this, the spacetime metric can be decomposed as
\be
g_{\mu\nu}\de x^\mu\de x^\nu=\de r^2+q_{ij}(r)\de y^i\de y^j,
\ee
where in cylindrical coordinates
\be
q_{ij}(r)\de y^i\de y^j=-\de t^2+r^2\de\phi^2.
\ee
Placing the time-like boundary $\partial M$ at $r=\ell$, we have that the induced metric at the boundary is $g_{ij}\big|_{\partial M}=q_{ij}(\ell)\eqqcolon q_{ij}$. With this radial decomposition, all the total derivatives are along the directions $y^i$, and can therefore be discarded because of our choice of cylindrical topology $M=\mathbb{R}\times D$. We will therefore freely integrate by parts over $t$ and $\phi$.

Similarly to the standard Hamiltonian analysis with respect to time $t$, the bulk Maxwell action in \eqref{M extended action} can be written in radial Hamiltonian form as
\be
S_M=\int_M\sqrt{|g|}\big(\Pi^i\partial_rA_i-H\big),
\ee
where the conjugate momentum to $A_i$ is $\Pi^i\coloneqq-F^{ri}$, and the Hamiltonian is
\be
H=\f{1}{4}F^{ij}F_{ij}-\f{1}{2}\Pi^i\Pi_i.
\ee
We now use a Hodge decomposition of the gauge field and the momenta by writing
\be
A_i=\partial_i\alpha+\eps_{ij}\partial^j\beta\coloneqq\partial_i\alpha+\beta_i,\q\q\Pi^i=\partial^i\xi+\eps^{ij}\partial_j\lambda\coloneqq\partial^i\xi+\lambda^i,
\ee
where $\beta_i$ and $\lambda^i$ are the divergence-free parts $\partial^i\beta_i=0=\partial_i\lambda^i$. The constraint enforced by the radial gauge fixing is that
\be\label{radial constraint}
\partial_i\Pi^i=\partial^2\xi=0,
\ee
where we define $\partial^2\coloneqq\partial_i\partial^i$ to be the Laplace operator on the slices of constant radius $r$. In terms of this Hodge decomposition, the canonical term of the action and the terms of the Hamiltonian can be decomposed as
\begin{subequations}
\be
\Pi^i\partial_rA_i&=\partial^i\xi\partial_i\partial_r\alpha+\lambda^i\partial_i\partial_r\alpha+\partial^i\xi\partial_r\beta_i+\lambda^i\partial_r\beta_i=-\partial^2\xi\partial_r\alpha+\lambda^i\partial_r\beta_i,\\
\Pi^i\Pi_i&=\partial^i\xi\partial_i\xi+2\partial^i\xi\lambda_i+\lambda^i\lambda_i=-\xi\partial^2\xi+\lambda^i\lambda_i,\\
\f{1}{2}F^{ij}F_{ij}&=\partial^i\beta^j\partial_i\beta_j-\partial^i\beta^j\partial_j\beta_i=-\beta^i\partial^2\beta_i.
\ee
\end{subequations}
With this, the bulk Maxwell action becomes
\be
S_M=\int_M\sqrt{|g|}\left(-\left(\partial_r\alpha+\f{1}{2}\xi\right)\partial^2 \xi+\lambda^i\partial_r\beta_i+\f{1}{2}\beta^i\partial^2\beta_i+\f{1}{2}\lambda^i\lambda_i\right).
\ee
The first term vanishes once the constraint \eqref{radial constraint} is imposed. Path integrating over the momentum variable $\lambda$ then yields
\be\label{radial Maxwell effective bulk}
S_M=\f{1}{2}\int_M\sqrt{|g|}\,\big(\beta^i\partial^2\beta_i-\partial_r\beta^i\partial_r\beta_i\big),
\ee
so one can see that the bulk contribution is determined by the divergence-free part of the Hodge decomposition. Notice that if we write explicitly the path integral with the Hodge decomposition, taking into account the change of measure
\be
\mathscr{D}[A,\Pi]=\mathscr{D}[\alpha,\beta,-\partial^2\xi,\lambda]=\det(-\partial^2)\mathscr{D}[\alpha,\beta,\xi,\lambda]
\ee
and the change of variables in the Gauss constraint using
\be
\delta\big(\partial^2\xi\big)=\det(-\partial^2)^{-1}\delta(\xi),
\ee
the two determinant factors cancel out. Furthermore, there is no determinant factor produced when integrating over $\lambda$, and we are therefore left with the path integral over $\beta$ of \eqref{radial Maxwell effective bulk}. This is a crucial difference with the Hodge decomposition of the path integral in the temporal gauge, which produces determinant factors as explained in appendix \ref{appendix: MCS}.

We can now add to this the boundary action containing the edge mode field $a$ to obtain the full bulk + boundary action \eqref{M extended action}. More precisely, using the radial gauge with the Hodge decomposition, imposing the constraint \eqref{radial constraint} and integrating out the momenta $\lambda$, we obtain the extended Maxwell action\footnote{We deliberately include the boundary metric into the boundary volume form. This does not alter the analysis of the extended phase space symplectic structure, as it can be viewed as the redefinition of $j$.}
\be\label{3d Maxwell beta action}
S=\f{1}{2}\int_M\sqrt{|g|}\big(\beta^i\partial^2\beta_i-\partial_r\beta^i\partial_r\beta_i\big)+\int_{\partial M}\sqrt{|q|}\,\eps^{im}j_i(\beta_m+\partial_m\varphi),
\ee
where we have introduced the gauge-invariant scalar $\varphi\coloneqq a+\alpha$. Variation with respect to $\beta_i$ gives in the bulk the flat massless Klein--Gordon equation in cylindrical coordinates,
\be
\f{1}{\sqrt{|g|}}\partial_\mu\big(\sqrt{|g|}\partial^\mu\beta^i\big)=-\partial^2_t\beta^i+\f{1}{r}\partial_r(r\partial_r\beta^i)+\f{1}{r^2}\partial^2_\phi\beta^i=0, 
\ee
and on the boundary the boundary condition
\be
\partial_r\beta^i(\ell,y)=-\eps^{im}j_m(y),
\ee
where $\beta^i(\ell,y)=\beta^i(r,y)\big|_{\partial M}$. Similarly to the calculation \eqref{M path integral}, path integrating over $\beta$ in the bulk produces an operator determinant for the massless scalar, and, recalling that on-shell the bulk action is a boundary term, the total boundary action (i.e. the initial one in \eqref{M extended action} plus the piece coming on-shell from the bulk) becomes
\be
S_\text{edge}=\int_{\partial M}\sqrt{|q|}\left(\eps^{im}j_i(\beta_m+\partial_m\varphi)-\f{1}{2}\beta_i\partial_r\beta^i\right)=\int_{\partial M}\sqrt{|q|}\eps^{im}j_i\left(\f{1}{2}\beta_m[j]+\partial_m\varphi\right),
\ee
where we have used the boundary condition. This expression is the effective boundary action in \eqref{M path integral}.

The boundary value $\beta_i[j]$ is now determined by solving the bulk equation of motion with respect to a Neumann-type boundary condition, i.e.
\be
\Box\beta^i=0,\q\q\partial_r\beta^i(\ell,y)=-\eps^{im}j_m(y),
\ee
which is the 3-dimensional version of \eqref{M A and j relation} in the radial gauge. The bulk equation is solved by going to momentum space, with $k_i=(k_t,k_\phi)$, as
\be
\beta^i(r,y)=\int\f{\de^2k}{2\pi}J_{k_\phi}(rk_t)\tilde{\beta}^i(k)e^{\i k\cdot y},
\ee
where $k\cdot y=k_iy^i$, and $J_n$ denotes the Bessel function of integer order. Note that $k_\phi$ has a discrete spectrum due to the compactness of the $\phi$ direction. Now, writing
\be
j_i(y)=\int\f{\de^2k}{2\pi}\tilde{j}_i(k)e^{\i k\cdot y},\q\q\varphi(y)=\int\f{\de^2k}{2\pi}\tilde{\varphi}(k)e^{\i k\cdot y},
\ee
the boundary condition translates into
\be
\tilde{\beta}^i(k)=-\f{1}{k_t\partial_rJ_{k_\phi}(\ell k_t)}\eps^{im}\tilde{j}_m(k).
\ee
We therefore obtain $\beta$ at the boundary in the form
\be
\beta^i(\ell,y)=-\eps^{im}\int \f{\de^2 k}{2\pi}\tilde{G}(k)\tilde{j}_m(k)e^{\i k\cdot y},
\ee
where $\tilde{G}(k)$ represents the momentum space Green function. In the present case, we have
\be
\tilde{G}(k)=\f{J_{k_\phi}(\ell k_t)}{k_t\partial_rJ_{k_\phi}(\ell k_t)}.
\ee
Putting this together, the effective edge mode action becomes
\be
S_\text{edge}=\int\de^2k\left(\f{1}{2}\tilde{j}^i(k)\tilde{G}(k)\tilde{j}_i(-k)+\eps^{im}\tilde{j}_i(k)(\i k_m)\tilde{\varphi}(-k)\right).
\ee
As expected, this action is quadratic in $\tilde{j}$. We can now choose to integrate over $\tilde{\varphi}$ to obtain the condition
\be
\eps^{im}\tilde{j}_i(k)k_m=0\q\Rightarrow\q\tilde{j}_i(k)\sim k_i,
\ee
which is equivalent to the condition that boundary current is conserved, i.e. $\de j=0$. The path integral then reduces to an integral over all conserved currents of the quadratic action written above. Alternatively, we can integrate out the current $\tilde{j}_i$ in order to get the quadratic action for $\tilde{\varphi}$ which is
\be\label{effective Maxwell momentum action}
S_\text{edge}=\f{1}{2}\int\de^2k\,k^2\tilde{\varphi}(k)\tilde{G}(k)^{-1}\tilde{\varphi}(-k).
\ee
This is the effective boundary action in momentum space.

\section{Maxwell--Chern--Simons theory in temporal gauge}
\label{appendix: MCS}

Here we present the detailed calculations of the Hodge decomposition of MCS theory, keeping carefully all the boundary terms, together with the manipulations of the path integral with and without boundaries. We recall that the spacetime has the topology $M=\mathbb{R}\times D$ of an infinite cylinder with $x^\mu=(t,r,\phi)$ such that $\eps^{tr\phi}=1$, so total derivatives $\partial_t$ and $\partial_\phi$ can be ignored. We will constantly use this fact to freely move these derivatives around. For simplicity, even though we are using cylindrical-looking coordinates, our choice of metric will be $g_{\mu\nu}=\text{diag}(-1,1,1)$, so we will allow ourselves to write the expressions below with all spatial indices downstairs for simplicity, and drop the determinant $\sqrt{|g|}$ in the integrals. This does not affect the results of this appendix. We denote the spatial Levi--Civita tensor by $\eps_{ab}\coloneqq{\eps^t}_{ab}$, with $\eps_{ab}\eps_{ac}=\delta_{bc}$ and $\eps_{r\phi}=1$. Finally, the spatial Laplacian will be denoted by $\Delta\coloneqq\partial^a\partial_a$, the wave operator by $\Box=-\partial_t^2+\Delta$, and the time derivative $\partial_t\alpha$ by a dot $\dot{\alpha}$.

With a $2+1$ decomposition identifying the momentum, the Gauss constraint, and the Hamiltonian, the bulk part of the action in \eqref{MCS extended action} can be written as
\be
S_M=\int_M\Pi_a\dot{A}_a+A_tG-H-\int_{\partial M}A_t\Pi_r.
\ee
The momentum conjugated to the gauge field is
\be\label{MCS momentum to A}
\Pi_a=-F^{ta}+m\eps_{ab}A_b=F_{ta}+m\eps_{ab}A_b,
\ee
where one should notice that the first term has picked up a sign because we have lowered the indices. The Gauss constraint is
\be
G=\partial_a(\Pi_a+m\eps_{ab}A_b),
\ee
and the Hamiltonian is
\be
H=\f{1}{4}(F_{ab})^2+\f{1}{2}(\Pi_a-m\eps_{ab}A_b)^2\eqqcolon\f{1}{4}(F_{ab})^2+\f{1}{2}(E_a)^2.
\ee
We are now going to rewrite these quantities using the Hodge decomposition
\be
A_a=\partial_a\alpha+\eps_{ab}\partial_b\beta,\q\q\Pi_a=\partial_a\xi+\eps_{ab}\partial_b\lambda.
\ee
In various expressions, we will only keep total derivatives in $r$ since the ones in $\phi$ vanish when going to the boundary. In these total derivative, which will give boundary terms, we will furthermore use \eqref{MCS momentum to A} and the temporal gauge $A_t=0$ to rewrite
\be\label{Lagrangian Hodge variables}
\xi=\dot{\alpha}-m\beta,\q\q\lambda=\dot{\beta}+m\alpha.
\ee
This is justified since on the boundary it is $A$ (and therefore $\alpha$ and $\beta$) which is conjugated to the edge mode field $a$. The Hodge decomposition gives
\begin{subequations}\label{Hodge}
\be
\Pi_a\dot{A}_a
&=-\xi\Delta\dot{\alpha}-\lambda\Delta\dot{\beta}+\partial_r(\xi\partial_r\dot{\alpha}+\lambda\partial_r\dot{\beta}+\xi\partial_\phi\dot{\beta}-\lambda\partial_\phi\dot{\alpha})\nn\\
&=-\xi\Delta\dot{\alpha}-\lambda\Delta\dot{\beta}+\partial_r\big(\dot{\alpha}\partial_r\dot{\alpha}+\dot{\beta}\partial_r\dot{\beta}+2\dot{\alpha}\partial_\phi\dot{\beta}-m(\alpha\partial_\phi\dot{\alpha}+\beta\partial_\phi\dot{\beta})-m\partial_r(\dot{\alpha}\beta)\big),\\
G&=\Delta(\xi-m\beta)\eqqcolon\Delta\tilde{G},\\
\f{1}{2}(F_{ab})^2&=\beta\Delta^2\beta+\partial_r^2(\partial_r\beta\partial_r\beta-\beta\partial_r^2\beta),\\
(E_a)^2
&=-\xi\Delta\xi-\lambda\Delta\lambda-2m(\xi\Delta\beta-\lambda\Delta\alpha)-m^2(\alpha\Delta\alpha+\beta\Delta\beta)\nn\\
&\phantom{=\ }+\partial_r\big(\xi\partial_r\xi+\lambda\partial_r\lambda+2\xi\partial_\phi\lambda-2m(\lambda\partial_r\alpha-\xi\partial_r\beta+\xi\partial_\phi\alpha+\lambda\partial_\phi\beta)+m^2(\alpha\partial_r\alpha+\beta\partial_r\beta+2\alpha\partial_\phi\beta)\big)\nn\\
&=-\xi\Delta\xi-\lambda\Delta\lambda-2m(\xi\Delta\beta-\lambda\Delta\alpha)-m^2(\alpha\Delta\alpha+\beta\Delta\beta)+\partial_r(\dot{\alpha}\partial_r\dot{\alpha}+\dot{\beta}\partial_r\dot{\beta}+2\dot{\alpha}\partial_\phi\dot{\beta}).
\ee
\end{subequations}
From this, we can now derive several results.

First, let us focus on the bulk contributions to explain how the path integral for the massive scalar field arises. We will perform manipulations at the level of the Lorentzian path integral, but simply write the actions alone in order to avoid unnecessary notational cluttering. Working in the temporal gauge $A_t=0$ we have to impose the Gauss constraint, which we write here in the form $\tilde{G}=\xi-m\beta=0$. This transforms all the bulk terms above according to
\be
S_M
&=\f{1}{2}\int_M-2\xi\Delta\dot{\alpha}-2\lambda\Delta\dot{\beta}-\beta\Delta^2\beta+\xi\Delta\xi+\lambda\Delta\lambda+2m(\xi\Delta\beta-\lambda\Delta\alpha)+m^2(\alpha\Delta\alpha+\beta\Delta\beta)\nn\\
&=\f{1}{2}\int_M-2m\beta\Delta\dot{\alpha}-2\lambda\Delta\dot{\beta}-\beta\Delta^2\beta+\lambda\Delta\lambda-2m\lambda\Delta\alpha+m^2\alpha\Delta\alpha+4m^2\beta\Delta\beta,
\ee
where for the second equality we have used the constraint enforced by the temporal gauge. Performing now the Gaussian integral over $\lambda$ leads to
\be\label{MCS massive beta}
S_M=\f{1}{2}\int_M\beta(-\Delta)(\Box-4m^2)\beta-\f{m}{2}\int_{\partial M}\alpha\partial_r\dot{\beta}+\beta\partial_r\dot{\alpha}.
\ee
Let us now keep track of the various determinants which have been produced by these manipulations in the path integral. First, when using the Hodge decomposition, the measure on phase space changes as
\be\label{MCS Hodge symplectic}
\mathscr{D}[A,\Pi]=\mathscr{D}[-\Delta\alpha,-\Delta\beta,\xi,\lambda]=(\det\Delta)^2\mathscr{D}[\alpha,\beta,\xi,\lambda].
\ee
Using the identity
\be\label{MCS Hodge Gauss}
\delta(G)=\delta(\Delta\tilde{G})=(\det\Delta)^{-1}\delta(\tilde{G}),
\ee
we have also picked up a factor $(\det\Delta)^{-1}$ when imposing the Gauss law with a delta function in the path integral. Then, the Gaussian integral over $\lambda$ has produced a factor of $(\det\Delta)^{-1/2}$. Putting all these factors together, assuming that there is no boundary, and performing a final Gaussian integral over $\beta$ in \eqref{MCS massive beta}, we finally get that
\be\label{MCS path integral steps}
\Z_M
&=\int\mathscr{D}[A,\Pi]\delta(G)\exp(\i S_M)\nn\\
&=(\det\Delta)^2\int\mathscr{D}[\alpha,\beta,\xi,\lambda]\delta(G)\exp(\i S_M)\nn\\
&=\det\Delta\int\mathscr{D}[\alpha,\beta,\xi,\lambda]\delta(\tilde{G})\exp(\i S_M)\nn\\
&=(\det\Delta)^{1/2}\int\mathscr{D}[\beta]\exp\left(\f{\i}{2}\int_M\beta(-\Delta)(\Box-4m^2)\beta\right)\nn\\
&=\big(\det(\Box-4m^2)\big)^{-1/2},
\ee
where we have dropped gauge volume factors (which can be absorbed by properly normalizing the path integral). As expected, we recover the evaluation of the path integral for a massive scalar field, and all the factors of $\det\Delta$ have cancelled out.

We can now look more carefully at all the boundary contributions coming from the bulk action $S_M$, we will will denote by $\partial S_M$. More precisely, these contributions come from the decomposition \eqref{Hodge} and from \eqref{MCS massive beta}. On the boundary, we will use the relation \eqref{Lagrangian Hodge variables} to write the Gauss law as
\be\label{MCS boundary Gauss}
\tilde{G}\big|_{\partial{M}}=\dot{\alpha}-2m\beta=0.
\ee
With this the boundary term in \eqref{MCS massive beta} is actually vanishing. More precisely, the boundary contributions are
\be\label{MCS boundary action piece 1}
\partial S_M
&=\f{1}{2}\int_{\partial M}\dot{\alpha}\partial_r\dot{\alpha}+\dot{\beta}\partial_r\dot{\beta}+2\dot{\alpha}\partial_\phi\dot{\beta}-2m(\alpha\partial_\phi\dot{\alpha}+\beta\partial_\phi\dot{\beta})-2m\partial_r(\dot{\alpha}\beta)-\partial_r(\partial_r\beta\partial_r\beta-\beta\partial_r^2\beta)\nn\\
&\phantom{=\f{1}{2}\int_{\partial M}}-m(\alpha\partial_r\dot{\beta}+\beta\partial_r\dot{\alpha})\nn\\
&=\f{1}{2}\int_{\partial M}\dot{\beta}\partial_r\dot{\beta}+2m\beta\partial_\phi\dot{\beta}-4m^2\beta\partial_r\beta-4m^2\alpha\partial_\phi\beta-\partial_r(\partial_r\beta\partial_r\beta-\beta\partial_r^2\beta).
\ee
In order to get the dynamics of the edge modes, we have to compute the path integral for this boundary theory coupled to the boundary action in \eqref{MCS extended action}, in which we have to take into account the constraint
\be\label{MCS j phi}
j_\phi=\Pi_r+m\partial_\phi a=\partial_r\dot{\alpha}+\partial_\phi\dot{\beta}+m\big(\partial_\phi(a+\alpha)-\partial_r\beta\big)=m\partial_r\beta+\partial_\phi\dot{\beta}+m\partial_\phi(a+\alpha)
\ee
imposed by the temporal gauge $A_t=0$. With this the boundary action \eqref{MCS extended action} becomes
\be\label{MCS boundary action piece 2}
S_{\partial M}
&=\int_{\partial M}(j_\phi+mA_\phi)\dot{a}-j_t(\partial_\phi a+A_\phi)+h\nn\\
&=\int_{\partial M}\big(j_\phi+m(\partial_\phi\alpha-\partial_r\beta)\big)\dot{a}-j_t\big(\partial_\phi(a+\alpha)-\partial_r\beta\big)+h\nn\\
&=\int_{\partial M}\big(\partial_\phi\dot{\beta}+m\partial_\phi\varphi+m\partial_\phi\alpha\big)\dot{a}-j_t(\partial_\phi\varphi-\partial_r\beta)+h,
\ee
where we have introduced $\varphi\coloneqq a+\alpha$. Combining the two boundary actions \eqref{MCS boundary action piece 1} and \eqref{MCS boundary action piece 2} into a total boundary action
\be
S_\text{edge}\coloneqq\partial S_M+S_{\partial M}
\ee
now leads to
\be
S_\text{edge}
&=\int_{\partial M}\f{1}{2}\dot{\beta}\partial_r\dot{\beta}+m\beta\partial_\phi\dot{\beta}+2m^2\beta\partial_r\beta-\f{1}{2}\partial_r(\partial_r\beta\partial_r\beta-\beta\partial_r^2\beta)\nn\\
&\phantom{=\ \int_{\partial M}}+\dot{\varphi}\partial_\phi\dot{\beta}+m\dot{\varphi}\partial_\phi\varphi-j_t(\partial_\phi\varphi-\partial_r\beta)+h.
\ee
With a further change of variables $\chi\coloneqq\varphi+\dot{\beta}/(2m)$ we finally get
\be
S_\text{edge}=\int_{\partial M}B[\beta]+m\dot{\chi}\partial_\phi\chi-j_t\left(\partial_\phi\chi-\f{1}{2m}\partial_\phi\dot{\beta}-\partial_r\beta\right)+h,
\ee
where
\be\label{B of beta}
B[\beta]\coloneqq\f{1}{2}\dot{\beta}\partial_r\dot{\beta}+m\beta\partial_\phi\dot{\beta}+2m^2\beta\partial_r\beta-\f{1}{2}\partial_r(\partial_r\beta\partial_r\beta-\beta\partial_r^2\beta)-\f{1}{4m}\ddot{\beta}\partial_\phi\dot{\beta}.
\ee
We then have to choose a boundary Hamiltonian $h$ and integrate over $j_t$ in order to get the final form of the effective boundary action. In the main text we use the Hamiltonian
\be
h=\f{1}{m}(j_t\mp j_\phi)j_\phi.
\ee
Noticing that the expression \eqref{MCS j phi} for $j_\phi$ can be written in terms of $\chi$ as
\be\label{MCS j phi chi}
j_\phi=m\partial_r\beta+\f{1}{2}\partial_\phi\dot{\beta}+m\partial_\phi\chi,
\ee
with this Hamiltonian we get
\be
S_\text{edge}=\int_{\partial M}B[\beta]+m\dot{\chi}\partial_\phi\chi+j_t\left(\f{1}{m}\partial_\phi\dot{\beta}+2\partial_r\beta\right)\mp\f{1}{m}\left(m\partial_r\beta+\f{1}{2}\partial_\phi\dot{\beta}+m\partial_\phi\chi\right)^2,
\ee
which can be path integrated over $j_t$ to finally obtain
\be
S_\text{edge}=m\int_{\partial M}\dot{\chi}\partial_\phi\chi\mp(\partial_\phi\chi)^2.
\ee
Note that this last step involves the fact that, under the constraint imposed by $j_t$, we have
\be
B[\beta]\Big|_{\big(2m\partial_r\beta+\partial_\phi\dot{\beta}\,=\,0\big)}=0,
\ee
as one can easily check.

\section{Boundary conditions as boundary sources}
\label{appendix: BCs}

In this appendix we show how the boundary conditions obtained from the extended boundary + boundary action can equivalently be treated as boundary sources. To see this in a simpler setting, we will use the 3-dimensional radial gauge formulation of Maxwell theory introduced in appendix \ref{appendix: M radial gauge}, where the coordinates are $x^\mu=(t,r,\phi)=(r,y^i)$ with $y^i=(t,\phi)$. We have shown in \eqref{3d Maxwell beta action} that in this case the extended action is given by
\be
S=\f{1}{2}\int_M\sqrt{|g|}\big(\beta^i\partial^2\beta_i-\partial_r\beta^i\partial_r\beta_i\big)+\int_{\partial M}\sqrt{|q|}\,\eps^{im}j_i(\beta_m+\partial_m\varphi),
\ee
where $A_i=\partial_i\alpha+\beta_i$ and $\varphi=a+\alpha$. Variation with respect to $\beta_i$ leads to the bulk and boundary equations of motion
\be
\Box\beta^i=0,\q\q\partial_r\beta^i(\ell,y)=-\eps^{im}j_m(y),
\ee
where $r=\ell$ is the location of the boundary $\partial M$. Alternatively, we can use a Dirac delta to rewrite this action as
\be
S=\f{1}{2}\int_M\sqrt{|g|}\big(\beta^i\partial^2\beta_i-\partial_r\beta^i\partial_r\beta_i+2\eps^{im}j_i\beta_m\delta(r-\ell)\big)+\int_{\partial M}\sqrt{|q|}\,\eps^{im}j_i\partial_m\varphi,
\ee
which leads to the massless Klein--Gordon equation with boundary sources
\be
\Box\beta^i(r,y)=\eps^{im}j_m(y)\delta(r-\ell),\q\q\partial_r\beta^i(\ell,y)=0.
\ee
To solve this boundary problem we use the decomposition
\be
\beta^i(r,y)=\beta^i_0(r,y)+\int\de^2y'\sqrt{|g|}\,G(r-\ell,y-y')\eps^{im}j_m(y'),
\ee
where $\beta^i_0$ is the homogeneous solution with Dirichlet boundary condition, i.e.
\be
\Box\beta^i_0(r,y)=0,\q\q\beta^i_0(\ell,y)=0,\q\q\partial_r\beta^i_0(\ell,y)=0,
\ee
and the Green function satisfies
\be
\Box G(x-x')=\delta(x-x'),\q\q\partial_rG(r-\ell,y-y')\big|_{r=\ell}=0.
\ee
Using this ansatz, the extended action can be written as 
\be
S=&-\f{1}{2}\int_M\sqrt{|g|}\,\beta^i_0\Box\beta_{i0}+\int_{\partial M}\de^2y\,\sqrt{|q|}\left(\eps^{im}j_i\partial_m\varphi+\f{1}{2}\int_{\partial M}\de^2y'\,\sqrt{|q|}\,j^i(y)G(0,y-y')j_i(y')\right).
\ee
Finally, focusing on the boundary piece (the bulk gives a factor of $(\det\Box)^{-1/2}$ computed with Dirichlet boundary conditions) and path integrating over $j_i$ leads to the effective action
\be
S_\text{edge}=-\f{1}{2}\int_{\partial M}\de^2y\,\sqrt{|q|}\int_{\partial M}\de^2y'\,\sqrt{|q|}\,\partial^i\varphi(y)G(0,y-y')^{-1}\partial_i\varphi(y').
\ee
One can verify that in momentum space this effective edge mode action coincides with \eqref{effective Maxwell momentum action}.

We have shown that the boundary current $j$ can be treated either as a boundary condition or as a boundary source. In general, this is valid for any theory whose equations of motion involve a Laplace-type operator. In order to see this, let us consider two functions $\varphi$ and $\psi$ defined over $M$ and satisfying
\be
\Box\varphi=0,\q\q\Box\psi=-j\delta(x-x|_{\partial M}),
\ee
with the boundary conditions
\be
n[\varphi]=j,\q\q n[\psi]=0,
\ee
where $n=n^\mu\partial_\mu$ is a unit normal vector to the boundary. From Green's second identity
\be
\int_M\big(\varphi\Box\psi-\psi\Box\varphi\big)\de V=\int_{\partial M}\big(\varphi n[\psi]-\psi n[\varphi]\big)\de S
\ee
we get
\be
\int_{\partial M}j(\varphi-\psi)\de S=0,
\ee
which therefore means that $\varphi|_{\partial M}=\psi|_{\partial M}$.

\section{Hamiltonian quantization of the chiral bosons}
\label{appendix:CS-EE}

For abelian CS theory, as we have seen in \eqref{chiral boundary action}, the edge mode dynamics is that of a free 2-dimensional chiral boson with action
\be
S_{\sl,\sr}=\f{\k}{4\pi}\int_{\partial M}\de t\,\de\phi\,\Big(\partial_t\varphi\partial_\phi\varphi\mp(\partial_\phi\varphi)^2\Big),
\ee
where the right moving and left moving bosons are to be identified respectively with the minus and plus signs. Taking into account the chiral action coming from both sides of the entangling surface, we can express the total action as
\be
S=\f{\k}{4\pi}\int_{\partial M}\de t\,\de\phi\,\Big(\partial_t\varphi^\sl\partial_\phi\varphi^\sl-\partial_t\varphi^\sr\partial_\phi\varphi^{\text{\tiny{R}}}-(\partial_\phi\varphi^\sl)^2-(\partial_\phi\varphi^\sr)^2-16\pi^2\lambda\big[1-\cos(\varphi^\sl+\varphi^\sr)\big]\Big),
\ee
where once again the different signs come from the different orientations of $\partial M$ and $\partial\bar{M}$. In addition to the left-right chiral bosons, we have also introduced here a gluing action (known in condensed matter as a tunneling term), and the gluing constraint along the entangling surface can be enforced by taking $\lambda\rightarrow\infty$. This term provides a coupling between left and right moving fields.

\paragraph{Mode expansion.} Since the boundary is a circle of length $\ell$, we can write the fields $\varphi^\sl$ and $\varphi^\sr$ in a mode expansion
\begin{subequations}
\be
\varphi^\sr(t,\phi)&=\varphi_0^\sr(t)+\f{2\pi N^\sr}{\ell}\phi+\sum_{n>0}\left(\f{\alpha_n(t)}{\sqrt{|n|}}\exp\left(\f{2\pi\i n\phi}{\ell}\right)+\f{\alpha_n^\dagger(t)}{\sqrt{|n|}}\exp\left(-\f{2\pi\i n\phi}{\ell}\right)\right),\\
\varphi^\sl(t,\phi)&=\varphi_0^\sl(t)+\f{2\pi N^\sl}{\ell}\phi+\sum_{n<0}\left(\f{\alpha_n(t)}{\sqrt{|n|}}\exp\left(\f{2\pi\i n\phi}{\ell}\right)+\f{\alpha_n^\dagger(t)}{\sqrt{|n|}}\exp\left(-\f{2\pi\i n\phi}{\ell}\right)\right),\\
\partial_t\varphi(t,\phi)&=\partial_t\varphi_0(t)+\sum_n\left(\f{\partial_t\alpha_n(t)}{\sqrt{|n|}}\exp\left(\f{2\pi\i n\phi}{\ell}\right)+\f{\partial_t\alpha_n^\dagger(t)}{\sqrt{|n|}}\exp\left(-\f{2\pi\i n\phi}{\ell}\right)\right),\\
\partial_\phi\varphi(t,\phi)&=\f{2\pi N}{\ell}+\f{2\pi\i}{\ell}\sum_n\left(\f{n\alpha_n(t)}{\sqrt{|n|}}\exp\left(\f{2\pi\i n\phi}{\ell}\right)-\f{n\alpha_n^\dagger(t)}{\sqrt{|n|}}\exp\left(-\f{2\pi\i n\phi}{\ell}\right)\right).
\ee
\end{subequations}
With this the canonical term of the above action becomes
\be
\f{\k}{4\pi}\int_{\partial M}\de t\,\de\phi\,\big(\partial_t\varphi^\sl\partial_\phi\varphi^\sl-\partial_t\varphi^\sr\partial_\phi\varphi^\sr\big)=\f{\k}{2}\int\de t\,\Bigg(N^\sl\partial_t\varphi_0^\sl-N^\sr\partial_t\varphi_0^\sr+2\i\sum_{n\neq0}\alpha_n^\dagger\partial_t\alpha_n\Bigg),
\ee
showing that we have the canonical commutators
\be
\big[\varphi_0^\sr,N^\sr\big]=-\f{2\i}{\k},\q\q\big[\varphi_0^\sl,N^\sl\big]=\f{2\i}{\k},\q\q\big[\alpha_m,\alpha_n^\dagger\big]=\f{1}{\k}\delta_{mn}.
\ee
Expanding the gluing contraint to quadratic order, the Hamiltonian is given by
\be
H\approx\f{\k}{4\pi}\int_0^\ell\de\phi\,\big((\partial_\phi\varphi^\sr)^2+(\partial_\phi a^\sl)^2+8\pi^2\lambda(\varphi^\sl+\varphi^\sr)^2\big)=H_0+H_\alpha,
\ee
and is the sum of a zero mode and an oscillator part. In order to compute this Hamiltonian we first use
\be
\f{\k}{4\pi}\int_0^\ell\de\phi\,(\partial_\phi\varphi)^2=\f{2\pi\k}{\ell}\left(\f{1}{2}N^2+\sum_n|n|\left(\alpha_n^\dagger\alpha_n+\f{1}{2\k}\right)\right).
\ee
This leads to
\be
\f{\k}{4\pi}\int_0^\ell\de\phi\,\big((\partial_\phi\varphi^\sr)^2+(\partial_\phi\varphi^\sl)^2\big)=\f{2\pi\k}{\ell}\left(\f{1}{4}(N^\sr-N^\sl)^2+\sum_{n\neq0}|n|\left(\alpha_n^\dagger\alpha_n+\f{1}{2\k}\right)\right),
\ee
where we have used $(N^\sr)^2+(N^\sl)^2=\big((N^\sr-N^\sl)^2+(N^\sr+N^\sl)^2\big)/2$ and the fact that $N^\sr+N^\sl$ annihilates the ground state. Then we use this annihilation condition again to get
\be
\f{\k}{4\pi}\int_0^\ell\de\phi\,8\pi^2\lambda(\varphi^\sr+\varphi^\sl)^2=\f{2\pi\k}{\ell}\left(\lambda\ell^2(\varphi_0^\sr+\varphi_0^\sl)^2+\lambda\ell^2\sum_{n\neq0}\f{1}{|n|}\left(2\alpha_n^\dagger\alpha_n+\f{1}{\k}+\alpha_n\alpha_{-n}+\alpha_n^\dagger\alpha_{-n}^\dagger\right)\right).
\ee
Putting this together, we get that the zero mode Hamiltonian is given by
\be
H_0=\f{2\pi\k}{\ell}\left(\f{1}{4}(N^\sr-N^\sl)^2+\lambda\ell^2(\varphi_0^\sr+\varphi_0^\sl)^2\right),
\ee
while the oscillator Hamiltonian is given by
\be
H_\alpha=\f{2\pi\k}{\ell}\sum_{n\neq0}\left(|n|\left(\alpha_n^\dagger\alpha_n+\f{1}{2\k}\right)+\f{\lambda\ell^2}{|n|}\left(2\alpha_n^\dagger\alpha_n+\f{1}{\k}+\alpha_n\alpha_{-n}+\alpha_n^\dagger\alpha_{-n}^\dagger\right)\right).
\ee

\paragraph{Zero mode density matrix.} Introducing the discrete harmonic oscillator variables
\be
X=\f{\sqrt{\k}}{2}(N^\sr-N^\sl),\q P=\f{\sqrt{\k}}{2}(\varphi_0^\sr+\varphi_0^\sl),\q m=\f{1}{16\pi\lambda\ell},\q\omega=8\pi\sqrt{\lambda},
\ee
which are such that $[X,P]=\i$, we can rewrite the zero mode Hamiltonian as
\be
H_0=\f{1}{2}\left(\f{P^2}{m}+m\omega^2X^2\right).
\ee
In the low energy limit $m\rightarrow0$ or $\ell\rightarrow\infty$, we can treat this as a continuum system. The unnormalized ground state is given in terms of the eigenvalues $\nu$ of the operators $N$ as
\be
|\psi_0\rangle=\sum_{\nu^\sr,\nu^\sl\in\mathbb{Z}}\exp\left(-\f{1}{2}m\omega X^2\right)|\nu^\sr,\nu^\sl\rangle=\sum_{\nu^\sr,\nu^\sl\in\mathbb{Z}}\exp\left(-\f{\k}{16\ell\sqrt{\lambda}}(\nu^\sr-\nu^\sl)^2\right)|\nu^\sr,\nu^\sl\rangle.
\ee
The constraint $(N^\sr+N^\sl)|\psi_0\rangle=0$ then leads to
\be
|\psi_0\rangle=\sum_{\nu\in\mathbb{Z}}\exp\left(-\f{\k\nu^2}{4\ell\sqrt{\lambda}}\right)|\nu,-\nu\rangle,
\ee
and the reduced density matrix is given by
\be
\rho_0^{\sr,\sl}=\Tr^{\sl,\sr}|\psi_0\rangle\langle\psi_0|=\sum_{\nu\in\mathbb{Z}}\exp\left(-\f{\k\nu^2}{2\ell\sqrt{\lambda}}\right)|\nu\rangle\langle\nu|.
\ee

\paragraph{Oscillator density matrix.} Using the Bogoliubov transformation
\be
\begin{pmatrix}
\alpha_n\\
\alpha_{-n}^\dagger
\end{pmatrix}
=
\begin{pmatrix}
\cosh(\theta_n)&\sinh(\theta_n)\\
\sinh(\theta_n)&\cosh(\theta_n)
\end{pmatrix}
\begin{pmatrix}
\beta_n\\
\beta_{-n}^\dagger
\end{pmatrix},\q
\begin{pmatrix}
\beta_n\\
\beta_{-n}^\dagger
\end{pmatrix}
=
\begin{pmatrix}
\cosh(\theta_n)&-\sinh(\theta_n)\\
-\sinh(\theta_n)&\cosh(\theta_n)
\end{pmatrix}
\begin{pmatrix}
\alpha_n\\
\alpha_{-n}^\dagger
\end{pmatrix},
\ee
with
\begin{subequations}
\be
\cosh(2\theta_n)&=\cosh^2(\theta_n)+\sinh^2(\theta_n)=1+2\sinh^2(\theta_n)=\f{1}{E_n}\f{2\pi\k}{\ell}\left(|n|+\f{2\lambda\ell^2}{|n|}\right),\\
\sinh(2\theta_n)&=2\cosh(\theta_n)\sinh(\theta_n)=-\f{1}{E_n}\f{2\pi\k}{\ell}\f{2\lambda\ell^2}{|n|},
\ee
\end{subequations}
and
\be
E_n=\f{2\pi\k}{\ell}\sqrt{|n|^2+4\lambda\ell^2},
\ee
we can diagonalise the oscillator Hamiltonian and write it in the simple form
\be
H_\alpha=\sum_{n\neq0}E_n\left(\beta_n^\dagger\beta_n+\f{1}{2\k}\right).
\ee
The ground state reduced density matrix (say for the right-moving sector) can be written formally in terms of an entanglement Hamiltonian as
\be\label{oscillator entanglement H}
\rho_\alpha^\sr=\f{1}{Z_\alpha^\sr}\exp(-H_\alpha^\sr),\q\q H_\alpha^\sr=\sum_{n>0}h_n\left(\alpha_n^\dagger\alpha_n+\f{1}{2\k}\right),
\ee
where the unknowns are the so-called entanglement energies $h_n$. From this expression and the fact that $\Tr^\sr\rho_\alpha^\sr=1$ we get that
\be\label{oscillator entanglement Z}
Z_\alpha^\sr=\Tr^\sr\exp(-H_\alpha^\sr)=\prod_{n>0}\f{1}{2}\,\text{csch}\left(\f{h_n}{2\k}\right).
\ee
The two-point correlation function can be written equivalently as an expectation value in the oscillator ground state $|\psi_\alpha\rangle$, or as a trace using the reduced density matrix, so we have the equality
\be
\langle\psi_\alpha|\alpha_n^\dagger\alpha_n|\psi_\alpha\rangle=\Tr^\sr(\rho_\alpha^\sr\alpha_n^\dagger\alpha_n).
\ee
In the oscillator ground state $|\psi_\alpha\rangle$, which is annihilated by $\beta_n$, we have that
\be
\langle\psi_\alpha|\alpha_n^\dagger\alpha_n|\psi_\alpha\rangle=\langle\psi_\alpha|\beta_{-n}\beta_{-n}^\dagger\sinh^2(\theta_n)|\psi_\alpha\rangle=\f{1}{\k}\sinh^2(\theta_n).
\ee
On the other hand from \eqref{oscillator entanglement H} and \eqref{oscillator entanglement Z} we get that
\be
\Tr^\sr(\rho_\alpha^\sr\alpha_n^\dagger\alpha_n)=-\f{\partial}{\partial h_n}\ln Z_\alpha^\sr-\f{1}{2\k}=\f{1}{2\k}\coth\left(\f{h_n}{2\k}\right)-\f{1}{2\k}=\f{1}{\exp(h_n)-1}.
\ee
Using the above equality then leads to
\be
\coth\left(\f{h_n}{2\k}\right)=1+2\sinh^2(\theta_n)=\cosh(2\theta_n),
\ee
which finally implies
\be
h_n=2\k\coth^{-1}\big(\cosh(2\theta_n)\big)=\k\ln\left(\f{\cosh(2\theta_n)+1}{\cosh(2\theta_n)-1}\right).
\ee
In the low energy limit the entanglement energy is given by
\be
h_n\stackrel{\ell\rightarrow\infty\vphantom{\f{1}{2}}}{=}\f{2\k|n|}{\ell\sqrt{\lambda}}-\f{\k|n|^3}{12\ell^3\lambda^{3/2}}+\dots.
\ee

\paragraph{Total entanglement Hamiltonian.} The full unnormalized reduced density matrix is now obtained by tensoring the reduced density matrices for the zero and the oscillator modes. In the low energy limit, this is
\be
\rho^\sr=\rho_0^\sr\otimes\rho_\alpha^\sr=\left[\sum_{\nu\in\mathbb{Z}}\exp\left(-\f{\k\nu^2}{2\ell\sqrt{\lambda}}\right)|\nu\rangle\langle\nu|\right]\otimes\left[\exp\left(-\sum_{n>0}\f{2\k n}{\ell\sqrt{\lambda}}\left(\alpha_n^\dagger\alpha_n+\f{1}{2\k}\right)\right)\right],
\ee
and the corresponding total entanglement Hamiltonian is therefore given by
\be
H^\sr=-\ln\rho^\sr=\f{\k}{\ell\sqrt{\lambda}}\left(\f{1}{2}(N^\sr)^2+2\sum_{n>0}n\left(\alpha_n^\dagger\alpha_n+\f{1}{2\k}\right)\right)=\f{\k}{\ell\sqrt{\lambda}}\left(\f{1}{2}(N^\sr)^2+2\sum_{n>0}n\alpha_n^\dagger\alpha_n-\f{1}{12\k}\right),
\ee
where we have used a $\zeta$-function regularization.

\paragraph{Thermal partition functions.} We define the total thermal partition function as
\be
Z^\sr(\tau)=\Tr^\sr\exp(-\beta H^\sr)=Z_0^\sr(\tau)Z_\alpha^\sr(\tau),
\ee
where the zero mode partition function is
\be
Z_0^\sr(\tau)=\sum_{\nu\in\mathbb{Z}}\exp\left(\f{\pi\i\nu^2\k\tau}{2}\right)=\theta\left(0,\f{\k\tau}{2}\right),
\ee
the oscillator partition function is
\be
Z_\alpha^\sr(\tau)=\exp\left(-\f{\pi\i\tau}{12}\right)\prod_{n>0}\f{1}{1-\exp(2\pi\i n\tau)}=\f{1}{\eta(\tau)},
\ee
and the modular parameter is
\be
\tau\coloneqq\f{\i\beta}{\pi\ell\sqrt{\lambda}}.
\ee
In the large $\ell$ limit, $\tau$ is small, the Jacobi function behaves as $\theta(0,\tau)\sim1/\sqrt{-\i\tau}$, and the Dedekind function behaves as $\eta(\tau)\sim e^{\i\pi/12\tau}/\sqrt{-\i\tau}$. We therefore get that
\be
Z_0^\sr(\tau)\stackrel{\ell\rightarrow\infty\vphantom{\f{1}{2}}}{\longrightarrow}\left(\f{2\pi\ell\sqrt{\lambda}}{\beta\k}\right)^{1/2},\q\q Z_\alpha^\sr(\tau)\stackrel{\ell\rightarrow\infty\vphantom{\f{1}{2}}}{\longrightarrow}\left(\f{\beta}{\pi\ell\sqrt{\lambda}}\right)^{1/2}\exp\left(\f{\pi^2\ell\sqrt{\lambda}}{12\beta}\right),
\ee
and the total partition function behaves as
\be
Z^\sr(\tau)\stackrel{\ell\rightarrow\infty\vphantom{\f{1}{2}}}{\longrightarrow}\sqrt{\f{2}{\k}}\exp\left(\f{\pi^2\ell\sqrt{\lambda}}{12\beta}\right).
\ee

\paragraph{Entanglement entropy.} The entanglement entropy is now given by the thermal entropy of the entanglement Hamiltonian $H^\sr$, i.e.
\be
\S_\text{CS}=\f{\partial}{\partial\beta^{-1}}(\beta^{-1}\log Z^\sr)\big|_{\beta=1}=\f{\bf{A}}{2\pi}\f{\pi^2\sqrt{\lambda}}{12}-\f{1}{2}\log\k+\O(\ell^{-1}),
\ee
in agreement with \eqref{CS EE}.

\section{Extended action and phase space for non-Abelian theories}
\label{appendix: non-Abelian}

In this appendix we present the extended actions for non-Abelian Chern--Simons, Yang--Mills and BF theories, and show that they lead as expected to the extended phase space structures which have been derived in \cite{Donnelly:2016auv,Geiller:2017xad}. We postpone the study of the effective boundary dynamics to future work.

Throughout this appendix, the gauge fields are 1-forms with values in the Lie algebra $\mathfrak{g}$, whose bracket is denoted by $[\cdot\,,\cdot]$. The non-Abelian covariant derivative and field strength are given by
\be
\de_AP=\de P+[A\wedge P],\q\q F=\de A+\f{1}{2}[A\wedge A].
\ee
For forms $P$ and $Q$ of respective degree $p$ and $q$ the bracket satisfies $[P\wedge Q]=(-1)^{pq+1}[Q\wedge P]$. For a group element $g$ we denote the finite gauge transformations by
\be
g^*A=g^{-1}(A+\de)g.
\ee
Finally, recall that all the expressions below should be understood with an implicit pairing between the Lie algebra elements, which we choose to drop for notational clarity, and which is furthermore invariant under the adjoint action of the group on its algebra.

\subsection{Chern--Simons theory}

In non-Abelian CS theory, the edge mode field which we need to introduce is now a group element. We will denote it by $u$. Under the action of finite gauge transformations, this edge mode field transforms as $g^*u=g^{-1}u$, and the current $j$ transform as $g^*j=g^{-1}jg$. With this, the extended bulk + boundary action naturally takes the form
\be
S=\int_MA\wedge\left(F-\f{1}{6}[A\wedge A]\right)-\f{1}{6}\de uu^{-1}\wedge[\de uu^{-1}\wedge\de uu^{-1}]+\int_{\partial M}A\wedge\de uu^{-1}+j\wedge(A+\de uu^{-1}),
\ee
where by comparison with the Abelian case we have now included the bulk NWZW term. The variation of this action can be written as
\be
\delta S
&=2\int_M\delta A\wedge F\nn\\
&\phantom{=\ }+\int_{\partial M}\delta A\wedge(A+\de uu^{-1}-j)+\delta j\wedge(A+\de uu^{-1})+u^{-1}\delta u\de\big(u^*(A+j)\big)-\de\big(\delta uu^{-1}(A+j)\big),
\ee
where
\be
u^*(A+j)=u^{-1}(A+j)u+u^{-1}\de u.
\ee
To obtain this form of the variation of the action, we have used several identities. The first one is the variation of the bulk WZNW term, which gives a boundary term according to
\be
\f{1}{6}\delta\big(\de uu^{-1}\wedge[\de uu^{-1}\wedge\de uu^{-1}]\big)=\f{1}{2}\de\big(\delta uu^{-1}[\de uu^{-1}\wedge\de uu^{-1}]\big).
\ee
The second one is $\delta(\de uu^{-1})=u\de(u^{-1}\delta u)u^{-1}$. Finally, we have also used the fact that
\be
u^{-1}\delta u\de(u^{-1}\de u)=u^{-1}\delta u\de u^{-1}u\wedge u^{-1}\de u=-\f{1}{2}\delta uu^{-1}[\de uu^{-1}\wedge\de uu^{-1}],
\ee
which comes from the invariance of the (implicit) pairing under the adjoint action of $u$.

To obtain the extended potential, we have to remember that the bulk NWZW term also brings a contribution. The total extended potential is therefore given by
\be
\theta_\text{e}
&=\delta A\wedge A-\f{1}{2}\delta uu^{-1}[\de uu^{-1}\wedge\de uu^{-1}]+\de\big(\delta uu^{-1}(A+j)\big)\nn\\
&\approx\delta A\wedge A-\f{1}{2}\delta uu^{-1}[\de uu^{-1}\wedge\de uu^{-1}]+\de\big(\delta uu^{-1}(2A+\de uu^{-1})\big)\nn\\
&=\delta A\wedge A+\de(\delta uu^{-1})\wedge\de uu^{-1}+2\de(A\delta uu^{-1}).
\ee
Upon taking a further variation the NWZW term gets pushed to the corner using
\be
\delta\big(\de(\delta uu^{-1})\wedge\de uu^{-1}\big)=\de\big(\de(\delta uu^{-1})\delta uu^{-1}\big),
\ee
and one finally obtains the extended symplectic structure
\be
\Omega=-\int_\Sigma\delta A\wedge A+\int_S\big(2\delta A+\de_A(\delta uu^{-1})\big)\delta uu^{-1},
\ee
in agreement with \cite{Geiller:2017whh}. As explained in this reference, similarly to what happens in Abelian Chern--Simons theory, we then have that the generators of $\delta_\alpha$ are integrable and vanishing on-shell, while the boundary symmetries acting as $\Delta_\alpha A=0$ and $\Delta_\alpha u=u\alpha$ have a generator $\Delta_\alpha\ipp\Omega$ which satisfies a non-Abelian current algebra.

\subsection{Yang--Mills theory}

To treat the case of 4-dimensional Yang--Mills theory, we need once again a group element $u$ and a current 2-form $j$, transforming respectively under gauge transformations as $g^*u=g^{-1}u$ and $g^*j=g^{-1}jg$. With this we can then form the extended action
\be
S=-\f{1}{2}\int_M\st F\wedge F+\int_{\partial M}j\wedge(A+\de uu^{-1}).
\ee
Its variation is given by
\be
\delta S=-\int_M\delta A\wedge\de_A\st F+\int_{\partial M}\delta A\wedge(j-\st F)+\delta j\wedge(A+\de uu^{-1})-u^{-1}\delta u\,\de(u^*j)+\de(j\delta uu^{-1}),
\ee
where the third term on the boundary can actually be rewritten using
\be
u^{-1}\delta u\,\de(u^*j)=\delta uu^{-1}(\de j-[\de uu^{-1}\wedge j]).
\ee
The two boundary equations of motion imposed by $\delta j$ and $\delta uu^{-1}$ imply that the boundary current is conserved, i.e. $\de_Aj=0$. The extended potential is given by
\be
\theta_\text{e}=-\delta A\wedge\st F-\de(j\delta uu^{-1})\approx-\delta A\wedge\st F-\de(\st F\delta uu^{-1}),
\ee
in agreement with \cite{Donnelly:2016auv} and with the Abelian limit \eqref{Maxwell extended potential}.

\subsection{BF theory}

For 3-dimensional non-Abelian BF theory, which is actually 3-dimensional first order gravity (here with a vanishing cosmological constant), the edge mode fields are a group element $u$ and a Lie algebra element $b$, transforming respectively as $g^*u=g^{-1}u$ and $g^*b=g^{-1}bg$. The extended action is
\be
S=\int_MB\wedge F+\int_{\partial M}bF+j\wedge(A+\de uu^{-1}),
\ee
and is of course invariant under the shift symmetry $\delta_\phi$ and the non-Abelian gauge transformation $\delta_\alpha$. The variation of this action is
\be
\delta S
&=\int_M\delta B\wedge F+\delta A\wedge\de_AB\nn\\
&\phantom{=\ }+\int_{\partial M}\delta A\wedge(B+\de_Ab-j)+\delta j\wedge(A+\de uu^{-1})+u^{-1}\delta u\,\de(u^*j)+\delta bF-\de(j\delta uu^{-1}-b\delta A).
\ee
From this we can read once again the bulk and boundary equations of motions, and the extended potential becomes
\be
\theta_\text{e}=\delta A\wedge B+\de(j\delta uu^{-1}-b\delta A)\approx\delta A\wedge B+\de\big((B+\de_Ab)\delta uu^{-1}-b\delta A\big),
\ee
in agreement with \cite{Geiller:2017xad}. The computation of the extended symplectic structure, the boundary observables, and their algebra, then follows the results of this reference.

\bibliography{Biblio.bib}

\providecommand{\href}[2]{#2}\begingroup\raggedright\begin{thebibliography}{100}

\bibitem{Banados:1998ta}
M.~Banados, T.~Brotz and M.~E. Ortiz, \emph{{Boundary dynamics and the
  statistical mechanics of the (2+1)-dimensional black hole}},
  \href{http://dx.doi.org/10.1016/S0550-3213(99)00069-3}{\emph{Nucl. Phys.}
  {\bfseries B545} (1999) 340--370},
  [\href{https://arxiv.org/abs/hep-th/9802076}{{\ttfamily hep-th/9802076}}].

\bibitem{Carlip:2005zn}
S.~Carlip, \emph{{Conformal field theory, (2+1)-dimensional gravity, and the
  BTZ black hole}},
  \href{http://dx.doi.org/10.1088/0264-9381/22/12/R01}{\emph{Class. Quant.
  Grav.} {\bfseries 22} (2005) R85--R124},
  [\href{https://arxiv.org/abs/gr-qc/0503022}{{\ttfamily gr-qc/0503022}}].

\bibitem{Afshar:2017okz}
H.~Afshar, D.~Grumiller, M.~M. Sheikh-Jabbari and H.~Yavartanoo, \emph{{Horizon
  fluff, semi-classical black hole microstates --- Log-corrections to BTZ
  entropy and black hole/particle correspondence}},
  \href{http://dx.doi.org/10.1007/JHEP08(2017)087}{\emph{JHEP} {\bfseries 08}
  (2017) 087}, [\href{https://arxiv.org/abs/1705.06257}{{\ttfamily
  1705.06257}}].

\bibitem{Wen:2004ym}
X.~G. Wen, \emph{{Quantum field theory of many-body systems: From the origin of
  sound to an origin of light and electrons}}.
\newblock 2004.

\bibitem{Wen:1992vi}
X.-G. Wen, \emph{{Theory of the edge states in fractional quantum Hall
  effects}}, \href{http://dx.doi.org/10.1142/S0217979292000840}{\emph{Int. J.
  Mod. Phys.} {\bfseries B6} (1992) 1711--1762}.

\bibitem{Tong:2016kpv}
D.~Tong, \emph{{Lectures on the Quantum Hall Effect}},
  [\href{https://arxiv.org/abs/1606.06687}{{\ttfamily 1606.06687}}].

\bibitem{Asante:2018kfo}
S.~K. Asante, B.~Dittrich and H.~M. Haggard, \emph{{Holographic description of
  boundary gravitons in (3+1) dimensions}},
  \href{http://dx.doi.org/10.1007/JHEP01(2019)144}{\emph{JHEP} {\bfseries 01}
  (2019) 144}, [\href{https://arxiv.org/abs/1811.11744}{{\ttfamily
  1811.11744}}].

\bibitem{Asante:2019ndj}
S.~K. Asante, B.~Dittrich and F.~Hopfmueller, \emph{{Holographic formulation of
  3D metric gravity with finite boundaries}},
  \href{http://dx.doi.org/10.3390/universe5080181}{\emph{Universe} {\bfseries
  5} (2019) 181}, [\href{https://arxiv.org/abs/1905.10931}{{\ttfamily
  1905.10931}}].

\bibitem{Dittrich:2018xuk}
B.~Dittrich, C.~Goeller, E.~R. Livine and A.~Riello, \emph{{Quasi-local
  holographic dualities in non-perturbative 3d quantum gravity}},
  \href{http://dx.doi.org/10.1088/1361-6382/aac606}{\emph{Class. Quant. Grav.}
  {\bfseries 35} (2018) 13LT01},
  [\href{https://arxiv.org/abs/1803.02759}{{\ttfamily 1803.02759}}].

\bibitem{Dittrich:2017hnl}
B.~Dittrich, C.~Goeller, E.~Livine and A.~Riello, \emph{{Quasi-local
  holographic dualities in non-perturbative 3d quantum gravity I: Convergence
  of multiple approaches and examples of Ponzano-Regge statistical duals}},
  \href{http://dx.doi.org/10.1016/j.nuclphysb.2018.06.007}{\emph{Nucl. Phys.}
  {\bfseries B938} (2019) 807--877},
  [\href{https://arxiv.org/abs/1710.04202}{{\ttfamily 1710.04202}}].

\bibitem{Dittrich:2017rvb}
B.~Dittrich, C.~Goeller, E.~R. Livine and A.~Riello, \emph{{Quasi-local
  holographic dualities in non-perturbative 3d quantum gravity II: From
  coherent quantum boundaries to BMS3 characters}},
  \href{http://dx.doi.org/10.1016/j.nuclphysb.2018.06.010}{\emph{Nucl. Phys.}
  {\bfseries B938} (2019) 878--934},
  [\href{https://arxiv.org/abs/1710.04237}{{\ttfamily 1710.04237}}].

\bibitem{Riello:2018anu}
A.~Riello, \emph{{Quantum edge modes in 3d gravity and 2+1d topological phases
  of matter}}, \href{http://dx.doi.org/10.1103/PhysRevD.98.106002}{\emph{Phys.
  Rev.} {\bfseries D98} (2018) 106002},
  [\href{https://arxiv.org/abs/1802.02588}{{\ttfamily 1802.02588}}].

\bibitem{Goeller:2019zpz}
C.~Goeller, E.~R. Livine and A.~Riello, \emph{{Non-Perturbative 3D Quantum
  Gravity: Quantum Boundary States and Exact Partition Function}},
  [\href{https://arxiv.org/abs/1912.01968}{{\ttfamily 1912.01968}}].

\bibitem{Grumiller:2016pqb}
D.~Grumiller and M.~Riegler, \emph{{Most general AdS$_{3}$ boundary
  conditions}}, \href{http://dx.doi.org/10.1007/JHEP10(2016)023}{\emph{JHEP}
  {\bfseries 10} (2016) 023},
  [\href{https://arxiv.org/abs/1608.01308}{{\ttfamily 1608.01308}}].

\bibitem{Donnelly:2016auv}
W.~Donnelly and L.~Freidel, \emph{{Local subsystems in gauge theory and
  gravity}}, \href{http://dx.doi.org/10.1007/JHEP09(2016)102}{\emph{JHEP}
  {\bfseries 09} (2016) 102},
  [\href{https://arxiv.org/abs/1601.04744}{{\ttfamily 1601.04744}}].

\bibitem{Grumiller:2017sjh}
D.~Grumiller, W.~Merbis and M.~Riegler, \emph{{Most general flat space boundary
  conditions in three-dimensional Einstein gravity}},
  \href{http://dx.doi.org/10.1088/1361-6382/aa8004}{\emph{Class. Quant. Grav.}
  {\bfseries 34} (2017) 184001},
  [\href{https://arxiv.org/abs/1704.07419}{{\ttfamily 1704.07419}}].

\bibitem{Geiller:2017xad}
M.~Geiller, \emph{{Edge modes and corner ambiguities in 3d Chern-Simons theory
  and gravity}},
  \href{http://dx.doi.org/10.1016/j.nuclphysb.2017.09.010}{\emph{Nucl. Phys.}
  {\bfseries B924} (2017) 312--365},
  [\href{https://arxiv.org/abs/1703.04748}{{\ttfamily 1703.04748}}].

\bibitem{Speranza:2017gxd}
A.~J. Speranza, \emph{{Local phase space and edge modes for
  diffeomorphism-invariant theories}},
  \href{http://dx.doi.org/10.1007/JHEP02(2018)021}{\emph{JHEP} {\bfseries 02}
  (2018) 021}, [\href{https://arxiv.org/abs/1706.05061}{{\ttfamily
  1706.05061}}].

\bibitem{Geiller:2017whh}
M.~Geiller, \emph{{Lorentz-diffeomorphism edge modes in 3d gravity}},
  \href{http://dx.doi.org/10.1007/JHEP02(2018)029}{\emph{JHEP} {\bfseries 02}
  (2018) 029}, [\href{https://arxiv.org/abs/1712.05269}{{\ttfamily
  1712.05269}}].

\bibitem{Freidel:2019ofr}
L.~Freidel, E.~R. Livine and D.~Pranzetti, \emph{{Kinematical Gravitational
  Charge Algebra}},  [\href{https://arxiv.org/abs/1910.05642}{{\ttfamily
  1910.05642}}].

\bibitem{Freidel:2018fsk}
L.~Freidel and D.~Pranzetti, \emph{{Electromagnetic duality and central
  charge}},  [\href{https://arxiv.org/abs/1806.03161}{{\ttfamily 1806.03161}}].

\bibitem{Freidel:2018pvm}
L.~Freidel and E.~R. Livine, \emph{{Bubble networks: framed discrete geometry
  for quantum gravity}},
  \href{http://dx.doi.org/10.1007/s10714-018-2493-y}{\emph{Gen. Rel. Grav.}
  {\bfseries 51} (2019) 9}, [\href{https://arxiv.org/abs/1810.09364}{{\ttfamily
  1810.09364}}].

\bibitem{Freidel:2019ees}
L.~Freidel, E.~R. Livine and D.~Pranzetti, \emph{{Gravitational edge modes:
  from KacMoody charges to Poincar networks}},
  \href{http://dx.doi.org/10.1088/1361-6382/ab40fe}{\emph{Class. Quant. Grav.}
  {\bfseries 36} (2019) 195014},
  [\href{https://arxiv.org/abs/1906.07876}{{\ttfamily 1906.07876}}].

\bibitem{Buividovich:2008gq}
P.~V. Buividovich and M.~I. Polikarpov, \emph{{Entanglement entropy in gauge
  theories and the holographic principle for electric strings}},
  \href{http://dx.doi.org/10.1016/j.physletb.2008.10.032}{\emph{Phys. Lett.}
  {\bfseries B670} (2008) 141--145},
  [\href{https://arxiv.org/abs/0806.3376}{{\ttfamily 0806.3376}}].

\bibitem{Donnelly:2011hn}
W.~Donnelly, \emph{{Decomposition of entanglement entropy in lattice gauge
  theory}}, \href{http://dx.doi.org/10.1103/PhysRevD.85.085004}{\emph{Phys.
  Rev.} {\bfseries D85} (2012) 085004},
  [\href{https://arxiv.org/abs/1109.0036}{{\ttfamily 1109.0036}}].

\bibitem{Casini:2013rba}
H.~Casini, M.~Huerta and J.~A. Rosabal, \emph{{Remarks on entanglement entropy
  for gauge fields}},
  \href{http://dx.doi.org/10.1103/PhysRevD.89.085012}{\emph{Phys. Rev.}
  {\bfseries D89} (2014) 085012},
  [\href{https://arxiv.org/abs/1312.1183}{{\ttfamily 1312.1183}}].

\bibitem{Casini:2014aia}
H.~Casini and M.~Huerta, \emph{{Entanglement entropy for a Maxwell field:
  Numerical calculation on a two dimensional lattice}},
  \href{http://dx.doi.org/10.1103/PhysRevD.90.105013}{\emph{Phys. Rev.}
  {\bfseries D90} (2014) 105013},
  [\href{https://arxiv.org/abs/1406.2991}{{\ttfamily 1406.2991}}].

\bibitem{Donnelly:2014fua}
W.~Donnelly and A.~C. Wall, \emph{{Entanglement entropy of electromagnetic edge
  modes}}, \href{http://dx.doi.org/10.1103/PhysRevLett.114.111603}{\emph{Phys.
  Rev. Lett.} {\bfseries 114} (2015) 111603},
  [\href{https://arxiv.org/abs/1412.1895}{{\ttfamily 1412.1895}}].

\bibitem{Donnelly:2015hxa}
W.~Donnelly and A.~C. Wall, \emph{{Geometric entropy and edge modes of the
  electromagnetic field}},
  \href{http://dx.doi.org/10.1103/PhysRevD.94.104053}{\emph{Phys. Rev.}
  {\bfseries D94} (2016) 104053},
  [\href{https://arxiv.org/abs/1506.05792}{{\ttfamily 1506.05792}}].

\bibitem{Agarwal:2016cir}
A.~Agarwal, D.~Karabali and V.~P. Nair, \emph{{Gauge-invariant Variables and
  Entanglement Entropy}},
  \href{http://dx.doi.org/10.1103/PhysRevD.96.125008}{\emph{Phys. Rev.}
  {\bfseries D96} (2017) 125008},
  [\href{https://arxiv.org/abs/1701.00014}{{\ttfamily 1701.00014}}].

\bibitem{Pretko:2018nsz}
M.~Pretko, \emph{{On the Entanglement Entropy of Maxwell Theory: A Condensed
  Matter Perspective}},  [\href{https://arxiv.org/abs/1801.01158}{{\ttfamily
  1801.01158}}].

\bibitem{Blommaert:2018rsf}
A.~Blommaert, T.~G. Mertens, H.~Verschelde and V.~I. Zakharov, \emph{{Edge
  State Quantization: Vector Fields in Rindler}},
  [\href{https://arxiv.org/abs/1801.09910}{{\ttfamily 1801.09910}}].

\bibitem{Lin:2018bud}
J.~Lin and D.~Radicevic, \emph{{Comments on Defining Entanglement Entropy}},
  [\href{https://arxiv.org/abs/1808.05939}{{\ttfamily 1808.05939}}].

\bibitem{Belin:2019mlt}
A.~Belin, N.~Iqbal and J.~Kruthoff, \emph{{Bulk entanglement entropy for
  photons and gravitons in AdS$_3$}},
  [\href{https://arxiv.org/abs/1912.00024}{{\ttfamily 1912.00024}}].

\bibitem{Strominger:2017zoo}
A.~Strominger, \emph{{Lectures on the Infrared Structure of Gravity and Gauge
  Theory}},  [\href{https://arxiv.org/abs/1703.05448}{{\ttfamily 1703.05448}}].

\bibitem{Freidel:2019ohg}
L.~Freidel, F.~Hopfmueller and A.~Riello, \emph{{Asymptotic Renormalization in
  Flat Space: Symplectic Potential and Charges of Electromagnetism}},
  \href{http://dx.doi.org/10.1007/JHEP10(2019)126}{\emph{JHEP} {\bfseries 10}
  (2019) 126}, [\href{https://arxiv.org/abs/1904.04384}{{\ttfamily
  1904.04384}}].

\bibitem{Blommaert:2018oue}
A.~Blommaert, T.~G. Mertens and H.~Verschelde, \emph{{Edge Dynamics from the
  Path Integral: Maxwell and Yang-Mills}},
  [\href{https://arxiv.org/abs/1804.07585}{{\ttfamily 1804.07585}}].

\bibitem{Blommaert:2018iqz}
A.~Blommaert, T.~G. Mertens and H.~Verschelde, \emph{{Fine Structure of
  Jackiw-Teitelboim Quantum Gravity}},
  [\href{https://arxiv.org/abs/1812.00918}{{\ttfamily 1812.00918}}].

\bibitem{Mathieu:2019lgi}
P.~Mathieu, L.~Murray, A.~Schenkel and N.~J. Teh, \emph{{Homological
  perspective on edge modes in linear Yang-Mills and Chern-Simons theory}},
  [\href{https://arxiv.org/abs/1907.10651}{{\ttfamily 1907.10651}}].

\bibitem{Witten:2018zxz}
E.~Witten, \emph{{Notes on Some Entanglement Properties of Quantum Field
  Theory}},  [\href{https://arxiv.org/abs/1803.04993}{{\ttfamily 1803.04993}}].

\bibitem{Buividovich_2009}
P.~V. Buividovich and M.~I. Polikarpov, \emph{Entanglement entropy in lattice
  gauge theories},
  \href{http://dx.doi.org/10.1088/1751-8113/42/30/304005}{\emph{Journal of
  Physics A: Mathematical and Theoretical} {\bfseries 42} (Jul, 2009) 304005}.

\bibitem{Delcamp:2016eya}
C.~Delcamp, B.~Dittrich and A.~Riello, \emph{{On entanglement entropy in
  non-Abelian lattice gauge theory and 3D quantum gravity}},
  \href{http://dx.doi.org/10.1007/JHEP11(2016)102}{\emph{JHEP} {\bfseries 11}
  (2016) 102}, [\href{https://arxiv.org/abs/1609.04806}{{\ttfamily
  1609.04806}}].

\bibitem{Fliss:2017wop}
J.~R. Fliss, X.~Wen, O.~Parrikar, C.-T. Hsieh, B.~Han, T.~L. Hughes et~al.,
  \emph{{Interface Contributions to Topological Entanglement in Abelian
  Chern-Simons Theory}},
  \href{http://dx.doi.org/10.1007/JHEP09(2017)056}{\emph{JHEP} {\bfseries 09}
  (2017) 056}, [\href{https://arxiv.org/abs/1705.09611}{{\ttfamily
  1705.09611}}].

\bibitem{Wong:2017pdm}
G.~Wong, \emph{{A note on entanglement edge modes in Chern Simons theory}},
  [\href{https://arxiv.org/abs/1706.04666}{{\ttfamily 1706.04666}}].

\bibitem{Donnelly:2018ppr}
W.~Donnelly and G.~Wong, \emph{{Entanglement branes, modular flow, and extended
  topological quantum field theory}},
  [\href{https://arxiv.org/abs/1811.10785}{{\ttfamily 1811.10785}}].

\bibitem{Fendley:2006gr}
P.~Fendley, M.~P.~A. Fisher and C.~Nayak, \emph{{Topological entanglement
  entropy from the holographic partition function}},
  \href{http://dx.doi.org/10.1007/s10955-006-9275-8}{\emph{J. Statist. Phys.}
  {\bfseries 126} (2007) 1111},
  [\href{https://arxiv.org/abs/cond-mat/0609072}{{\ttfamily
  cond-mat/0609072}}].

\bibitem{Dong:2008ft}
S.~Dong, E.~Fradkin, R.~G. Leigh and S.~Nowling, \emph{{Topological
  Entanglement Entropy in Chern-Simons Theories and Quantum Hall Fluids}},
  \href{http://dx.doi.org/10.1088/1126-6708/2008/05/016}{\emph{JHEP} {\bfseries
  05} (2008) 016}, [\href{https://arxiv.org/abs/0802.3231}{{\ttfamily
  0802.3231}}].

\bibitem{Das:2015oha}
D.~Das and S.~Datta, \emph{{Universal features of left-right entanglement
  entropy}},
  \href{http://dx.doi.org/10.1103/PhysRevLett.115.131602}{\emph{Phys. Rev.
  Lett.} {\bfseries 115} (2015) 131602},
  [\href{https://arxiv.org/abs/1504.02475}{{\ttfamily 1504.02475}}].

\bibitem{Wen:2016snr}
X.~Wen, S.~Matsuura and S.~Ryu, \emph{{Edge theory approach to topological
  entanglement entropy, mutual information and entanglement negativity in
  Chern-Simons theories}},
  \href{http://dx.doi.org/10.1103/PhysRevB.93.245140}{\emph{Phys. Rev.}
  {\bfseries B93} (2016) 245140},
  [\href{https://arxiv.org/abs/1603.08534}{{\ttfamily 1603.08534}}].

\bibitem{Kitaev:2005dm}
A.~Kitaev and J.~Preskill, \emph{{Topological entanglement entropy}},
  \href{http://dx.doi.org/10.1103/PhysRevLett.96.110404}{\emph{Phys. Rev.
  Lett.} {\bfseries 96} (2006) 110404},
  [\href{https://arxiv.org/abs/hep-th/0510092}{{\ttfamily hep-th/0510092}}].

\bibitem{Levin:2006zz}
M.~Levin and X.-G. Wen, \emph{{Detecting Topological Order in a Ground State
  Wave Function}},
  \href{http://dx.doi.org/10.1103/PhysRevLett.96.110405}{\emph{Phys. Rev.
  Lett.} {\bfseries 96} (2006) 110405}.

\bibitem{Kabat:1995eq}
D.~N. Kabat, \emph{{Black hole entropy and entropy of entanglement}},
  \href{http://dx.doi.org/10.1016/0550-3213(95)00443-V}{\emph{Nucl. Phys.}
  {\bfseries B453} (1995) 281--299},
  [\href{https://arxiv.org/abs/hep-th/9503016}{{\ttfamily hep-th/9503016}}].

\bibitem{Kabat:1995jq}
D.~N. Kabat, S.~H. Shenker and M.~J. Strassler, \emph{{Black hole entropy in
  the O(N) model}},
  \href{http://dx.doi.org/10.1103/PhysRevD.52.7027}{\emph{Phys. Rev.}
  {\bfseries D52} (1995) 7027--7036},
  [\href{https://arxiv.org/abs/hep-th/9506182}{{\ttfamily hep-th/9506182}}].

\bibitem{Kabat:2012ns}
D.~Kabat and D.~Sarkar, \emph{{Cosmic string interactions induced by gauge and
  scalar fields}},
  \href{http://dx.doi.org/10.1103/PhysRevD.86.084021}{\emph{Phys. Rev.}
  {\bfseries D86} (2012) 084021},
  [\href{https://arxiv.org/abs/1206.5642}{{\ttfamily 1206.5642}}].

\bibitem{Carlip:1995cd}
S.~Carlip, \emph{{Statistical mechanics and black hole entropy}},
  [\href{https://arxiv.org/abs/gr-qc/9509024}{{\ttfamily gr-qc/9509024}}].

\bibitem{Forman1987}
R.~Forman, \emph{Functional determinants and geometry},
  \href{http://dx.doi.org/10.1007/BF01391828}{\emph{Inventiones mathematicae}
  {\bfseries 88} (Oct, 1987) 447--493}.

\bibitem{BURGHELEA199234}
D.~Burghelea, L.~Friedlander and T.~Kappeler, \emph{{Meyer-Vietoris type
  formula for determinants of elliptic differential operators}},
  \href{http://dx.doi.org/10.1016/0022-1236(92)90099-5}{\emph{Journal of
  Functional Analysis} {\bfseries 107} (1992) 34 -- 65}.

\bibitem{doi:10.1063/1.4936074}
K.~Kirsten and Y.~Lee, \emph{{The Burghelea-Friedlander-Kappeler--gluing
  formula for zeta-determinants on a warped product manifold and a product
  manifold}}, \href{http://dx.doi.org/10.1063/1.4936074}{\emph{Journal of
  Mathematical Physics} {\bfseries 56} (2015) 123501}.

\bibitem{PARK_2006}
J.~Park, \emph{Gluing formulae of spectral invariants and cauchy data spaces},
  \href{http://dx.doi.org/10.1142/9789812773609_0002}{\emph{Analysis, Geometry
  and Topology of Elliptic Operators} (Apr, 2006) 23--38}.

\bibitem{Kirsten2018ThePA}
K.~Kirsten and Y.~Lee, \emph{{The Polynomial Associated with the BFK-Gluing
  Formula of the Zeta-Determinant on a Compact Warped Product Manifold}},
  {\emph{The Journal of Geometric Analysis} {\bfseries 28} (2018) 3856--3891}.

\bibitem{Kirsten:2019xel}
K.~Kirsten and Y.~Lee, \emph{{{The BFK-gluing formula and the curvature tensors
  on a 2-dimensional compact hypersurface}}},
  [\href{https://arxiv.org/abs/1912.11433}{{\ttfamily 1912.11433}}].

\bibitem{Campiglia:2017dpg}
M.~Campiglia, L.~Coito and S.~Mizera, \emph{{Can scalars have asymptotic
  symmetries?}},
  \href{http://dx.doi.org/10.1103/PhysRevD.97.046002}{\emph{Phys. Rev.}
  {\bfseries D97} (2018) 046002},
  [\href{https://arxiv.org/abs/1703.07885}{{\ttfamily 1703.07885}}].

\bibitem{Campiglia_2019}
M.~Campiglia, L.~Freidel, F.~Hopfmueller and R.~M. Soni, \emph{Scalar
  asymptotic charges and dual large gauge transformations},
  \href{http://dx.doi.org/10.1007/jhep04(2019)003}{\emph{Journal of High Energy
  Physics} {\bfseries 2019} (Apr, 2019) }.

\bibitem{Henneaux_2019}
M.~Henneaux and C.~Troessaert, \emph{Asymptotic structure of a massless scalar
  field and its dual two-form field at spatial infinity},
  \href{http://dx.doi.org/10.1007/jhep05(2019)147}{\emph{Journal of High Energy
  Physics} {\bfseries 2019} (May, 2019) }.

\bibitem{Balasubramanian:2018axm}
V.~Balasubramanian and O.~Parrikar, \emph{{Comments on Entanglement Entropy in
  String Theory}},
  \href{http://dx.doi.org/10.1103/PhysRevD.97.066025}{\emph{Phys. Rev.}
  {\bfseries D97} (2018) 066025},
  [\href{https://arxiv.org/abs/1801.03517}{{\ttfamily 1801.03517}}].

\bibitem{Setare:2018mii}
M.~R. Setare and H.~Adami, \emph{{Edge modes and Surface-Preserving Symmetries
  in Einstein-Maxwell Theory}},
  [\href{https://arxiv.org/abs/1808.03257}{{\ttfamily 1808.03257}}].

\bibitem{Iyer:1994ys}
V.~Iyer and R.~M. Wald, \emph{{Some properties of Noether charge and a proposal
  for dynamical black hole entropy}},
  \href{http://dx.doi.org/10.1103/PhysRevD.50.846}{\emph{Phys. Rev.} {\bfseries
  D50} (1994) 846--864}, [\href{https://arxiv.org/abs/gr-qc/9403028}{{\ttfamily
  gr-qc/9403028}}].

\bibitem{Jacobson:1993vj}
T.~Jacobson, G.~Kang and R.~C. Myers, \emph{{On black hole entropy}},
  \href{http://dx.doi.org/10.1103/PhysRevD.49.6587}{\emph{Phys. Rev.}
  {\bfseries D49} (1994) 6587--6598},
  [\href{https://arxiv.org/abs/gr-qc/9312023}{{\ttfamily gr-qc/9312023}}].

\bibitem{Regge:1974zd}
T.~Regge and C.~Teitelboim, \emph{{Role of Surface Integrals in the Hamiltonian
  Formulation of General Relativity}},
  \href{http://dx.doi.org/10.1016/0003-4916(74)90404-7}{\emph{Annals Phys.}
  {\bfseries 88} (1974) 286}.

\bibitem{Balachandran:1991dw}
A.~P. Balachandran, G.~Bimonte, K.~S. Gupta and A.~Stern, \emph{{Conformal edge
  currents in Chern-Simons theories}},
  \href{http://dx.doi.org/10.1142/S0217751X92002106}{\emph{Int. J. Mod. Phys.}
  {\bfseries A7} (1992) 4655--4670},
  [\href{https://arxiv.org/abs/hep-th/9110072}{{\ttfamily hep-th/9110072}}].

\bibitem{Balachandran:1992qg}
A.~P. Balachandran and P.~Teotonio-Sobrinho, \emph{{The Edge states of the BF
  system and the London equations}},
  \href{http://dx.doi.org/10.1142/S0217751X9300028X}{\emph{Int. J. Mod. Phys.}
  {\bfseries A8} (1993) 723--752},
  [\href{https://arxiv.org/abs/hep-th/9205116}{{\ttfamily hep-th/9205116}}].

\bibitem{Balachandran:1994up}
A.~P. Balachandran, L.~Chandar and A.~Momen, \emph{{Edge states in gravity and
  black hole physics}},
  \href{http://dx.doi.org/10.1016/0550-3213(95)00622-2}{\emph{Nucl. Phys.}
  {\bfseries B461} (1996) 581--596},
  [\href{https://arxiv.org/abs/gr-qc/9412019}{{\ttfamily gr-qc/9412019}}].

\bibitem{Balachandran:1995qa}
A.~P. Balachandran, L.~Chandar and A.~Momen, \emph{{Edge states in canonical
  gravity}},  in \emph{{17th Annual MRST (Montreal-Rochester-Syracuse-Toronto)
  Meeting on High-energy Physics Rochester, New York, May 8-9, 1995}}, 1995.
\newblock \href{https://arxiv.org/abs/gr-qc/9506006}{{\ttfamily
  gr-qc/9506006}}.

\bibitem{Husain:1997fm}
V.~Husain and S.~Major, \emph{{Gravity and BF theory defined in bounded
  regions}}, \href{http://dx.doi.org/10.1016/S0550-3213(97)00371-4}{\emph{Nucl.
  Phys.} {\bfseries B500} (1997) 381--401},
  [\href{https://arxiv.org/abs/gr-qc/9703043}{{\ttfamily gr-qc/9703043}}].

\bibitem{Witten:1988hf}
E.~Witten, \emph{{Quantum Field Theory and the Jones Polynomial}},
  \href{http://dx.doi.org/10.1007/BF01217730}{\emph{Commun.Math.Phys.}
  {\bfseries 121} (1989) 351}.

\bibitem{MR1025431}
S.~Elitzur, G.~Moore, A.~Schwimmer and N.~Seiberg, \emph{Remarks on the
  canonical quantization of the {C}hern-{S}imons-{W}itten theory},
  \href{http://dx.doi.org/10.1016/0550-3213(89)90436-7}{\emph{Nuclear Phys. B}
  {\bfseries 326} (1989) 108--134}.

\bibitem{Carlip:1991zm}
S.~Carlip, \emph{{Inducing Liouville theory from topologically massive
  gravity}}, \href{http://dx.doi.org/10.1016/0550-3213(91)90558-F}{\emph{Nucl.
  Phys.} {\bfseries B362} (1991) 111--124}.

\bibitem{Barnich:2001jy}
G.~Barnich and F.~Brandt, \emph{{Covariant theory of asymptotic symmetries,
  conservation laws and central charges}},
  \href{http://dx.doi.org/10.1016/S0550-3213(02)00251-1}{\emph{Nucl. Phys.}
  {\bfseries B633} (2002) 3--82},
  [\href{https://arxiv.org/abs/hep-th/0111246}{{\ttfamily hep-th/0111246}}].

\bibitem{Compere:2018aar}
G.~Compere and A.~Fiorucci, \emph{{Advanced Lectures on General Relativity}},
  \href{http://dx.doi.org/10.1007/978-3-030-04260-8}{\emph{Lect. Notes Phys.}
  {\bfseries 952} (2019) 150},
  [\href{https://arxiv.org/abs/1801.07064}{{\ttfamily 1801.07064}}].

\bibitem{Wieland:2017zkf}
W.~Wieland, \emph{{New boundary variables for classical and quantum gravity on
  a null surface}},
  \href{http://dx.doi.org/10.1088/1361-6382/aa8d06}{\emph{Class. Quant. Grav.}
  {\bfseries 34} (2017) 215008},
  [\href{https://arxiv.org/abs/1704.07391}{{\ttfamily 1704.07391}}].

\bibitem{Freidel:2020xyx}
L.~Freidel, M.~Geiller and D.~Pranzetti, \emph{{Edge modes of gravity - I:
  Corner potentials and charges}},
  [\href{https://arxiv.org/abs/2006.12527}{{\ttfamily 2006.12527}}].

\bibitem{Freidel:2020svx}
L.~Freidel, M.~Geiller and D.~Pranzetti, \emph{{Edge modes of gravity - II:
  Corner metric and Lorentz charges}},
  [\href{https://arxiv.org/abs/2007.03563}{{\ttfamily 2007.03563}}].

\bibitem{Harlow:2019yfa}
D.~Harlow and J.-Q. Wu, \emph{{Covariant phase space with boundaries}},
  [\href{https://arxiv.org/abs/1906.08616}{{\ttfamily 1906.08616}}].

\bibitem{Rubalcava-Garcia:2020bcp}
I.~Rubalcava-Garcia, \emph{{Constructing the theory at the boundary, its
  dynamics and degrees of freedom}},
  [\href{https://arxiv.org/abs/2003.06241}{{\ttfamily 2003.06241}}].

\bibitem{Kapustin:2010hk}
A.~Kapustin and N.~Saulina, \emph{{Topological boundary conditions in abelian
  Chern-Simons theory}},
  \href{http://dx.doi.org/10.1016/j.nuclphysb.2010.12.017}{\emph{Nucl. Phys.}
  {\bfseries B845} (2011) 393--435},
  [\href{https://arxiv.org/abs/1008.0654}{{\ttfamily 1008.0654}}].

\bibitem{Balachandran:1994ik}
A.~P. Balachandran, L.~Chandar and B.~Sathiapalan, \emph{{Duality and the
  fractional quantum Hall effect}},
  \href{http://dx.doi.org/10.1016/0550-3213(95)00122-9}{\emph{Nucl. Phys.}
  {\bfseries B443} (1995) 465--500},
  [\href{https://arxiv.org/abs/hep-th/9405141}{{\ttfamily hep-th/9405141}}].

\bibitem{Arcioni:2002vv}
G.~Arcioni, M.~Blau and M.~O'Loughlin, \emph{{On the boundary dynamics of
  Chern-Simons gravity}},
  \href{http://dx.doi.org/10.1088/1126-6708/2003/01/067}{\emph{JHEP} {\bfseries
  01} (2003) 067}, [\href{https://arxiv.org/abs/hep-th/0210089}{{\ttfamily
  hep-th/0210089}}].

\bibitem{Cano:2014pya}
J.~Cano, T.~L. Hughes and M.~Mulligan, \emph{{Interactions along an
  Entanglement Cut in 2+1D Abelian Topological Phases}},
  \href{http://dx.doi.org/10.1103/PhysRevB.92.075104}{\emph{Phys. Rev.}
  {\bfseries B92} (2015) 075104},
  [\href{https://arxiv.org/abs/1411.5369}{{\ttfamily 1411.5369}}].

\bibitem{Dong_2008}
S.~Dong, E.~Fradkin, R.~G. Leigh and S.~Nowling, \emph{Topological entanglement
  entropy in chern-simons theories and quantum hall fluids},
  \href{http://dx.doi.org/10.1088/1126-6708/2008/05/016}{\emph{Journal of High
  Energy Physics} {\bfseries 2008} (May, 2008) 016--016}.

\bibitem{Ishibashi:1988kg}
N.~Ishibashi, \emph{{The Boundary and Crosscap States in Conformal Field
  Theories}}, \href{http://dx.doi.org/10.1142/S0217732389000320}{\emph{Mod.
  Phys. Lett.} {\bfseries A4} (1989) 251}.

\bibitem{Carlip_1990}
S.~Carlip, M.~Clements, S.~DellaPietra and V.~DellaaPietra, \emph{Sewing
  polyakov amplitudes i: Sewing at a fixed conformal structure},
  \href{http://dx.doi.org/10.1007/bf02096756}{\emph{Communications in
  Mathematical Physics} {\bfseries 127} (Feb, 1990) 253--271}.

\bibitem{balach1993MCS}
A.~P. Balachandran, L.~Chandar, E.~Ercolessi, T.~R. Govindarajan and
  R.~Shankar, \emph{{Maxwell-Chern-Simons Electrodynamics on a Disk}},
  \href{http://dx.doi.org/10.1142/S0217751X94001357}{\emph{Int. J. Mod. Phys.}
  {\bfseries A09} (1993) 3417--3441},
  [\href{https://arxiv.org/abs/cond-mat/9309051}{{\ttfamily
  cond-mat/9309051}}].

\bibitem{Blasi:2010gw}
A.~Blasi, N.~Maggiore, N.~Magnoli and S.~Storace, \emph{{Maxwell-Chern-Simons
  Theory With Boundary}},
  \href{http://dx.doi.org/10.1088/0264-9381/27/16/165018}{\emph{Class. Quant.
  Grav.} {\bfseries 27} (2010) 165018},
  [\href{https://arxiv.org/abs/1002.3227}{{\ttfamily 1002.3227}}].

\bibitem{Maggiore:2018bxr}
N.~Maggiore, \emph{{Holographic reduction of Maxwell-Chern-Simons theory}},
  \href{http://dx.doi.org/10.1140/epjp/i2018-12130-y}{\emph{Eur. Phys. J. Plus}
  {\bfseries 133} (2018) 281},
  [\href{https://arxiv.org/abs/1807.09960}{{\ttfamily 1807.09960}}].

\bibitem{Andrade:2011sx}
T.~Andrade, J.~I. Jottar and R.~G. Leigh, \emph{{Boundary Conditions and
  Unitarity: the Maxwell-Chern-Simons System in $AdS_3/CFT_2$}},
  \href{http://dx.doi.org/10.1007/JHEP05(2012)071}{\emph{JHEP} {\bfseries 05}
  (2012) 071}, [\href{https://arxiv.org/abs/1111.5054}{{\ttfamily 1111.5054}}].

\bibitem{Andrade:2005ur}
T.~Andrade, M.~Banados, R.~Benguria and A.~Gomberoff, \emph{{The 2+1 charged
  black hole in topologically massive electrodynamics}},
  \href{http://dx.doi.org/10.1103/PhysRevLett.95.021102}{\emph{Phys. Rev.
  Lett.} {\bfseries 95} (2005) 021102},
  [\href{https://arxiv.org/abs/hep-th/0503095}{{\ttfamily hep-th/0503095}}].

\bibitem{Cho:2010rk}
G.~Y. Cho and J.~E. Moore, \emph{{Topological BF field theory description of
  topological insulators}},
  \href{http://dx.doi.org/10.1016/j.aop.2010.12.011}{\emph{Annals Phys.}
  {\bfseries 326} (2011) 1515--1535},
  [\href{https://arxiv.org/abs/1011.3485}{{\ttfamily 1011.3485}}].

\bibitem{Chen_2016}
X.~Chen, A.~Tiwari and S.~Ryu, \emph{Bulk-boundary correspondence in
  (3+1)-dimensional topological phases},
  \href{http://dx.doi.org/10.1103/physrevb.94.045113}{\emph{Physical Review B}
  {\bfseries 94} (Jul, 2016) }.

\bibitem{Chen_2017}
X.~Chen, A.~Tiwari, C.~Nayak and S.~Ryu, \emph{Gauging (3+1)-dimensional
  topological phases: An approach from surface theories},
  \href{http://dx.doi.org/10.1103/physrevb.96.165112}{\emph{Physical Review B}
  {\bfseries 96} (Oct, 2017) }.

\bibitem{Lin:2018xkj}
J.~Lin, \emph{{Entanglement entropy in Jackiw-Teitelboim Gravity}},
  [\href{https://arxiv.org/abs/1807.06575}{{\ttfamily 1807.06575}}].

\bibitem{Mertens:2018fds}
T.~G. Mertens, \emph{{The Schwarzian Theory - Origins}},
  [\href{https://arxiv.org/abs/1801.09605}{{\ttfamily 1801.09605}}].

\bibitem{Gonzalez:2018enk}
H.~A. Gonz{\'a}lez, D.~Grumiller and J.~Salzer, \emph{{Towards a bulk
  description of higher spin SYK}},
  \href{http://dx.doi.org/10.1007/JHEP05(2018)083}{\emph{JHEP} {\bfseries 05}
  (2018) 083}, [\href{https://arxiv.org/abs/1802.01562}{{\ttfamily
  1802.01562}}].

\bibitem{Harlow:2018tqv}
D.~Harlow and D.~Jafferis, \emph{{The Factorization Problem in
  Jackiw-Teitelboim Gravity}},
  [\href{https://arxiv.org/abs/1804.01081}{{\ttfamily 1804.01081}}].

\bibitem{Momen:1996dg}
A.~Momen, \emph{{Edge dynamics for BF theories and gravity}},
  \href{http://dx.doi.org/10.1016/S0370-2693(97)00010-5}{\emph{Phys. Lett.}
  {\bfseries B394} (1997) 269--274},
  [\href{https://arxiv.org/abs/hep-th/9609226}{{\ttfamily hep-th/9609226}}].

\bibitem{Blasi_2012}
A.~Blasi, A.~Braggio, M.~Carrega, D.~Ferraro, N.~Maggiore and N.~Magnoli,
  \emph{Non-abelian bf theory for 2 + 1 dimensional topological states of
  matter}, \href{http://dx.doi.org/10.1088/1367-2630/14/1/013060}{\emph{New
  Journal of Physics} {\bfseries 14} (Jan, 2012) 013060}.

\bibitem{Dupuis:2017otn}
M.~Dupuis, L.~Freidel and F.~Girelli, \emph{{Discretization of 3d gravity in
  different polarizations}},
  \href{http://dx.doi.org/10.1103/PhysRevD.96.086017}{\emph{Phys. Rev.}
  {\bfseries D96} (2017) 086017},
  [\href{https://arxiv.org/abs/1701.02439}{{\ttfamily 1701.02439}}].

\bibitem{Delcamp:2018sef}
C.~Delcamp, L.~Freidel and F.~Girelli, \emph{{Dual loop quantizations of 3d
  gravity}},  [\href{https://arxiv.org/abs/1803.03246}{{\ttfamily
  1803.03246}}].

\bibitem{Freidel:2018pbr}
L.~Freidel, F.~Girelli and B.~Shoshany, \emph{{2+1D Loop Quantum Gravity on the
  Edge}}, \href{http://dx.doi.org/10.1103/PhysRevD.99.046003}{\emph{Phys. Rev.}
  {\bfseries D99} (2019) 046003},
  [\href{https://arxiv.org/abs/1811.04360}{{\ttfamily 1811.04360}}].

\bibitem{Balachandran:1995iq}
A.~P. Balachandran, L.~Chandar and A.~Momen, \emph{{Edge states and
  entanglement entropy}},
  \href{http://dx.doi.org/10.1142/S0217751X97000578}{\emph{Int. J. Mod. Phys.}
  {\bfseries A12} (1997) 625--642},
  [\href{https://arxiv.org/abs/hep-th/9512047}{{\ttfamily hep-th/9512047}}].

\bibitem{Kapec:2015ena}
D.~Kapec, M.~Pate and A.~Strominger, \emph{{New Symmetries of QED}},
  [\href{https://arxiv.org/abs/1506.02906}{{\ttfamily 1506.02906}}].

\bibitem{Afshar:2019axx}
H.~Afshar, H.~A. Gonzlez, D.~Grumiller and D.~Vassilevich, \emph{{Flat space
  holography and complex SYK}},
  [\href{https://arxiv.org/abs/1911.05739}{{\ttfamily 1911.05739}}].

\bibitem{Barnich:2019xhd}
G.~Barnich, \emph{{Black hole entropy from nonproper gauge degrees of freedom:
  The charged vacuum capacitor}},
  \href{http://dx.doi.org/10.1103/PhysRevD.99.026007}{\emph{Phys. Rev.}
  {\bfseries D99} (2019) 026007},
  [\href{https://arxiv.org/abs/1806.00549}{{\ttfamily 1806.00549}}].

\bibitem{Coussaert:1995zp}
O.~Coussaert, M.~Henneaux and P.~van Driel, \emph{{The Asymptotic dynamics of
  three-dimensional Einstein gravity with a negative cosmological constant}},
  \href{http://dx.doi.org/10.1088/0264-9381/12/12/012}{\emph{Class. Quant.
  Grav.} {\bfseries 12} (1995) 2961--2966},
  [\href{https://arxiv.org/abs/gr-qc/9506019}{{\ttfamily gr-qc/9506019}}].

\bibitem{Carlip:2016lnw}
S.~Carlip, \emph{{The Dynamics of Supertranslations and Superrotations in 2+1
  Dimensions}},  [\href{https://arxiv.org/abs/1608.05088}{{\ttfamily
  1608.05088}}].

\bibitem{Speranza:2019hkr}
A.~J. Speranza, \emph{{Geometrical tools for embedding fields, submanifolds,
  and foliations}},  [\href{https://arxiv.org/abs/1904.08012}{{\ttfamily
  1904.08012}}].

\bibitem{Gelca_1997}
R.~Gelca, \emph{Topological quantum field theory with corners based on the
  kauffman bracket},
  \href{http://dx.doi.org/10.1007/s000140050013}{\emph{Commentarii Mathematici
  Helvetici} {\bfseries 72} (Sep, 1997) 216--243}.

\bibitem{tsumura20132categorical}
Y.~Tsumura, \emph{{A 2-categorical extension of the Reshetikhin-Turaev
  theory}},  [\href{https://arxiv.org/abs/1309.3630}{{\ttfamily 1309.3630}}].

\bibitem{Carqueville:2017fmn}
N.~Carqueville and I.~Runkel, \emph{{Introductory lectures on topological
  quantum field theory}},  [\href{https://arxiv.org/abs/1705.05734}{{\ttfamily
  1705.05734}}].

\bibitem{2010arXiv1004.1533K}
J.~{Kirillov}, Alexander and B.~{Balsam}, \emph{{Turaev-Viro invariants as an
  extended TQFT}}, {\emph{arXiv e-prints} (Apr, 2010) arXiv:1004.1533},
  [\href{https://arxiv.org/abs/1004.1533}{{\ttfamily 1004.1533}}].

\bibitem{Dittrich:2016typ}
B.~Dittrich and M.~Geiller, \emph{{Quantum gravity kinematics from extended
  TQFTs}}, \href{http://dx.doi.org/10.1088/1367-2630/aa54e2}{\emph{New J.
  Phys.} {\bfseries 19} (2017) 013003},
  [\href{https://arxiv.org/abs/1604.05195}{{\ttfamily 1604.05195}}].

\bibitem{Gervais:1980bz}
J.-L. Gervais and D.~Zwanziger, \emph{{Derivation From First Principles of the
  Infrared Structure of Quantum Electrodynamics}},
  \href{http://dx.doi.org/10.1016/0370-2693(80)90903-X}{\emph{Phys. Lett.}
  {\bfseries 94B} (1980) 389--393}.

\bibitem{Balachandran:2018jwf}
A.~P. Balachandran and V.~P. Nair, \emph{{An Action for the Infrared Regime of
  Gauge Theories and the Problem of Color Transformations}},
  [\href{https://arxiv.org/abs/1804.07214}{{\ttfamily 1804.07214}}].

\bibitem{Gomes:2016mwl}
H.~Gomes and A.~Riello, \emph{{The Observer's Ghost: a field-space
  connection-form and its application to gauge theories and general
  relativity}},  [\href{https://arxiv.org/abs/1608.08226}{{\ttfamily
  1608.08226}}].

\bibitem{Gomes:2018dxs}
H.~Gomes, F.~Hopfm{\"u}ller and A.~Riello, \emph{{A unified geometric framework
  for boundary charges and dressings: non-Abelian theory and matter}},
  [\href{https://arxiv.org/abs/1808.02074}{{\ttfamily 1808.02074}}].

\bibitem{Gomes:2018shn}
H.~Gomes and A.~Riello, \emph{{A Unified Geometric Framework for Boundary
  Charges and Particle Dressings}},
  [\href{https://arxiv.org/abs/1804.01919}{{\ttfamily 1804.01919}}].

\bibitem{Gomes:2019xto}
H.~Gomes and A.~Riello, \emph{{The quasilocal degrees of freedom of Yang-Mills
  theory}},  [\href{https://arxiv.org/abs/1910.04222}{{\ttfamily 1910.04222}}].

\bibitem{Gomes:2019xhu}
H.~Gomes, \emph{{Gauging the Boundary in Field-space}},
  \href{http://dx.doi.org/10.1016/j.shpsb.2019.04.002}{\emph{Stud. Hist. Phil.
  Sci.} {\bfseries B67} (2019) 89--110},
  [\href{https://arxiv.org/abs/1902.09258}{{\ttfamily 1902.09258}}].

\bibitem{Gomes:2019otw}
H.~Gomes, \emph{{Holism as the significance of gauge symmetries}},
  [\href{https://arxiv.org/abs/1910.05330}{{\ttfamily 1910.05330}}].

\bibitem{Gomes:2019rgg}
H.~Gomes and A.~Riello, \emph{{Notes on a few quasilocal properties of
  Yang-Mills theory}},  [\href{https://arxiv.org/abs/1906.00992}{{\ttfamily
  1906.00992}}].

\end{thebibliography}\endgroup
\bibliographystyle{Biblio}

\end{document}